\documentclass[floatfix,aps,twocolumn,nofootinbib,superscriptaddress]{revtex4-2}

\usepackage[utf8]{inputenc}

\usepackage[lining,semibold]{libertine} 
\usepackage[libertine, cmintegrals, bigdelims, vvarbb]{newtxmath}

\usepackage{amsmath}
\usepackage{amsfonts}
\usepackage{mathrsfs}
\usepackage{gensymb}
\usepackage{bbm}
\usepackage{dsfont}

\usepackage{kbordermatrix}
\renewcommand{\kbldelim}{(}
\renewcommand{\kbrdelim}{)}

\usepackage{chemformula}
\usepackage[caption=false]{subfig}
\usepackage{chemfig}
\usepackage[version=3]{mhchem}

\usepackage{tikz}
\usetikzlibrary{matrix,positioning,decorations.pathreplacing}

\usepackage{soul}
\usepackage{xcolor}

\usepackage{scalerel} 

\definecolor{webgreen}{rgb}{0,.5,0}
\definecolor{webbrown}{rgb}{.6,0,0}
\definecolor{grigio}{rgb}{.85,.85,.85} 
\definecolor{RoyalBlue}{rgb}{0.0, 0.14, 0.4}
\definecolor{skyblue1}{rgb}{0.45,0.62,0.81}
\definecolor{skyblue2}{rgb}{0.2,0.39,0.64}
\definecolor{skyblue3}{rgb}{0.13,0.29,0.53}
\definecolor{scarlet1}{rgb}{0.93,0.16,0.16}
\definecolor{scarlet2}{rgb}{0.8,0,0}
\definecolor{scarlet3}{rgb}{0.64,0,0}

\definecolor{g}{gray}{0.50}

\usepackage{hyperref}
\hypersetup{%
    colorlinks=true, linktocpage=true, pdfstartpage=1, pdfstartview=FitV,%
    breaklinks=true, pdfpagemode=UseNone, pageanchor=true, pdfpagemode=UseOutlines,%
    plainpages=false, bookmarksnumbered, bookmarksopen=true, bookmarksopenlevel=1,%
    hypertexnames=true, pdfhighlight=/O,
    urlcolor=webbrown, linkcolor=RoyalBlue, citecolor=webgreen, 
    pdftitle={},%
    pdfauthor={Francesco Avanzini},%
    pdfsubject={},%
    pdfkeywords={},%
    pdfcreator={pdfLaTeX},%
    pdfproducer={LaTeX REVTeX}%
}

\usepackage{amsmath,amsfonts}
\usepackage{bm}
\usepackage{mathtools}
\usepackage{graphicx}
\usepackage{printlen}
\usepackage{nicematrix} 
\usepackage{svg}
\usepackage[version=3]{mhchem}
\usepackage{chemformula}
\usepackage{mathtools}
\usepackage{siunitx}
\usepackage{blkarray}
\usepackage{eufrak}
\makeatletter
\def\maketag@@@#1{\hbox{\m@th\normalfont\normalsize#1}}
\makeatother

\DeclareMathAlphabet{\mathpzc}{OT1}{pzc}{m}{it}

\begin{document}
\title{Characterizing the conditions for indefinite growth in open chemical reaction networks}
\newcommand\unilu{\affiliation{Complex Systems and Statistical Mechanics, Department of Physics and Materials Science, University of Luxembourg, L-1511 Luxembourg City, Luxembourg}}
\newcommand\unipdchim{\affiliation{Department of Chemical Sciences, University of Padova, Via F. Marzolo, 1, I-35131 Padova, Italy}}

\author{Shesha Gopal Marehalli Srinivas}
\email{shesha.marehalli@uni.lu}
\unilu
\author{Francesco Avanzini}
\email{francesco.avanzini@unipd.it}
\unilu
\unipdchim
\author{Massimiliano Esposito}
\email{massimiliano.esposito@uni.lu}
\unilu

\begin{abstract}
The thermodynamic and dynamical conditions necessary to observe indefinite growth in homogeneous open chemical reaction networks (CRNs) satisfying mass action kinetics were presented in Srinivas et al. (2023): Unimolecular CRNs can only accumulate equilibrium concentrations of species while multimolecular CRNs are needed to produce indefinite growth with nonequilibrium concentrations. Within multimolecular CRNs, pseudo-unimolecular CRNs produce nonequilibrium concentrations with zero efficiencies. Nonequilibrium growth with finite efficiencies requires dynamically nonlinear CRNs. In this paper, we provide a detailed analysis supporting these results.
Mathematical proofs are provided for growth in unimolecular and pseudo-unimolecular CRNs. For multimolecular CRNs, four models displaying very distinctive topological properties are extensively studied, both numerically and partly analytically. 

\end{abstract}

\maketitle

\section{Introduction}

Open chemical reaction networks (CRNs) are known to exhibit rich dynamical behaviors such as nonequilibrium steady states~\cite{gaspard2004}, chemical oscillations \cite{andrieux2008}, chaotic dynamics \cite{gaspard2020},
kinetic non-invertibility~\cite{Srinivas2023a}, patterns \cite{falasco2018turing, Avanzini2020a,timur2023} and waves \cite{Avanzini2019a,kumar2021}.
Recent progress in nonequilibrium thermodynamics of open CRNs~\cite{qian2005,schmiedl2007,Polettini2014,Rao2016,Rao2018b,Avanzini2021,avanzini2022flux} nowadays allows to characterize the energetics of these complex dynamics.
In this work, we study the dynamics and energetics of indefinite chemical growth in open CRNs: an indefinite increase in the concentrations of species.
The literature on this topic has focused on the dynamics of CRNs made of irreversible reactions which prevent a consistent thermodynamics description~\cite{Lin2020,Szathmary1991,Wills1998,Iyerbiswas2014, AngeliSontag2009, Nandori2022}.
Recently, it was proven numerically that under certain chemosttating conditions, fully reversible open CRNs can undergo growth~\cite{avanzini2022flux}.
Building on this work, in a companion paper~\cite{Comp2}, we argued that growth needs an influx of species into the system at a constant rate and that growth with the accumulation of nonequilibrium concentrations of species is only possible in open multimolecular CRNs.
In this paper, we not only provide extensive analytical and numerical support for these results in a self-contained way,
but we also extend their scope by considering additional models.

The plan of the paper is as follows. 
In Sec.\ref{Sec:Supplementary_Basics} we define the dynamics and thermodynamics of open CRNs, as well as different chemostatting mechanisms (resp. concentration, flux and mixed control) and indefinite growth. 
In Sec.\ref{Sec:linear_full_SI}, we consider unimolecular CRNs. We prove that they only grow under flux control and that their concentrations always remains close to equilibrium values, implying a vanishing dissipation and an optimal efficiency of growth tending to one. 
In Sec.\ref{Sec:Pseudo_linear} we turn to pseudo-unimolecular CRNs, i.e., a subclass of multimolecular CRNs displaying linear dynamics.
We find that similarly to unimolecular CRNs, they can only grow under flux control, but unlike them, concentrations can be far from equilibrium. As a result, the dissipation scales extensively in time and the efficiency of growth goes to zero. 
In Sec.~\ref{Sec:nonlinear_full_SI}, we consider four multimolecular CRNs displaying nonlinear dynamics. 
Each of them have different topological properties and are studied (numerically and partly analytically) under the three chemosttating procedures.
We find that growing multimolecular CRNs can show striking differences in the growth dynamics compared to unimolecular and pseudo-unimolecular CRNs. Their growing concentrations scale nonlinearly with time and growth regimes depend on their initial concentrations.
Furthermore, we find that multimolecular CRNs can grow with nonequilibrium concentrations under flux control and mixed control with an efficiency strictly between zero and one. 
Conclusions are drawn in Sec.\ref{Sec:Discussion}.

\section{Chemical Reaction Networks}
\label{Sec:Supplementary_Basics}
\subsection{Dynamics}
\label{Sec:Supp_dynamics}
We consider CRNs in ideal dilute solutions.
The chemical species, labeled $\alpha \in Z$, are interconverted via elementary~\cite{Svehla1993}, reversible, mass-balanced chemical reactions $\rho$ of the form, 
\begin{equation}
      \boldsymbol{\nu}_{+\rho} \cdot \boldsymbol{\alpha} \xrightleftharpoons[-\rho]{+\rho}  \boldsymbol{\nu}_{-\rho} \cdot \boldsymbol{\alpha}\,.
      \label{eq:chemeq}
\end{equation}
Here, $\boldsymbol{\alpha} = (\dots,\alpha,\dots)^{\intercal}$ and $\boldsymbol{\nu}_{+ \rho} = (\dots,\nu_{\alpha,+ \rho},\dots)^\intercal$ (resp. $\boldsymbol{\nu}_{-\rho}= (\dots,\nu_{\alpha,- \rho},\dots)^\intercal$) is the vector collecting the stoichiometric coefficients of the forward $+\rho$ (resp. backward $-\rho$) reaction. 
The topology of CRNs is encoded in the stoichiometric matrix $\mathbb{S}$ whose columns are given by $\mathbb{S}_{\rho} = \boldsymbol{\nu}_{-\rho}-\boldsymbol{\nu}_{+\rho}$.

We consider open CRNs: some species are chemostatted, namely, they are exchanged with the surroundings.
Hence, the concentrations of the chemical species $\boldsymbol{z} = (\dots,[\alpha],\dots)^\intercal$ evolve according to the following rate equation:
\begin{equation}
     d_{t} \boldsymbol{z} = \mathbb{S}\boldsymbol{j}(\boldsymbol{z}) + \boldsymbol{I}(\boldsymbol{z})\,.
     \label{eq:req}
\end{equation}
The vector $\boldsymbol{I}(\boldsymbol{z}) =(\dots, I_\alpha,\dots)$ accounts for the chemostatting procedure by collecting the net fluxes (influx minus outflux) of each species $\alpha$ {between the CRN and} the surroundings;
$\mathbb{S}\boldsymbol{j}(\boldsymbol{z})$ accounts for the change of concentrations due to the chemical reactions. 
The vector $\boldsymbol{j}(\boldsymbol{z}) = (\dots,{j}_{\rho}(\boldsymbol{z}),\dots)^\intercal$ collects the reaction currents which are given by the difference between the forward and the backward fluxes, ${j}_{\rho}(\boldsymbol{z})= j_{+\rho}(\boldsymbol{z})-j_{-\rho}(\boldsymbol{z})$, satisfying mass action kinetics, i.e.,
\begin{equation}\label{Eq:mass_action_law}
    j_{\pm \rho}(\boldsymbol{z}) = {k_{\pm \rho}\prod_{\alpha \in Z}[\alpha]^{\nu_{\alpha,\pm \rho}}} \equiv k_{\pm \rho}\boldsymbol{z}^{\boldsymbol{\nu}_{\pm \rho}}  \,.
\end{equation}
where  we introduce the notation~$\boldsymbol{a}^{\boldsymbol{b}} = \Pi_{i}a_{i}^{b_{i}}$. 
We can partition the set of species $Z$ into the disjoint subsets of internal $X$ and chemostatted species $Y$ whose concentrations are given by $\boldsymbol{x}$ and $\boldsymbol{y}$, respectively. By definition, the concentrations of the internal species change only due to the reactions while the concentrations of chemostatted species change due to the reactions and the chemostatting procedure.
By applying the same splitting to the stoichiometric matrix
\begin{equation}
    \mathbb{S}=
    \begin{pmatrix}
        \mathbb{S}^{X} \\ 
        \mathbb{S}^{Y}\\
    \end{pmatrix}\,,
\end{equation}
the rate equation~\eqref{eq:req} can be rewritten as
\begin{align}
    d_{t}\boldsymbol{x} &= \mathbb{S}^{X}\boldsymbol{j}(\boldsymbol{z})\label{Eqn:Open_internal_dynamics}\,,\\
    d_{t}\boldsymbol{y} &= \mathbb{S}^{Y}\boldsymbol{j}(\boldsymbol{z}) + \boldsymbol{I}^{Y}(\boldsymbol{z})\,,
\label{Eqn:open_exchange_dynamics}
\end{align}
where the vector~$\boldsymbol{I}^{Y}(\boldsymbol{z})$ is the restriction of the vector~$\boldsymbol{I}(\boldsymbol{z})$ to the set of chemostatted species $Y$. 
We consider three chemostatting procedures: flux control, mixed control, and concentration control. 
Under \textit{flux control}, the influx/outflux of species is constant,
\begin{equation}\label{Eq:flux_control_defn}
\boldsymbol{I}^{Y}(\boldsymbol{z}) = \Tilde{\boldsymbol{I}}\,.
\end{equation}
Under \textit{mixed control}, the influx is constant, while the outflux is proportional to the concentration, i.e.,
\begin{equation}\label{Eq:mixed_control_defn}
  \boldsymbol{I}^{Y}(\boldsymbol{z}) = -\Tilde{\mathbb{D}}\boldsymbol{y} + \Tilde{\boldsymbol{I}}\,,  
\end{equation}
where the matrix $\Tilde{\mathbb{D}}$, called the extraction matrix, is a diagonal matrix with entries being the extraction rates $\{ k^{e}_{\alpha}\}$ with $\alpha\in Y$ and $ \Tilde{\boldsymbol{I}}\geq0$.
Under \textit{concentration control}, the concentration of the $Y$ species is held constant, implying,
\begin{equation}\label{Eq:CC_defn}
    \boldsymbol{I}^{Y}(\boldsymbol{z}) = -\mathbb{S}^{Y}\boldsymbol{j}(\boldsymbol{z}).
\end{equation}

\paragraph*{Steady state.}
Note that the rate equation~\eqref{eq:req} can admit two kinds of steady states:
Equilibrium steady states~$\boldsymbol{z}_{\text{eq}}$ that satisfy $\boldsymbol{j}\left(\boldsymbol{z}_{\text{eq}}\right) = 0$ and nonequilibrium steady states~$\boldsymbol{z}_{\text{ss}}$ that satisfy $\boldsymbol{j}\left(\boldsymbol{z}_{\text{ss}}\right) \neq 0$, but $\mathbb{S}^{X}\boldsymbol{j}(\boldsymbol{z}_{\text{ss}}) = 0$ and $\mathbb{S}^{Y}\boldsymbol{j}(\boldsymbol{z}_{\text{ss}}) + \boldsymbol{I}^{Y}(\boldsymbol{z}_{\text{ss}}) = 0$.
Closed CRNs always relax towards an equilibrium steady state. Open CRNs, if they relax towards a steady state, generally reach a nonequilibrium steady state.

\paragraph*{Coarse-grained dynamics.}
When a time scale separation arises between the dynamics of the concentrations of the internal species $\boldsymbol{x}$ and the dynamics of chemostatted species $\boldsymbol y$, the former quickly relax to a steady state $\boldsymbol x_{\text{ss}}(\boldsymbol y)$ (if it exists) determined by $\boldsymbol y$. 
The reaction currents are then given by
\begin{equation}\label{eq:ss_curr}
    \bar{\boldsymbol{j}}(\boldsymbol{y}) \equiv {\boldsymbol{j}}(\boldsymbol x_{\text{ss}}(\boldsymbol{y}), \boldsymbol{y})\,,
\end{equation}
namely, the steady state current of Eq.~\eqref{Eqn:Open_internal_dynamics}: 
$\mathbb{S}^{X}\bar{\boldsymbol{j}}(\boldsymbol{y}) = 0$.
Equation~\eqref{Eqn:open_exchange_dynamics} can then be coarse-grained~\cite{Avanzini2020b, Avanzini2023circuit} into a closed dynamical equation for the chemostatted species: 
\begin{equation}\label{Eq:Coarse_grained_dynamics}
    d_{t}\boldsymbol{y} = \mathbb{S}^{Y}\bar{\boldsymbol{j}}(\boldsymbol{y}) + \boldsymbol{I}^{Y}(\boldsymbol{y})\,.
\end{equation}
In Sec.~\ref{Sec:nonlinear_full_SI}, we will make use of this time scale separation in growing multimolecular CRNs and compare the solutions of the rate equation~\eqref{Eqn:open_exchange_dynamics}, i.e., $\boldsymbol y(t)$, with that of the coarse-grained rate equation~\eqref{Eq:Coarse_grained_dynamics}, i.e.,  $\boldsymbol y_{\text{cg}}(t)$.

\subsection{Conservation Laws}\label{Sec:conservation_laws}
The linearly-independent left null eigenvectors of the stoichiometric matrix, $\boldsymbol{\ell}^{\lambda}\cdot\mathbb{S} = 0$, are called conservation laws:
they identify parts of (or entire) molecules, called moieties, that are preserved by the reactions. 
Indeed, their concentrations, defined as $L^{\lambda} = \boldsymbol{\ell}^{\lambda}\cdot\boldsymbol{z}$, would be conserved if CRNs were closed, i.e.,  $d_{t} L^{\lambda} = \boldsymbol{\ell}^{\lambda}\cdot d_{t} \boldsymbol{z}= 0$ (using Eq.~\eqref{eq:req} with $\boldsymbol{I}(\boldsymbol{z})=0$).
Since we consider CRNs with all mass-balanced chemical reactions, 
there is always a conservation law, called the mass conservation law and denoted $\boldsymbol{\ell}^{m}$, which involves all the species: 
$\boldsymbol{\ell}^{m} = (\dots,\ell^{m}_{\alpha},\dots)$ with $\ell^{m}_{\alpha}\geq1$. 
The corresponding concentration,
\begin{equation}
    L^{m} = \boldsymbol{\ell}^{m}\cdot\boldsymbol{z}\,,
    \label{eq:massdensity}
\end{equation}
is the mass density.
In open CRNs, some moieties are exchanged with the surroundings and their corresponding concentrations are not conserved anymore:
\begin{equation}
   d_{t}L^{\lambda} = \sum_{\alpha \in Y}\ell^{\lambda}_{\alpha}~{I}_{\alpha}\neq 0\,,
\label{Eqn:Open_conserv_laws}   
\end{equation}
where we used Eq.~\eqref{eq:req} and the definition of conservation laws.
This happens when $\ell^{\lambda}_{\alpha} \neq 0$ for at least one $\alpha \in Y$.
The corresponding conservation laws are said to be broken and labeled $\{\boldsymbol\ell^{\lambda_{b}}\}$ hereafter.
In open CRNs, the mass conservation law is always broken.
On the other hand, the concentrations of some moieties might still be conserved, i.e., $d_{t}L^{\lambda} = 0$, 
which happens when $\ell^{\lambda}_{\alpha} =  0$ for all $\alpha \in Y$.
The corresponding conservation laws are said to be unbroken and labeled $\{\boldsymbol\ell^{\lambda_{u}}\}$ hereafter.

Note that the representation (or set) of the conservation laws $\{\boldsymbol\ell^\lambda\}$ is not unique. 
Different representations identify different moieties.

\subsection{Thermodynamics}
\label{sec:SI_thermo_intro}
The theory of nonequilibrium thermodynamics of CRNs is based on two assumptions.
First, all the degrees of freedom, except for the concentrations, are at equilibrium~\cite{Rao2016,Avanzini2021} 
(the temperature $T$ is set by the solvent and the solution is assumed incompressible). 
Thus, all the thermodynamic quantities have the same form as in equilibrium thermodynamics but are evaluated at nonequilibrium concentrations. 
The (vector of) chemical potentials, $\boldsymbol{\mu} = (\dots,\mu_{\alpha},\dots)^{\intercal}$, reads
\begin{equation}
    \boldsymbol{\mu} = \boldsymbol{\mu}^{0} + RT~\text{ln}\left(\boldsymbol{z}\right)\,,
\label{Eqn:chem_pot_defn}
\end{equation}
where $\boldsymbol{\mu}^{0}= (\dots,\mu^{0}_{\alpha},\dots)^{\intercal}$ is the vector of standard chemical potentials 
and $\text{ln}\left(\boldsymbol{z}\right) = (\dots,\text{ln}([\alpha]),\dots)^{\intercal}$. 
Second, the local detailed balance assumption establishes a correspondence between dynamics and thermodynamics according to
\begin{equation}
    {RT}~\text{ln}\left(\frac{k_{+\rho}}{k_{-\rho}}\right) = -\boldsymbol{\mu}^{0} \cdot \mathbb{S}_{\rho}\,.
\label{eqn:LDB_SI}    
\end{equation}
In this framework, the second law reads~\cite{Rao2016}
\begin{equation}
 {T}{\dot{\Sigma}} = {\dot{w}}_{\text{c}}-d_{t} G~\geq 0 \,.
\label{Eq:second_law_form1}    
\end{equation}
Here, $G(\boldsymbol{z})$ is the Gibbs free energy
\begin{equation}
    G(\boldsymbol{z}) = \boldsymbol{\mu} \cdot \boldsymbol{z} - RT||\boldsymbol z|| \,, 
    \label{eq:gibbs}
\end{equation}
where $||\bullet||$ denotes the 1-norm of a vector and hence $||\boldsymbol z|| = \sum_{\alpha \in Z} [\alpha]$. 
$\dot{\Sigma}$ is the entropy production rate (EPR),
\begin{equation}
\begin{split}
T{\dot{\Sigma}} = -\boldsymbol{\mu}\cdot\mathbb{S}\boldsymbol{j}(\boldsymbol{z}) = RT\sum_{\rho} j_{\rho}(\boldsymbol{z})~\text{ln}\left(\frac{j_{+\rho}(\boldsymbol{z})}{j_{-\rho}(\boldsymbol{z})}\right) \geq 0 \,,
\end{split}
\label{Eqn:EPR_two_defn}
\end{equation}
quantifying the dissipated free energy, 
and ${\dot{w}}_{\text{c}}$ is the chemical work rate,
\begin{equation}
     {\dot{w}}_{\text{c}} = \sum_{\alpha \in Y} \mu_{\alpha}{I}_{\alpha},
\label{Eqn:chemical_work_rate}     
\end{equation}
accounting for the free energy supplied from the surroundings via the chemostats.

In this paper, we exploit the decomposition of the chemical work rate into two contributions: 
the moiety work rate $\dot{w}_{\text{m}}$ quantifying the energetic cost supplied to change the concentrations of the exchanged moieties in the CRN, 
and the nonconservative work rate $\dot{w}_{\text{nc}}$ accounting for the energetic cost to drive currents across the CRN. 
This splitting is obtained by first recognizing that chemostatting a species does not always break a conservation law~\cite{Polettini2014,Rao2016,Avanzini2021}. 
Thus, we divide the set of chemostatted species $Y$ into the set of  species $Y_{p} \subseteq Y$ that break conservation laws, called potential species,
and the remaining species $Y_{f} = Y\setminus Y_{p}$, called force species.
With this identification, 
we can associate a moiety to a single $Y_p$ species so that the moiety work rate has the form
\begin{equation}
    \dot{w}_{\text{m}} = \boldsymbol{\mu}_{Y_{p}} \cdot d_{t}\boldsymbol{m}\,,
\label{eqn:moiety_work_defn}    
\end{equation}
where $\boldsymbol{\mu}_{Y_{p}}$ is the vector of chemical potentials of the $Y_{p}$ species and $\boldsymbol{m}$ is the concentration vector of the corresponding exchanged moieties. The latter is given by:
\begin{equation}\label{Eq:moiety_defn}
    \boldsymbol{m} = \big(\mathbb{L}^{{b}}_{Y_{p}}\big)^{-1}\mathbb{L}^{{b}}\boldsymbol z\,,
\end{equation}
where we introduce the matrix $\mathbb{L}^{{b}}$ whose rows are the broken conservation laws and the matrix $\mathbb{L}_{Y_{p}}^{{b}}$ that is the (invertible) submatrix of $\mathbb{L}^{{b}}$ with only the columns corresponding to the $Y_{p}$ species.
Indeed, $\boldsymbol{m}$ quantifies the concentration of the moieties defined by a specific representation of the conservation laws where $\mathbb{L}_{Y_{p}}^{{b}}$ is the identity matrix 
namely, a representation where each potential species carries only one specific moiety~\cite{avanzini2022flux}. 

On the other hand, the force species carry moieties that are already carried by the potential species. This leads to an energetic cost for chemostatting different species that carry the same moiety~\cite{Avanzini2021, avanzini2022flux} which is captured by the nonconservative work rate~ $\dot{w}_{\text{nc}} = \dot{w}_{\text{c}}-\dot{w}_{\text{m}}$, 
\begin{equation}
       \dot{w}_{\text{nc}} = \left(\boldsymbol{\mu}_{Y}-\boldsymbol{\mu}_{Y_{p}}\left(\mathbb{L}^{b}_{Y_{p}}\right)^{-1}\mathbb{L}^{b}_{Y}\right)\cdot\boldsymbol{I}^{Y} \equiv \mathcal{F}_{Y}\cdot\boldsymbol{I}^{Y}\,,
\label{Eqn:nonconserv_work_defn}     
\end{equation}
where we used Eqs.~\eqref{Eqn:chemical_work_rate},~\eqref{eqn:moiety_work_defn} and \eqref{Eq:moiety_defn}.
We emphasize that, although the summation in Eq.~\eqref{Eqn:nonconserv_work_defn} is over all the $Y$ species, since the submatrix of $\big(\mathbb{L}^{b}_{Y_{p}}\big)^{-1}\mathbb{L}^{b}_{Y}$ for the $Y_{p}$ species is the identity matrix, the entries of the vector $\mathcal{F}_{Y}$ (also called nonconservative force) are zero for all the potential species $Y_{p}$. Thus, $\dot{w}_{\text{nc}} = 0$ if only the potential species were chemostatted. However, the entries of the vector $\mathcal{F}_{Y}$ for the force species $Y_{f}$ are given by the differences between the chemical potentials of species carrying the same moiety and thus $\dot{w}_{\text{nc}} \neq 0$ when the force species are chemostatted.

By combining Eq.~\eqref{eqn:moiety_work_defn}, \eqref{Eqn:nonconserv_work_defn} with \eqref{Eq:second_law_form1},  the second law becomes:
\begin{equation}
    T\dot{\Sigma} =  {\dot{w}}_{\text{nc}} + \dot{w}_{\text{m}} - d_{t}G~\geq 0\,.
\label{Eqn:Second_law_form2}   
\end{equation}
Note that, in closed CRNs, $\dot{w}_{\text{c}} = \dot{w}_{\text m} = \dot{w}_{\text{nc}} = 0$, implying $d_{t}G = -T\dot{\Sigma} \leq 0$. 
Thus, the Gibbs free energy monotonously decreases and can be shown to act as a Lyapunov function~\cite{Rao2016}. 
At equilibrium, $d_{t}G = T\dot{\Sigma} = 0$. 
In open CRNs, at a nonequilibrium steady state, the terms $d_{t}G$ and $\dot{w}_{\text m}$ in Eq \eqref{Eqn:Second_law_form2} vanish as they contain total time derivatives while the constant EPR is balanced by the nonconservative work rate, i.e., $T\dot{\Sigma} = \dot{w}_{\text{nc}} > 0$.

\subsection{Growth} 
\label{sec:SI_growth_intro}

We define growth as a limiting state of CRNs with unbounded concentrations: 
$\lim_{t \to \infty} \rVert \boldsymbol{z}(t)-\boldsymbol{z}(0) \rVert = \infty$. 
When CRNs grow, also the mass density $L^{m}$ in Eq.~\eqref{eq:massdensity} and the Gibbs free energy $G(\boldsymbol{z})$ in Eq.~\eqref{eq:gibbs} are unbounded by definition. 
The converse is also true. 
If the mass density $L^{m}$ and the Gibbs free energy $G(\boldsymbol{z})$ are unbounded, then there is at least one species whose concentration is unbounded. 
Indeed, if all the concentrations are bounded from above by some constant $M$, i.e., $[\alpha](t) \leq M~\forall~\alpha$, then, $L^{m}(t) \leq |Z|M$ and ${G(\boldsymbol{z})}/{RT} \leq CM~\text{ln}(M)$, where $|Z|$ is the number of species and $C$ is a constant.

We are now in the position to make three general statements about growth for any CRN based on the chemostatting procedure.

First, closed CRNs cannot grow  since the mass density is conserved.

Second, we consider CRNs under mixed control such that the extraction rates of \textit{all} species are finite while the constant influx $\Tilde{\boldsymbol{I}}$ is arbitrary and may act only on a subset of species. Continuous-flow Stirred Tank Reactors (CSTR) constitute a special case where all extraction rates are the same~\cite{blokhuis2018, Liu2023}.
Then,
\begin{equation}
\begin{split}
    d_{t} L^{m} &=  -\boldsymbol{\ell}^{m}\cdot\Tilde{\mathbb{D}}\boldsymbol{z} + \boldsymbol{\ell}^{m}\cdot\Tilde{\boldsymbol{I}}\,,\\
    &=  -\sum_{\alpha \in Z} k^{e}_{\alpha}\ell^{m}_{\alpha}[\alpha] + \sum_{\beta \in Z} \ell^{m}_{\beta}\Tilde{I}_{\beta}\,, 
\end{split}    
\label{eq:mass_evol_mean}    
\end{equation}
where we used Eqs.~\eqref{Eqn:Open_conserv_laws} and \eqref{Eq:mixed_control_defn}.
Let ${k}^{e}$ be the smallest extraction rate, then
\begin{equation}
     d_{t} L^{m} \leq - {k}^{e} L^{m} +   \boldsymbol{\ell}^{m} \cdot \Tilde{\boldsymbol{I}}\,, 
\label{Eq:Mass_evol_mean2}
\end{equation}
which implies
\begin{equation}
    L^{m}(t) \leq L^{m}(0)e^{-{k}^{e}t} + \left(\boldsymbol{\ell}^{m} \cdot \Tilde{\boldsymbol{I}}\right)\frac{\left(1-e^{-{k}^{e}t}\right)}{{k}^{e}}\,.
\label{Eq:total_mass_mean_inequal}    
\end{equation}
Thus, the mass density is bounded in time.

Third, flux control always leads to growth  when $\boldsymbol{\ell}^{m}\cdot\Tilde{\boldsymbol{I}}> 0$, namely, when the net influx of species is larger than the net outflux. Indeed, the evolution equation for the mass density becomes~$d_{t} L^{m} =  \boldsymbol{\ell}^{m}\cdot\Tilde{\boldsymbol{I}}$ which  leads to the divergence of the mass density linearly in time. Consequently, the concentrations can grow at most linearly in time.

\section{Unimolecular CRNs}
\label{Sec:linear_full_SI}

CRNs are said to be unimolecular when they are exclusively composed of reactions of the form ${\alpha \xrightleftharpoons{} \beta}$.
Each vector $\boldsymbol\nu_{+\rho}$ (resp. $\boldsymbol\nu_{-\rho}$) in Eq.~\eqref{eq:chemeq} has only one non-zero entry, that is equal to $+1$, corresponding to the reactant (resp. product) of reaction $\rho$.
Unimolecular CRNs can thus be represented as graphs by mapping species into nodes and reactions into edges. 
Without loss of generality, we assume that the graphs have a single connected component. 
Physically, this means that for any two species $\alpha$ and $\beta$, there is a sequence of reactions connecting them.
This assumption is not restrictive as different components correspond to independent unimolecular CRNs.
The stoichiometric matrix $\mathbb{S}$ becomes the incidence matrix of the graph and admits only one conservation law, or equivalently one moiety, namely, the mass conservation law $\boldsymbol{\ell}^{m}$ with $\ell^{m}_\alpha =1 $ $\forall \alpha \in Z$.

\subsection{Dynamics of Closed Unimolecular CRNs}\label{Sec:linear_closed}
For unimolecular CRNs, mass action kinetics (Eq.~\eqref{Eq:mass_action_law}) implies that the reaction currents are linear functions of the concentrations, i.e.,  $\boldsymbol{j}(\boldsymbol{z}) = \Gamma \boldsymbol{z}$, where the entries of the matrix $\Gamma$ are of the form
\begin{equation}\label{Eq:conversion_mat_linear}
    \Gamma_{\rho,\alpha} = \begin{cases}
        k_{+\rho}~~\text{if}~{\nu_{\alpha,+\rho} = 1}\\
        -k_{-\rho}~\text{if}~{\nu_{\alpha,-\rho} = 1}\\
        0~ \text{else}
    \end{cases}
\end{equation}
Thus, for closed unimolecular CRNs, Eq. (\ref{eq:req}) becomes
\begin{equation}
    d_{t}\boldsymbol{z} = \mathbb{W}\boldsymbol{z}\,,
\label{Eq:Linear_evol}    
\end{equation}
with $\mathbb{W} = \mathbb{S} \Gamma$ being an irreducible rate matrix~\cite{vankampen}.
Indeed, the off-diagonal elements of $\mathbb W$ read
\begin{equation}\label{Eq:W_def}
  \mathbb{W}_{\alpha,\beta} = \sum_{\rho}\mathbb{W}^{(\rho)}_{\alpha,\beta} = \sum_{\rho}\begin{cases}
                         k_{+\rho}~\text{if}~\nu_{\alpha,-\rho} = 1{~\text{and}~\nu_{\beta,+\rho} = 1} \\
                         k_{-\rho}~\text{if}~\nu_{\alpha,+\rho} = 1{~\text{and}~\nu_{\beta,-\rho} = 1}\\
                        0 ~\text{else}
     \end{cases}
\end{equation}
where we accounted for (possibly) multiple reactions interconverting the same species,
while the diagonal elements are given by
\begin{equation}\label{Eq:W_def_diagonal}
    \mathbb{W}_{\alpha,\alpha} = -\sum_{\beta \neq \alpha} \mathbb{W}_{\beta,\alpha} \,.
\end{equation}
From the Perron Frobenius theorem, $\mathbb{W}$ admits one zero eigenvalue while all the other eigenvalues have negative real parts. 
Because of the local detailed balance condition Eq.~\eqref{eqn:LDB_SI},
all $\{\mathbb W^{(\rho)}\}$ matrices, and consequently $\mathbb W$ too, are detailed-balanced: 
the eigenvector corresponding to the zero eigenvalue $\boldsymbol{\pi}^{\text{eq}} = (\dots, {\pi}^{\text{eq}}_{\alpha}, \dots)^\intercal$ satisfies
\begin{equation}
    \mathbb{W}_{\alpha,\beta}\pi^{\text{eq}}_{\beta} =\mathbb{W}_{\beta,\alpha}\pi^{\text{eq}}_{\alpha}\,,
    \label{eq:db_linear}
\end{equation}
and must be of the form
\begin{equation}
    \boldsymbol{\pi}^{\text{eq}} = \frac{\exp(-\frac{\boldsymbol{\mu}^{0}}{RT})}{\sum_{\alpha \in Z} \exp{\frac{-\mu^{0}_{\alpha}}{RT}}}\,.
    \label{eq:eqssLCRNs}
\end{equation}
The equilibrium steady state of Eq.~\eqref{Eq:Linear_evol} can be written as
\begin{equation}\label{eq:Linear_closed_eq}    
     \boldsymbol z_{\text{eq}}=L^{m}(0) \boldsymbol{\pi}^{\text{eq}}
     = (\dots,  [\alpha]_{\text{eq}}(L^{m}(0)), \dots)^\intercal\,,
\end{equation}
since the mass density is a conserved moiety, where we also introduced the function $[\alpha]_{\text{eq}}(L^{m}(0)) \equiv L^{m}(0){\pi}^{\text{eq}}_{\alpha}$ to stress that the equilibrium concentration $[\alpha]_{\text{eq}}$ is a function of the mass density $L^{m}(0)$.

\subsection{Dynamics of Open Unimolecular CRNs}
\label{sec:SI_linear_chemo_general}
For open unimolecular CRNs under any chemostatting procedure, Eq.~\eqref{eq:req} can be recast in the form
\begin{equation}
    d_{t}\boldsymbol{a} = \mathbb{V}\boldsymbol{a} + \bar{\boldsymbol{I}}_{a}\,,
\label{Eq:linear_general_evol}    
\end{equation}
where $\boldsymbol{a}$ is the vector of the dynamical variables,  $\mathbb{V}$ is an appropriately chosen (real) matrix, and $\bar{\boldsymbol{I}}_{a}$ is a constant vector. 
For the case of flux (resp. mixed) control, Eq.~\eqref{Eq:linear_general_evol} follows from Eq.~\eqref{Eq:flux_control_defn} (resp. Eq.~\eqref{Eq:mixed_control_defn}) along with mapping the vector of concentrations $\boldsymbol{z}$ into $\boldsymbol{a}$,  the detailed balanced rate matrix $\mathbb{W}$ (resp. the matrix difference~$\mathbb{W} - \mathbb{D}$) into $\mathbb{V}$ and $\Tilde{\boldsymbol{I}}$ into $\bar{\boldsymbol{I}}_{a}$.
To show that Eq.~\eqref{Eq:linear_general_evol} can also represent concentration control (defined in~\eqref{Eq:CC_defn}), we write Eqs.~\eqref{Eqn:Open_internal_dynamics} and~\eqref{Eqn:open_exchange_dynamics} for the internal species and chemostatted species as
\begin{align}
    d_{t}\boldsymbol{x} &= \mathbb{W}_{XX}\,\boldsymbol{x} + \mathbb{W}_{XY}\,\boldsymbol{y}\,,\label{Eq:conc_control_internal}\\
    d_{t}\boldsymbol{y} &= \mathbb{W}_{YX}\,\boldsymbol{x} + \mathbb{W}_{YY}\,\boldsymbol{y} + \boldsymbol{I}^{Y}({ \boldsymbol{x}, \boldsymbol{y}})  = \boldsymbol{0}\,,
\label{Eqn:conc_control_expanded}    
\end{align}
where the matrices $\{\mathbb{W}_{XX},\mathbb{W}_{XY},\mathbb{W}_{YX},\mathbb{W}_{YY}\}$ result from applying the splitting $Z = X \cup Y$ to $\mathbb{W}$.
Since the concentrations~$\boldsymbol{y}$ are constant, only the concentrations $\boldsymbol{x}$ are dynamical variables. 
By introducing the new detailed-balanced rate matrix $\hat{\mathbb{W}}$ with off-diagonal entries
\begin{equation}
     \hat{\mathbb{W}}_{\alpha,\beta} = \mathbb{W}_{\alpha,\beta}\,,
     \label{eq:nrm}
\end{equation}
with $\alpha,\beta \in X$ and diagonal entries $\hat{\mathbb{W}}_{\alpha,\alpha} = -\sum_{\beta\in X}\hat{\mathbb{W}}_{\beta,\alpha}$,
the rate equation~\eqref{Eq:conc_control_internal} becomes
\begin{equation}
    d_{t}\boldsymbol{x} = \hat{\mathbb{W}}\boldsymbol{x}-\mathbb{D}\boldsymbol{x}+\bar{\boldsymbol{I}}\,,
\label{Eq:effective_evol_conc_control}
\end{equation}
with the diagonal matrix $\mathbb{D}$ and constant vector $\bar{\boldsymbol{I}}$ given by,
\begin{align}
    \mathbb{D} &= {\hat{\mathbb{W}} - \mathbb{W}_{XX}}\,,\label{eq:ccD}\\
    \bar{\boldsymbol{I}} &= \mathbb{W}_{XY}\,\boldsymbol{y}\,.
\end{align}
We further assume without loss of generality that $\hat{\mathbb{W}}$ is irreducible (if not, different blocks of $\hat{\mathbb{W}}$ would decouple in Eq.~\eqref{Eq:effective_evol_conc_control} and evolve independently). 
Finally, by mapping $\boldsymbol{x}$ into $\boldsymbol{a}$, $\hat{\mathbb{W}}-\mathbb{D}$ into $\mathbb{V}$ and $\bar{\boldsymbol{I}}$ into $\bar{\boldsymbol{I}}_{a}$, Eq.~\eqref{Eq:effective_evol_conc_control}
takes the form of Eq.~\eqref{Eq:linear_general_evol}.

\subsection{ No Growth Under Mixed and Concentration Control}
\label{Sec:SI_Linear_Mixed chemo}
The dynamics of open unimolecular CRNs under mixed control (resp. concentration control) is given by Eq. \eqref{Eq:linear_general_evol} with $\mathbb V = \mathbb W - \mathbb{D}$ (resp. $\hat{\mathbb W} - \mathbb{D}$), $\boldsymbol{a} = \boldsymbol{z}$ (resp. $\boldsymbol{a} = \boldsymbol{x}$) and $\bar{\boldsymbol{I}}_{a} = \Tilde{\boldsymbol{I}}$ (resp. $\bar{\boldsymbol{I}}_{a} =\bar{\boldsymbol{I}}$). In this case, we prove in Appendix~\ref{sec:spectral} that all the eigenvalues of the matrix $\mathbb{V}$ have negative real parts which also implies that $\mathbb{V}$ is invertible. Then, at long times, $\boldsymbol{a}(t)$ relaxes to the steady state $-\mathbb{V}^{-1}\bar{\boldsymbol{I}}_{a}$, as shown in Appendix \ref{Sec:Appendix_I}. We thus conclude that unimolecular CRNs, under mixed or concentration control, cannot grow.

\subsection{ Growth Under Flux Control}
\label{sec:linear_flux_control_SI}
\subsubsection{Dynamics}
\label{Sec:linear_growth_SI}

The dynamics of open unimolecular CRNs under flux control is given by Eq. \eqref{Eq:linear_general_evol} with $\mathbb V = \mathbb W$, $\boldsymbol{a} = \boldsymbol{z}$, and $\bar{\boldsymbol{I}}_{a} = \Tilde{\boldsymbol{I}}$. We solve Eq. \eqref{Eq:linear_general_evol} 
in the basis of eigenvectors of $\mathbb{W}$ (as detailed balanced rate matrices are diagonalizable~\cite{vankampen}) by introducing the invertible matrix of eigenvectors $\mathbb{P}$ such that
\begin{equation}
    \mathbb{P}^{-1}\mathbb{W}\mathbb{P} = \mathbb{\Lambda}\,,
\end{equation}
where $\mathbb{\Lambda}_{\alpha,\beta} = \delta_{\alpha,\beta}\lambda_{\alpha}$ with $\delta_{\alpha,\beta}$ being the Kronecker delta and $\{\lambda_\alpha\}$ being the eigenvalues of $\mathbb W$.
From the Perron Frobenius theorem, $\mathbb{W}$ has a unique zero eigenvalue, i.e.,  $\lambda_{1} = 0$ and the other eigenvalues satisfy $\lambda_{\alpha}< 0$  for $\alpha\neq 1$. 
Furthermore, $\mathbb{P}_{\alpha,1} = \pi^{\text{eq}}_{\alpha}$ (given in Eq.~\eqref{eq:eqssLCRNs}),
and $\mathbb{P}^{-1}_{1,\alpha} = 1$.
Then, by defining the vector $\boldsymbol{u} = \mathbb{P}^{-1}\boldsymbol{z}$, 
Eq. \eqref{Eq:linear_general_evol} becomes
\begin{equation}\label{Eq:growth_eigenvec}
    d_{t}\boldsymbol{u} = \mathbb{\Lambda}\boldsymbol{u} + \mathbb{P}^{-1}\Tilde{\boldsymbol{I}}\,,
\end{equation}
whose solution componentwise reads 
\begin{align}
    u_{1}(t) &= L^{m}(t) =  L^{m}(0) + \left(\sum_{\alpha \in Z}\Tilde{I}_{\alpha}\right)t\,,\label{Eq:growth_mass_linear}\\
    u_{\alpha}(t) &= u_{\alpha}(0)e^{\lambda_{\alpha}t} + \left(\frac{(\mathbb{P}^{-1}\Tilde{\boldsymbol{I}})_{\alpha}}{-\lambda_{\alpha}}\right)\left(1-e^{\lambda_{\alpha}t}\right)\,. 
\label{Eq:growth_eigen}    
\end{align}
Hence, the concentrations $\boldsymbol{z} = \mathbb{P}\boldsymbol{u}$ in the long time limit read
\begin{subequations}
\begin{align}
    [\alpha](t) &= \mathbb{P}_{\alpha,1}L^{m}(t) + \sum_{\beta>1}\mathbb{P}_{\alpha,\beta} \left(\frac{(\mathbb{P}^{-1}\Tilde{\boldsymbol{I}})_{\beta}}{-\lambda_{\beta}}\right)\,, \\
    & = \pi^{\text{eq}}_{\alpha}L^{m}(t) + c_{\alpha}(\Tilde{\boldsymbol{I}})\,,
\label{Eq:flux_control_general_soln}
\end{align}
\end{subequations}
where we collected all the time-independent contributions into $c_{\alpha}(\Tilde{\boldsymbol{I}})$, 
and used Eq.~\eqref{eq:Linear_closed_eq}
to identify $\pi^{\text{eq}}_{\alpha}L^{m}(t)$ as the equilibrium concentration $[\alpha]_{\text{eq}}\left(L^{m}(t)\right)$ of the corresponding closed CRN with mass density $L^{m}(t)$.
Equations~\eqref{Eq:growth_mass_linear} and~\eqref{eq:Linear_closed_eq} show that the mass density, as well as the equilibrium concentrations $[\alpha]_{\text{eq}}\left(L^{m}(t)\right)$, grow linearly in time.
This, together with Eq.~\eqref{Eq:flux_control_general_soln}, implies that also the instantaneous concentrations $[\alpha](t)$ grow linearly in time.
Indeed, the relative (resp. absolute) difference between 
$[\alpha](t)$ and $[\alpha]_{\text{eq}}\left(L^{m}(t)\right)$ decreases in time (resp. is constant).
Thus, we call this type of growth \textit{equilibrium growth}.

We now examine the reaction currents of equilibrium growth.
Let us denote the current of the reaction $\beta \xrightleftharpoons[]{+\rho}\alpha$ as $j_{\rho}$.
Then, by using Eqs.~\eqref{Eq:flux_control_general_soln} and~\eqref{eq:db_linear}, in the long time limit, the current reads
\begin{align}
         j_{\rho}(t) &=~\mathbb{W}^{(\rho)}_{\alpha,\beta}[\beta](t)- \mathbb{W}^{(\rho)}_{\beta,\alpha}[\alpha](t)\,, \nonumber\\ 
        &=~\mathbb{W}^{(\rho)}_{\alpha,\beta}c_{\beta}(\Tilde{\boldsymbol{I}})-\mathbb{W}^{(\rho)}_{\beta,\alpha}c_{\alpha}(\Tilde{\boldsymbol{I}})\equiv J_{\rho}
        \label{Eq:linear_growth_currents}\,.
\end{align}
Notice that while the concentrations grow linearly in time close to their equilibrium value, 
the reaction currents become constant and do not vanish because the difference $c_{\alpha}(\Tilde{\boldsymbol{I}}) = [\alpha](t) -[\alpha]_{\text{eq}}(L^m(t))$ is constant.

\subsubsection{Thermodynamics}
\label{sec:linear_thermo_growth_SI}
We now analyze the energetics of equilibrium growth at long times using the two formulations of the second laws given in Eqs.~\eqref{Eq:second_law_form1} and~\eqref{Eqn:Second_law_form2}, respectively. 
We start by considering the chemical potentials.
By using Eqs.~\eqref{Eq:flux_control_general_soln} and~\eqref{eq:eqssLCRNs} in Eq. \eqref{Eqn:chem_pot_defn}, the chemical potentials read 
\begin{equation}\label{Eq:chem_pot_linear_growth}
\begin{aligned}
    \mu_{\alpha}(t)     =~  &RT\text{ln}\left(L^{m}(t)\right) - RT~\text{ln}\left(\sum_{\beta\in Z}e^{\frac{-\mu^{0}_{\beta}}{{RT}}}\right)\\
     &+RT\text{ln}\left(1 + \frac{c_\alpha}{\pi^{\text{eq}}_{\alpha}L^{m}(t)}\right)\,,
\end{aligned}
\end{equation}
which grows logarithmically in time.
This allows us to write the vector of chemical potentials as

\begin{equation}\label{Eqn:chem_pot_linear_vec}
\begin{aligned}
   \boldsymbol{\mu} = &\left[RT~\text{ln}\left(L^{m}(t)\right) - RT~\text{ln}\left(\sum_{\beta \in Z}e^{\frac{-\mu^{0}_{\beta}}{{RT}}}\right)\right]\boldsymbol{\ell}^{m}\\ &+ RT\frac{1}{L^{m}(t)}\left(\frac{\boldsymbol{c}}{\boldsymbol{\pi}^{\text{eq}}}\right)\,,
\end{aligned}   
\end{equation}
 by using $\text{ln}(1+x) \approx x$ for $x\ll1$ and introducing the vector $\boldsymbol{c}/{\boldsymbol{\pi}^{\text{eq}}} = \left(\dots,{c_\alpha}/{\pi^{\text{eq}}_{\alpha}},\dots\right)^{\intercal}$. 
Thus, by plugging Eq.~\eqref{Eqn:chem_pot_linear_vec} and Eq.~\eqref{Eq:linear_growth_currents} in  Eq.~\eqref{Eqn:EPR_two_defn} and using $\boldsymbol{\ell}^{m}\cdot\mathbb{S} = 0$, the EPR becomes
\begin{equation}\label{Eqn:EPR_linear_growth}
    T\dot{\Sigma} = -RT \left(\frac{\boldsymbol{c}}{\boldsymbol{\pi}^{\text{eq}}}\cdot \mathbb{S}\boldsymbol{J}\right) \frac{1}{L^{m}(t)} = \mathcal{O}(t^{-1})\,,
\end{equation}
with $\boldsymbol{J} \equiv \big(\dots,J_{\rho},\dots\big)^{\intercal}$.
Physically, Eq.~\eqref{Eqn:EPR_linear_growth} shows a monotonic decay of the EPR in time, which results from the decreasing relative difference between the instantaneous concentrations $[\alpha](t)$ and the equilibrium concentrations $[\alpha]_{\text{eq}}\left(L^{m}(t)\right)$ (see Eq.~\eqref{Eq:flux_control_general_soln}). 
On the other hand, by using Eq.~\eqref{Eqn:chem_pot_linear_vec} in Eq.~\eqref{Eqn:chemical_work_rate} and Eq.~\eqref{eq:gibbs}, the chemical work rate  and the rate of change of the Gibbs free energy can be expressed using  $\boldsymbol{\ell}^{m}\cdot d_{t}\boldsymbol{z}=\boldsymbol{\ell}^{m}\cdot{\Tilde{\boldsymbol{I}}}$ as
\begin{align}
 \begin{aligned}\label{Eqn:chem_work_growth_linear}
     \dot{w}_{c} = &\left[RT~\text{ln}\left(L^{m}(t)\right) - RT~\text{ln}\left(\sum_{\beta\in Z}e^{\frac{-\mu^{0}_{\beta}}{{RT}}}    \right)\right] (\boldsymbol{\ell}^{m}\cdot\Tilde{\boldsymbol{I}})\\
     &+ RT\left(\frac{\boldsymbol{c}}{\boldsymbol{\pi}^{\text{eq}}}\cdot\Tilde{\boldsymbol{I}}\right)\frac{1}{L^{m}(t)} =\mathcal{O}(\text{ln}(t))\,, 
 \end{aligned}\\
\begin{aligned}\label{Eqn:gibbs_energy_change_linear_SI} 
    d_{t}G  = &\left[RT~\text{ln}\left(L^{m}(t)\right) - RT~\text{ln}\left(\sum_{\beta\in Z}e^{\frac{-\mu^{0}_{\beta}}{{RT}}}    \right)\right](\boldsymbol{\ell}^{m}\cdot \Tilde{\boldsymbol{I}})\\ &+ RT\left(\frac{\boldsymbol{c}}{\boldsymbol{\pi}^{\text{eq}}}\cdot d_{t}\boldsymbol{z}\right)\frac{1}{L^{m}(t)} = \mathcal{O}(\text{ln}(t))\,, 
\end{aligned}        
\end{align}
showing a logarithmic increase in the long time limit.
Note that the EPR in Eq.~\eqref{Eqn:EPR_linear_growth}, the chemical work in Eq.~\eqref{Eqn:chem_work_growth_linear} and the time derivative of the Gibbs free energy  in Eq.~\eqref{Eqn:gibbs_energy_change_linear_SI} satisfy the second law~\eqref{Eq:second_law_form1}.
Indeed,
\begin{equation}
  \begin{split}
      \dot{w}_c - d_{t}G = RT\left(\frac{\boldsymbol{c}}{\boldsymbol{\pi}^{\text{eq}}}\cdot(\Tilde{\boldsymbol{I}}-d_{t}\boldsymbol{z})\right)\frac{1}{L^{m}(t)}
          = T\dot{\Sigma}\,.
  \end{split}      
\end{equation}
This physically means that the chemical work is fully converted into the Gibbs free energy in the long time limit when the EPR vanishes.
Thus, the efficiency of growth, defined in analogy with Ref.~\cite{penocchio2019eff} as the ratio between the free energy stored in the system and the chemical work,
\begin{equation}\label{Eq:effiency_defn}
    \eta = \frac{d_{t}G}{\dot{w}_{c}}\,,
\end{equation}
goes to one in the long time limit and therefore equilibrium growth becomes a thermodynamically reversible process.

We now turn to the formulation of the second law given in Eq.~\eqref{Eqn:Second_law_form2}.
Since unimolecular CRNs have only one moiety, namely, the mass density, 
they admit only one potential species (hereafter labeled $p$).
Therefore, the moiety work rate~\eqref{eqn:moiety_work_defn} and the nonconservative work rate~\eqref{Eqn:nonconserv_work_defn} become
\begin{align}
\begin{split}\label{Eqn:moeity_work_linear_SI}
   \dot{w}_{\text{m}} &= \bigg[RT\text{ln}\left(L^{m}(t)\right) - RT~\text{ln}\left(\sum_{\beta\in Z}e^{\frac{-\mu^{0}_{\beta}}{{RT}}}\right)\bigg]\left(\boldsymbol{\ell}^{m}\cdot \Tilde{\boldsymbol{I}}\right)\\
   &~~~+RT\frac{c_p}{\pi^{\text{eq}}_{p}L^{m}(t)}(\boldsymbol{\ell}^{m}\cdot \Tilde{\boldsymbol{I}}) = \mathcal{O}\left(\text{ln}(t)\right)\,,
\end{split}\\
   \dot{w}_{\text{nc}}  &= RT\left[\sum_{\alpha \in Y}\left(\frac{c_\alpha}{\pi^{\text{eq}}_{\alpha}} - \frac{c_p}{\pi^{\text{eq}}_{p}}\right)\Tilde{I}_{\alpha}\right]\frac{1}{L^{m}(t)} = \mathcal{O}(t^{-1})\,,\label{Eqn:nonconserv_linear_growth}
\end{align}
where we used Eq.~\eqref{Eqn:chem_pot_linear_vec} together with the fact that $\mathbb{L}^{{b}}_{Y_{p}}$ in Eq.~\eqref{Eq:moiety_defn} is the scalar $1$.

This allows us to stress two main points.
First, the difference between $\dot{w}_{\text m}-d_{t}G$ (obtained using Eqs.~\eqref{Eqn:moeity_work_linear_SI} and~\eqref{Eqn:gibbs_energy_change_linear_SI}) is of order $\mathcal{O}(t^{-1})$ implying that all the moiety work done to change the mass density is stored in the Gibbs free energy in the long time limit.
In the same limit, $\dot{w}_{\text m}$ and $d_{t}G$ grow logarithmically in time:
\begin{equation}\label{Eq:Gibbs_moeity_linear}
    d_{t}{G} \sim \dot{w}_{\text{m}} = \mathcal{O}\left(\text{ln}\left(t\right)\right)\,.
\end{equation}
Second, the nonconservative work rate~\eqref{Eqn:nonconserv_linear_growth}, namely, the amount of free energy spent to keep CRNs out of equilibrium, decreases in time at the same rate as the EPR~\eqref{Eqn:EPR_linear_growth}:
\begin{equation}\label{Eq:EPR_worknconc_linear}
    T\dot{\Sigma} \sim \dot{w}_{\text{nc}} = \mathcal{O}\left({t}^{-1}\right)\,.
\end{equation}
in agreement with the interpretation of this process as equilibrium growth.

\subsection{Summary}

From a dynamical standpoint, growth in open unimolecular CRNs can occur only under flux control.
In this case, 
the mass density increases linearly in time;
the instantaneous concentrations approach the equilibrium steady-state concentrations defined by the growing mass density (see Eq.~\eqref{Eq:flux_control_general_soln}); 
 all  concentrations scale linearly in time.


From a thermodynamic standpoint,
the work done in increasing the mass density and the change in free energy scale similarly in time (see Eq.~\eqref{Eq:Gibbs_moeity_linear}),
while the EPR, as well as the nonconservative work, monotonically decay (see Eq.~\eqref{Eq:EPR_worknconc_linear}). 
Thus, the efficiency of growth~\eqref{Eq:effiency_defn}
 goes to one at long times, implying that growth is a thermodynamically reversible process.

\section{Pseudo-unimolecular CRNs}
\label{Sec:Pseudo_linear}
Open CRNs are here said to be pseudo-unimolecular when two conditions are satisfied.
First, the set of species $Z$ can be divided into the subsets $Z = Z_{l}\cup Y_{h}$ such that every reaction is of the form
\begin{equation}\label{Reac:PSL}
    \alpha + \sum_{\gamma \in Y_{h}} \gamma\, \nu_{\gamma,+\rho} \xrightleftharpoons[-\rho] {+\rho} \beta + \sum_{\gamma \in Y_{h}} \gamma\,\nu_{\gamma,-\rho}\,,
\end{equation}
for some $\alpha$ and $\beta$ in $Z_{l}$.
Second, the species $Y_{h}$ undergo concentration control
while the $Z_l$ species can be internal or chemostatted.

From a topological standpoint, pseudo-unimolecular CRNs can be represented as graphs (like unimolecular CRNs), by mapping only the $Z_{l}$ species into nodes and reactions into edges. 
The corresponding incidence matrix is given by the substoiochiometric matrix $\mathbb{S}^{l}$ obtained by applying the splitting $Z = Z_{l}\cup Y_{h}$ to $\mathbb{S}$, i.e., $\mathbb{S} = \left(\mathbb{S}^{l},\mathbb{S}^{h}\right)^{\intercal}$.
Notice that this implies that 
i) pseudo-unimolecular CRNs always admit the conservation law $\boldsymbol{\ell}^{ml}$ 
with $\ell^{ml}_{\alpha} = 1$ for $\alpha \in Z_{l}$ and
$\ell^{ml}_{\alpha} = 0$ for $\alpha \in Y_h$, representing the mass conservation law for the $Z_{l}$ species, and ii) $\boldsymbol \ell^{ml}$ is the \textit{only} conservation law with null entries for the $Y_h$ species. 
Thus, the matrix $\mathbb{L}$ whose rows are the conservation laws can always be written as
\begin{equation}\label{Eqn:CLaw:PSL}
\renewcommand*{\arraystretch}{1.3}
    \mathbb{L} = \begin{pmatrix}
             \boldsymbol{\ell}^{ml}\\
             \Tilde{\mathbb{L}} 
        \end{pmatrix}
  = \kbordermatrix{
     &\color{g}Z_{l} & \color{g}{Y}_{{h}}\\
    &\boldsymbol{\ell}^{ml}_{l_{\mathrm{ }}}   & \boldsymbol{0}\\
    &\Tilde{\mathbb{L}}_{l} &
    \Tilde{\mathbb{L}}_{h}} \;.
\end{equation}    
Note that there is at least one conservation law involving the $Y_h$ species, i.e., the mass conservation law of the whole CRN. Furthermore, the conservation laws involving the $Y_h$ species are always broken since the $Y_h$ species undergo concentration control.

From a dynamical standpoint, mass action kinetics~\eqref{Eq:mass_action_law} implies that the reaction currents are \textit{linear} functions of the concentrations $\boldsymbol z_l = (\dots,[\alpha],\dots)^\intercal_{\alpha\in Z_l}$ since the constant concentrations $\boldsymbol y_h = (\dots,[\alpha],\dots)^\intercal_{\alpha\in Y_h}$ can be absorbed (i.e., \textit{hidden}) in the kinetic constants.
Indeed, the reaction currents $\boldsymbol{j}$ can be written as $\boldsymbol{j} = \Gamma \boldsymbol{z}_{l}$ with the  $\boldsymbol{y}_{h}$-dependent entries of the matrix $\Gamma$ reading

\begin{equation}\label{Eq:conversion_mat_plinear} 
    \Gamma_{\rho,\alpha} = \begin{cases}
        \bar{k}_{+\rho} \equiv k_{+\rho}{\displaystyle\prod_{\beta \in Y_{h_{}}}}[\beta]^{\nu_{\beta,+\rho}}~~\text{if}~{\nu_{\alpha,+\rho} = 1}\,,\\
        -\bar{k}_{-\rho} \equiv -k_{-\rho}{\displaystyle \prod_{\beta \in Y_{h_{}}}}[\beta]^{\nu_{\beta,-\rho}}~\text{if}~{\nu_{\alpha,-\rho} = 1}\,,\\
        0~ \text{else}\,.
    \end{cases}
\end{equation}
We thus call the species $Z_l$ and $Y_h$ dynamically linear and hidden, respectively.

\subsection{Pseudo-Closed Setup} \label{Sec:Pseudo_linear_closed}
When all the $Z_l$ species are internal, the rate equation~\eqref{eq:req} for the concentrations $\boldsymbol z_l$ becomes
\begin{equation}\label{Eq:W_pL}
    d_{t}\boldsymbol{z}_{l} = \mathbb{W}\boldsymbol{z}_{l}\,,
\end{equation}
where $\mathbb{W} \equiv \mathbb{S}^{l}\Gamma$ is an irreducible rate matrix whose diagonal elements are given by Eq.~\eqref{Eq:W_def} after replacing the kinetic constants $\{k_{\pm\rho}\}$ with $\{\bar{k}_{\pm\rho}\}$ defined in Eq.~\eqref{Eq:conversion_mat_plinear}.
Since Eq.~\eqref{Eq:W_pL} is similar to the rate equation~\eqref{Eq:Linear_evol} of closed unimolecular CRNs, we call this setup pseudo-closed.
However, $\mathbb{W}$ in  Eq.~\eqref{Eq:W_pL} is, in general, nondetailed balanced due to the contribution of the hidden species in $\{\bar{k}_{\pm\rho}\}$
unlike $\mathbb{W}$ in  Eq.~\eqref{Eq:Linear_evol}. 
We consider hereafter only non-detailed balanced rate matrices $\mathbb{W}$  as the case of detail balanced rate matrices $\mathbb{W}$ is mathematically equivalent to unimolecular CRNs and is thus already examined in Sec.~\ref{Sec:linear_closed}.
 
Equation~\eqref{Eq:W_pL} together with the Perron Frobenius theorem implies that $\boldsymbol z_l$ always reaches a steady state (as $\mathbb{W}$ admits one zero eigenvalue while all the other eigenvalues have negative real parts). 
By denoting the eigenvector of $\mathbb{W}$ corresponding to the zero eigenvalue by $\boldsymbol{\pi}=(\dots,\pi_{\alpha},\dots)_{\alpha \in Z_l}^{\intercal}$, the steady state of Eq.~\eqref{Eq:W_pL} reads
\begin{equation}\label{eq:PLinear_closed_eq}    
     \boldsymbol{z}^{\text{ss}}_{l}=L^{ml}(0) \boldsymbol{\pi}
     = (\dots,  [\alpha]_{\text{ss}}({L^{ml}(0))}, \dots)^\intercal\,,
\end{equation}
with $L^{ml}(0) = \boldsymbol{\ell}^{ml}_{l}\cdot \boldsymbol z_l$. 
Unlike $\boldsymbol{z}_{\text{eq}}$ in Eq.~\eqref{eq:Linear_closed_eq}, $\boldsymbol{z}^{\text{ss}}_{l}$ is a nonequilibrium steady state since $\mathbb W$ is not detailed balanced. The corresponding steady-state reaction currents,
labeled $j_{\rho}^{\mathrm{ss}}$
for the $\rho$ reaction interconverting $\beta$ into $\alpha$,
are given by 
\begin{equation}\label{Eq:currents_PL_closed}
    j_{\rho}^{\mathrm{ss}} = \mathbb{W}^{(\rho)}_{\alpha,\beta}[\beta]_{\text{ss}}-\mathbb{W}^{(\rho)}_{\beta,\alpha}[\alpha]_{\text{ss}} = L^{ml}(0)\bar{j}_{\rho}\,, 
\end{equation}
where we used Eq.~\eqref{eq:PLinear_closed_eq} and defined $\bar{j}_{\rho} \equiv \mathbb{W}^{(\rho)}_{\alpha,\beta}\pi_{\beta}-\mathbb{W}^{(\rho)}_{\beta,\alpha}\pi_{\alpha}$.

We now briefly discuss the thermodynamics of the nonequilibrium steady state $\boldsymbol z_{\mathrm{ss}}$ in Eq.~\eqref{eq:PLinear_closed_eq} as it will be also used in Subs.~\ref{sec_opne_PL}. 
By plugging  Eq.~\eqref{Eq:currents_PL_closed} in Eq~\eqref{Eqn:EPR_two_defn} and
using $\mathbb S^l\boldsymbol j^{\mathrm{ss}} = \mathbb{W}\boldsymbol{z}^{\text{ss}}_{l} = \boldsymbol{0}$ (with $\boldsymbol j^{\mathrm{ss}}=(\dots,j_{\rho}^{\mathrm{ss}},\dots)^\intercal = L^{ml}(0)\bar{\boldsymbol{j}}$), the steady-state EPR reads
\begin{equation}\label{Eqn:EPR_PL_closed}
    T\dot{\Sigma}^{\text{ss}} = -L^{ml}(0) \boldsymbol{\mu}_{h}\mathbb{S}^{h}\bar{\boldsymbol{j}}\,,
\end{equation}
where $\boldsymbol{\mu}_{h}$ is the vector of chemical potentials of the hidden species.
By identifying the potential hidden species $Y_{ph}$ and 
the force hidden species $Y_{fh}$ (as explained in general in Sec.~\ref{sec:SI_thermo_intro}), and by applying this splitting to 
the matrix of broken conservation laws in Eq.~\eqref{Eqn:CLaw:PSL}, i.e., $\Tilde{\mathbb{L}} = (\Tilde{\mathbb{L}}_l\,, \Tilde{\mathbb{L}}_{{fh}}\,, \Tilde{\mathbb{L}}_{{ph}})$,
and to the vector of chemical potentials, i.e., $\boldsymbol{\mu}_{{h}} = (\boldsymbol{\mu}_{{fh}}\,, \boldsymbol{\mu}_{{ph}})^\intercal $, the nonconservative work rate~\eqref{Eqn:nonconserv_work_defn} reads

\begin{equation}\label{Eqn:Nonconserv_PL_closed}
\begin{split}
    \dot{w}_{\text{nc}}^{\text{ss}} &=   -L^{ml}(0)\big(\boldsymbol{\mu}_{h}-\boldsymbol{\mu}_{{ph}}\big(\Tilde{\mathbb{L}}_{{ph}}\big)^{-1}\Tilde{\mathbb{L}}_{h}\big)\cdot(\mathbb{S}^{h}\bar{\boldsymbol{j}})\\
    &= -L^{ml}(0)~(\mathcal{F}_{h}\cdot\mathbb{S}^{h}\bar{\boldsymbol{j}}) \,,
\end{split}
\end{equation}
where we also used  $\boldsymbol{I}^{Y_{h}} = - L^{ml}(0)\mathbb{S}^{h}\bar{\boldsymbol{j}}$
and Eq.~\eqref{Eq:currents_PL_closed}. 
Equation~\eqref{Eqn:Nonconserv_PL_closed} defines the nonconservative forces $\mathcal{F}_{h}$. Finally, note that $ \dot{w}_{\text{nc}}^{\text{ss}} = T\dot{\Sigma}^{\text{ss}}$ in agreement with the second law Eq.~\eqref{Eqn:Second_law_form2} (this can be directly verified by using $\Tilde{\mathbb{L}}_{h}\cdot\mathbb{S}^{h}\bar{\boldsymbol{j}} = -\Tilde{\mathbb{L}}_{l}\cdot\mathbb{S}^{l}\bar{\boldsymbol{j}} = 0 $) which implies that $\mathcal{F}_{h}\cdot\mathbb{S}^{h}\bar{\boldsymbol{j}} = \boldsymbol{\mu}_{h}\mathbb{S}^{h}\bar{\boldsymbol{j}}$.

\subsection{Fully-Open setup}\label{sec_opne_PL}

We now consider the case where some of the $Z_l$ species are chemostatted. The set $Z_l$ splits into the internal $X_l$ and chemostatted $Y_l$ dynamically linear species. 

From a topological standpoint, chemostatting the $Z_l$ species breaks the conservation law $\boldsymbol{\ell}^{ml}$. Thus, the set of potential $Y_p$ (resp. force $Y_f$) species includes both dynamically linear and hidden species, i.e.,  $Y_p = {Y_{pl}} \cup Y_{ph}$ (resp. $Y_f = {Y_{fl}} \cup Y_{fh}$). The corresponding broken conservation laws and moieties are derived in Appendix \ref{Sec:PSL_open_topology}.

\subsubsection{Dynamics}
\label{Sec:PSL_open_dynamics}
The rate equation~\eqref{Eq:W_pL} for fully open pseudo-unimolecular CRNs under any chemostatting procedure can be written in the form of Eq.~\eqref{Eq:linear_general_evol}. 
For flux control (resp. mixed control), Eq.~\eqref{Eq:linear_general_evol} follows from the definition in Eq.~\eqref{Eq:flux_control_defn} (resp. Eq.~\eqref{Eq:mixed_control_defn}), and by mapping the vector of dynamically linear concentrations $\boldsymbol{z}_{l}$ into $ \boldsymbol{a}$, the nondetailed balanced rate matrix $\mathbb{W}$ (resp. the matrix difference~$\mathbb{W} - \mathbb{D}$) into $\mathbb{V}$, and $\Tilde{\boldsymbol{I}}$ into $\bar{\boldsymbol{I}}_{a}$.

Furthermore, by following the same reasoning as in Sec.\ref{sec:SI_linear_chemo_general},  Eq.~\eqref{Eq:linear_general_evol} can also represent concentration control by  mapping $\boldsymbol{x}$ into $\boldsymbol{a}$, $\hat{\mathbb{W}}-\mathbb{D}$ into $\mathbb{V}$, and $\bar{\boldsymbol{I}}$ into $\bar{\boldsymbol{I}}_{a}$.
We emphasize that the key difference between the dynamics (and thermodynamics) of open unimolecular CRNs and fully open pseudo-unimolecular CRNs is that the matrices  $\mathbb{W}$ and $\hat{\mathbb{W}}$ are  detailed balanced in the former case and nondetailed balanced in the latter case (see the discussion below Eq.~\eqref{Eq:W_pL}). 

\subsubsection{No Growth Under Mixed and Concentration Control}\label{subss:pseudo_dynamics_cc}

The dynamics of fully open pseudo-unimolecular CRNs under mixed control (resp. concentration control) is given by Eq. \eqref{Eq:linear_general_evol} with $\mathbb V = \mathbb W - \mathbb{D}$ (resp. ${\mathbb V = } \hat{\mathbb W} - \mathbb{D}$), $\boldsymbol{a} = \boldsymbol{z}_{l}$ (resp. $\boldsymbol{a} =\boldsymbol{x}_{l}$), and $\bar{\boldsymbol{I}}_{a} = \Tilde{\boldsymbol{I}}$ (resp. $\bar{\boldsymbol{I}}_{a} =\bar{\boldsymbol{I}}$).
According to Appendix~\ref{sec:spectral},
all the eigenvalues of $\mathbb{V}$ have negative real parts and $\mathbb{V}$ is invertible. 
Thus, in the long time limit, the vector $\boldsymbol{a}(t)$  relaxes to the steady state $-\mathbb{V}^{-1}\bar{\boldsymbol{I}}_{a}$ (see Appendix~\ref{Sec:Appendix_I}) implying that open pseudo-unimolecular CRNs do not grow if the dynamically linear species are chemostatted via mixed control or concentration control, similarly to unimolecular CRNs under mixed and concentration control (see Sec.~\ref{Sec:SI_Linear_Mixed chemo}).

\subsubsection{Growth Under Flux Control}\label{subss:pseudo_dynamics}

The dynamics of fully open pseudo-unimolecular CRNs under flux control is given by Eq.~\eqref{Eq:linear_general_evol} with $\mathbb V = \mathbb W$, $\boldsymbol{a} = \boldsymbol{z}$ and $\bar{\boldsymbol{I}}_{a} = \Tilde{\boldsymbol{I}}$. 
We assume here for simplicity
that $\mathbb{W}$ is diagonalizable
but the same dynamic and thermodynamic behavior emerges even for nondiagonalizable $\mathbb{W}$ as shown in Appendix~\ref{Sec:Appendix_NI}.  
By exactly following the same approach as in  Sec.\ref{Sec:linear_growth_SI},
we obtain that the mass density of the dynamically linear species $L^{ml}(t)$ reads
\begin{equation}\label{Eqn:growth_mass_PSL}
    L^{ml}(t) = L^{ml}(0) + \left(\sum_{\alpha \in Z_l}\Tilde{I}_{\alpha}\right)t\,,
\end{equation}
which is of the form of Eq.~\eqref{Eq:growth_mass_linear} and that
the concentrations in the long time limit read
\begin{subequations}
\begin{align}
    [\alpha](t) &= \mathbb{P}_{\alpha,1}L^{ml}(t) + \sum_{\beta>1}\mathbb{P}_{\alpha,\beta} \left(\frac{(\mathbb{P}^{-1}\Tilde{\boldsymbol{I}})_{\beta}}{-\lambda_{\beta}}\right)\,, \label{Eq:flux_control_PSL_general_soln_a}\\
    &= \pi_{\alpha}L^{ml}(t) + c_{\alpha}{(\Tilde{\boldsymbol{I}})} \,,
\label{Eq:flux_control_PSL_general_soln}
\end{align}
\end{subequations}
where $\mathbb{P}$ is the invertible matrix of eigenvectors of $\mathbb{W}$ written  by ordering the eigenvalues as $\lambda_{1} = 0$ and $\text{Re}(\lambda_{\alpha})< 0$  for $\alpha >1$, and the convention $\mathbb{P}_{\alpha,1} = \pi_{\alpha}$ and $\mathbb{P}^{-1}_{\alpha,1} = 1$. The time-independent contributions on the right-hand side of  Eq.~\eqref{Eq:flux_control_PSL_general_soln_a} are collected in $c_{\alpha}{(\Tilde{\boldsymbol{I}})}$ in Eq.~\eqref{Eq:flux_control_PSL_general_soln}. By using Eq.~\eqref{eq:PLinear_closed_eq}, we identify $\pi_{\alpha}L^{ml}(t)$ as the {nonequilibrium steady state} $[\alpha]_{\text{ss}}(L^{ml}(t))$ of the corresponding pseudo-closed CRN  with mass density $L^{ml}(t)$.  Equations~\eqref{Eqn:growth_mass_PSL} and~\eqref{eq:PLinear_closed_eq} show that the mass density of the dynamically linear species, $L^{ml}(t)$, as well as the steady state concentrations $[\alpha]_{\text{ss}}\left(L^{ml}(t)\right)$ grow linearly in time. This together with Eq.~\eqref{Eq:flux_control_PSL_general_soln} implies that the instantaneous concentrations $[\alpha](t)$ also grow linearly in time. 
Since the relative (resp. absolute) difference between 
$[\alpha](t)$ and the nonequilibrium steady state $[\alpha]_{\text{ss}}\left(L^{ml}(t)\right)$ decreases in time (resp. is constant) (see Eq.~\eqref{Eq:flux_control_PSL_general_soln},), we call this type of growth \textit{nonequilibrium growth}.

We now examine the corresponding reaction currents.
By replacing the steady-state concentrations in Eq.~\eqref{Eq:currents_PL_closed} with the 
concentrations in Eq.~\eqref{Eq:flux_control_PSL_general_soln},
the reaction current $j_{\rho}(t)$  in the long time limit reads
\begin{equation}\label{Eq:currents_PL_growth}
\begin{split}
   j_{\rho}(t)  &= \bar{j}_{\rho}L^{ml}(t) + \left(\mathbb{W}^{(\rho)}_{\alpha\beta}c_{\beta}-\mathbb{W}^{(\rho)}_{\beta\alpha}c_{\alpha}\right)\\
                         &= \bar{j}_{\rho}L^{ml}(t) + J_{\rho} = \mathcal{O}(t)\,, 
\end{split}
\end{equation}
where $J_{\rho} \equiv \mathbb{W}^{(\rho)}_{\alpha\beta}c_{\beta}-\mathbb{W}^{(\rho)}_{\beta\alpha}c_{\alpha}$. We note that in nonequilibrium growth both the dynamically linear concentrations and reaction currents grow linearly in time in contrast to equilibrium growth where only the former grow in time (compare Eqs.\eqref{Eq:flux_control_PSL_general_soln} and~\eqref{Eq:currents_PL_growth} with Eqs~\eqref{Eq:flux_control_general_soln} and~\eqref{Eq:linear_growth_currents}, respectively).

\subsubsection{Thermodynamics Under Flux Control}
\label{Sec:PSL_Flux_thermo}
We now analyze the thermodynamics of nonequilibrium growth. 
In the long time limit, by using Eq.~\eqref{Eq:flux_control_PSL_general_soln} in Eq. \eqref{Eqn:chem_pot_defn}, the chemical potentials of the dynamically linear species are given by
\begin{equation}\label{Eq:chempot_growth_PSL}
\begin{split}
    \mu_{\alpha} =~  &RT~\text{ln}\left(L^{ml}(t)\right) + RT~\text{ln}\left(\pi_{\alpha}e^{\frac{\mu^{0}_{\alpha}}{RT}}\right)\\
    &+ RT~\text{ln}\left(1 + \frac{c_\alpha}{\pi_{\alpha}L^{ml}(t)}\right)\,,
\end{split}    
\end{equation}
and are logarithmically increasing in time. 
The corresponding vector of chemical potentials $\boldsymbol{\mu}_{l}$ reads
\begin{equation}
\begin{split}
        \boldsymbol{\mu}_{l} =~ &RT~\text{ln}\left(L^{ml}(t)\right)\boldsymbol{\ell}^{ml}_{l} + RT~\text{ln}\left(\boldsymbol{\pi}\exp{ \frac{\boldsymbol{\mu^{0}}}{RT}}\right)\\
        &+ RT\frac{1}{L^{ml}(t)}\left(\frac{\boldsymbol{c}}{\boldsymbol{\pi}}\right)\,,\label{Eq:chempot_vec_growth_PSL}
\end{split}
\end{equation}
where we used $\text{ln}(1+x) \approx x$ for $x\ll1$, and introduced $\text{ln}\left(\boldsymbol{\pi}\exp{{\boldsymbol({\boldsymbol{\mu}^{0}}}/{RT}})\right) = \left(\dots,~\text{ln}\left(\pi_{\alpha}\exp({{\mu^{0}_{\alpha}}/{RT}})\right),\dots\right)^{\intercal}$ and ${\boldsymbol{c}}/{\boldsymbol{\pi}}= (\dots,c_{\alpha}/\pi_{\alpha},\dots)^{\intercal} $.
On the other hand, the chemical potentials of the hidden species $\boldsymbol{\mu}_{h}$ are constant as their concentrations are kept fixed.

By plugging Eqs.~\eqref{Eq:chempot_vec_growth_PSL} and \eqref{Eq:currents_PL_growth} in Eq.~\eqref{Eqn:EPR_two_defn} and using $\mathbb{S}^{l}\boldsymbol{j} = \mathbb{W}\boldsymbol{z}_{l} = \mathbb{W}\boldsymbol{c} $ (see Eq.~\eqref{Eq:flux_control_PSL_general_soln}), we find that the EPR at long times can be written as
\begin{equation}
\begin{split}
    T\dot{\Sigma} =  &\overbrace{L^{ml}(t)(-\boldsymbol{\mu}_{h}\mathbb{S}^{h}\bar{\boldsymbol{j}})}^{= T\dot{\Sigma}^{\text{ss}}(L^{ml}(t))}
    -\boldsymbol{\mu}_{h}\mathbb{S}^{h}{\boldsymbol{J}}\\ &- RT~\text{ln}\left(\boldsymbol{\pi}\exp{\frac{\boldsymbol{\mu^{0}}}{RT}}\right)\cdot\mathbb{W}\boldsymbol{c} = \mathcal{O}(t)\,,\label{Eq:PSL_EPR_growth}\\
\end{split}
\end{equation}
where $\boldsymbol{J} = \big(\dots,J_{\rho},\dots\big)^{\intercal}$ and we used Eq.~\eqref{Eqn:EPR_PL_closed} to identify the leading order term on the right hand side as the steady state EPR to which the corresponding pseudo-closed CRN with mass density $L^{ml}(t)$ would relax to. Equation~\eqref{Eq:PSL_EPR_growth} implies that nonequilibrium growth is a dissipative process with a monotonically increasing EPR resulting from the growing mass density of the dynamically linear species in Eq.~\eqref{Eqn:growth_mass_PSL}. 

By putting Eq.~\eqref{Eq:chempot_vec_growth_PSL}, $\boldsymbol{I}^{Y_l} = \Tilde{\boldsymbol{I}}$ and $\boldsymbol{I}^{Y_h} = -\mathbb{S}^{h}\boldsymbol{j}$ in Eq.~\eqref{Eqn:chemical_work_rate}, we find that the chemical work rate increases linearly in the long time limit:
\begin{align}
           \dot{w}_{\text{c}} =  &\overbrace{L^{ml}(t)(-\boldsymbol{\mu}_{h}\mathbb{S}^{h}\bar{\boldsymbol{j}}) -\boldsymbol{\mu}_{h}\mathbb{S}^{h}{\boldsymbol{J}} \vphantom{RT~\text{ln}\Big(\boldsymbol{\pi}\exp{\frac{\boldsymbol{\mu^{0}}}{RT}}\Big)\cdot\Tilde{\boldsymbol{I}}}}^{= \dot{w}_{\mathrm{h}}}\label{Eq:Chem_work_PSL_growth}\\ &+ \overbrace{{RT~\text{ln}\big(L^{ml}(t)\big)(\boldsymbol{\ell}^{ml}_{l}\cdot\Tilde{\boldsymbol{I}})
           + RT~\text{ln}\Big(\boldsymbol{\pi}\exp{\frac{\boldsymbol{\mu^{0}}}{RT}}\Big)\cdot\Tilde{\boldsymbol{I}}}}^{= \dot{w}_{\mathrm{l}}} = \mathcal{O}(t)\notag\,,
\end{align}
where we also split the chemical work rate into the hidden work rate $\dot{w}_{\mathrm{h}} \equiv \boldsymbol{\mu}_{h}\cdot \boldsymbol{I}^{Y_h}$ and the linear work rate $\dot{w}_{\mathrm{l}} \equiv \boldsymbol{\mu}_{l} \cdot \Tilde{\boldsymbol{I}}$. The hidden (resp. linear) work rate quantifies the contribution of the chemostatted hidden (resp. linear) species to the total chemical work rate. Furthermore, by using 
Eq.~\eqref{Eq:chempot_vec_growth_PSL} and $d_{t}\boldsymbol{y}_{h} = \boldsymbol{0}$ in Eq.~\eqref{eq:gibbs}, we find that $d_{t}G$ increases logarithmically in the long time limit:
\begin{equation}
\begin{split}
 d_{t}G  &= RT~\text{ln}\left(L^{ml}(t)\right)(\boldsymbol{\ell}^{ml}_{l}\cdot\Tilde{\boldsymbol{I}})\\
 &{}~~~+ RT~\text{ln}\left(\boldsymbol{\pi}\exp{\frac{\boldsymbol{\mu^{0}}}{RT}}\right)\cdot d_{t}\boldsymbol{z}_{l} = \mathcal{O}(\text{ln}(t))\label{Eq:Gibbs_PSL_growth}\,,   
\end{split}      
\end{equation}
where we used $\boldsymbol{\ell}^{ml}_{l}\cdot\Tilde{\boldsymbol{I}} = \boldsymbol{\ell}^{ml}_{l}\cdot d_{t}\boldsymbol{z}_l$ (see Eqs.~\eqref{Eq:flux_control_PSL_general_soln} and \eqref{Eqn:growth_mass_PSL}). Thus, by using Eqs.~\eqref{Eq:Gibbs_PSL_growth} and \eqref{Eq:Chem_work_PSL_growth}  in Eq.~\eqref{Eq:effiency_defn}, we find that the efficiency of nonequilibrium growth~Eq.\eqref{Eq:effiency_defn} goes to zero in the long time limit,
\begin{equation}\label{Eq:PSL_FC_efficiency}
    \lim_{t \to \infty} \eta = \lim_{t \to \infty}  \frac{RT~\text{ln}\left(L^{ml}(t)\right)}{L^{ml}(t)} = 0\,,
\end{equation}
implying that nonequilibrium growth is a thermodynamically irreversible process in contrast to equilibrium growth (see Subs.~\ref{sec:linear_thermo_growth_SI}).

We now turn to the moiety work in Eq.~\eqref{eqn:moiety_work_defn} and the nonconservative work in Eq.~ \eqref{Eqn:nonconserv_work_defn}.
Substituting Eqs.~\eqref{Eq:chempot_growth_PSL} and~\eqref{Eq:Moiety_PSL_opensetup} in Eq.~\eqref{eqn:moiety_work_defn}, the moiety work rate reads
\begin{equation}\label{Moeity_work_PSL_growth}
\begin{aligned}[b]
      \dot{w}_{\text m} =~ &RT~\text{ln}\left(L^{ml}(t)\right)(\boldsymbol{\ell}^{ml}_{l}\cdot\Tilde{\boldsymbol{I}})\\
      &+ RT~\text{ln}\left(\pi_{pl}\exp{\frac{\mu^{0}_{pl}}{RT}}\right)(\boldsymbol{\ell}^{ml}_{l}\cdot\Tilde{\boldsymbol{I}})\\
      &+ \boldsymbol{\mu}_{ph} \cdot \left(\overline{\mathbb{L}}_{X}\boldsymbol{\pi}_{X} + \overline{\mathbb{L}}_{fl}\boldsymbol{\pi}_{fl}\right)(\boldsymbol{\ell}^{ml}_{l}\cdot\Tilde{\boldsymbol{I}}) = \mathcal{O}(\text{ln}(t))\,,
\end{aligned}
\end{equation}
where we used $d_{t}\boldsymbol{x} = \boldsymbol{\pi}_{X}(\boldsymbol{\ell}^{ml}_{l}\cdot\Tilde{\boldsymbol{I}})$ and $d_{t}\boldsymbol{y}_{fl} = \boldsymbol{\pi}_{fl}(\boldsymbol{\ell}^{ml}_{l}\cdot\Tilde{\boldsymbol{I}})$ from Eq.~\eqref{Eq:flux_control_PSL_general_soln} and the matrices $\overline{\mathbb{L}}_{X}$, $\overline{\mathbb{L}}_{fl}$ given in Eq.~\eqref{Eq:Moiety_PSL_opensetup}. 
Note that the leading order term in Eq.~\eqref{Moeity_work_PSL_growth} is identical to the leading order terms of both $\dot{w}_{\mathrm{l}}$ in Eq.~\eqref{Eq:Chem_work_PSL_growth} and $d_{t}G$ in Eq.~\eqref{Eq:Gibbs_PSL_growth}. Namely,
\begin{equation}\label{Eq:Gibbs_moiety_PL}
   \dot{w}_{\mathrm{l}} \sim \dot{w}_{\text{m}} \sim d_{t}G = \mathcal{O}(\text{ln}(t))\,,
\end{equation}
which implies that the linear work, as well as the moiety work, is mainly converted into free energy. 
This results from the fact that the concentrations of the $Y_{h}$ species are constant in time and, therefore, they do not contribute to $d_{t}G$ (see Eq.~\eqref{eq:gibbs}). Only the concentrations of the dynamically linear species contribute to $d_{t}G$, implying that the linear work is mainly converted into free energy. Similarly, as the leading order contribution to $\dot{w}_{\text{m}}$  is the work done in changing $L^{ml}(t)$ (see Eq.~\eqref{Moeity_work_PSL_growth}), which does not involve any $Y_{h}$ species, the moiety work rate and the linear work rate scale similarly in time.

By using Eq.~\eqref{Eq:Broken_claw_PSL} and Eq.~\eqref{Eq:Yp_inv_PSL}, the nonconservative forces defined in Eq.~\eqref{Eqn:nonconserv_work_defn} become
\small
\begin{equation}\label{Eq:nonconforce_PSL_growth}
\mathcal{F}_{Y} = \kbordermatrix{    
     &\color{g}{Y}_{{pl}} & \color{g}{Y}_{fl} &\color{g}{Y}_{ph} &\color{g}{Y}_{fh} \\ 
      & 0~~ & \boldsymbol{\mu}_{fl} -\mu_{pl} (\ell^{ml}_{pl})^{-1} \boldsymbol{\ell}^{ml}_{fl} - \boldsymbol{\mu}_{ph}\cdot\overline{\mathbb{L}}_{fl}  &~~ \boldsymbol{0} &~~ \mathcal{F}_{h}}
\end{equation}
\normalsize
where $\mathcal{F}_{h}$ was defined in Eq.~\eqref{Eqn:Nonconserv_PL_closed}. By using the Eqs.~\eqref{Eq:currents_PL_growth}, \eqref{Eq:chempot_growth_PSL} and~\eqref{Eq:nonconforce_PSL_growth} in Eq.~\eqref{Eqn:nonconserv_work_defn}, the nonconservative work rate reads
\begin{align}
       \dot{w}_{\text{nc}} = &\overbrace{-L^{ml}(t)~(\mathcal{F}_{h}\cdot\mathbb{S}^{h}\bar{\boldsymbol{j}})}^{=\dot{w}_{\text{nc}}^{\text{ss}}(L^{ml}(t))}
       -\mathcal{F}_{h}\cdot\mathbb{S}^{h}{\boldsymbol{J}} \label{eq:nonconserv_work_PSL_open}\\
       &+\left(RT\text{ln}\left(\frac{\boldsymbol{\pi}_{fl}}{\pi_{pl}}\exp{\frac{\boldsymbol{\mu^{0}_{fl}}-\mu^{0}_{pl}}{RT}} \right) - \boldsymbol{\mu}_{ph}\cdot\overline{\mathbb{L}}_{fl}\right) \cdot \Tilde{\boldsymbol{I}}  = \mathcal{O}(t)\notag\,,
\end{align}
where 
\begin{align}
&RT\text{ln}(\boldsymbol{\pi}_{fl}/{\pi_{pl}}\exp{((\boldsymbol{\mu}^{0}_{fl}-\mu^{0}_{pl})/{RT})})\\
&=(\dots, RT\text{ln}(\pi_{\alpha}/{\pi_{pl}}\exp{((\mu^{0}_{\alpha}-\mu^{0}_{pl})/{RT})}),\dots)^{\intercal} \nonumber
\end{align}
for ${\alpha \in Y_{fl}}$ and we used Eq.~\eqref{Eqn:Nonconserv_PL_closed} to identify $\dot{w}_{\text{nc}}^{\text{ss}}(L^{ml}(t))$, namely, the nonconservative work at the steady state to which the corresponding pseudo-closed pseudo-unimolecular CRN with mass density $L^{ml}(t)$ would relax.
Note that the leading order terms of $\dot{w}_{\text{nc}}$ in Eq.~\eqref{eq:nonconserv_work_PSL_open}, the EPR in  Eqs.~\eqref{Eq:PSL_EPR_growth} and $\dot{w}_{\mathrm{h}}$ in Eq.~\eqref{Eq:Chem_work_PSL_growth} are equal. Namely, 
\begin{equation}\label{Eq:EPR_work_nconc_PSL}
  \dot{w}_{\mathrm{h}} \sim \dot{w}_{\text{nc}} \sim T\dot{\Sigma}  = \mathcal{O}(t)\,,
\end{equation}
which implies that the hidden work, as well as the nonconservative work rate, is mainly dissipated.

\subsection{Summary}
From a dynamical standpoint, growth in fully open pseudo-unimolecular CRNs
resembles growth in unimolecular CRNs.
Like in unimolecular CRNs,
it can occur only under flux control;
the mass density (of dynamically linear species) increases linearly in time;
the instantaneous concentrations approach the steady-state concentrations defined by the growing mass density (compare Eqs.~\eqref{Eq:flux_control_PSL_general_soln} and \eqref{Eq:flux_control_general_soln});
the concentrations of all (dynamically linear) species scale linearly in time.
However, the steady state is a nonequilibrium (resp. equilibrium) one 
for pseudo-unimolecular CRNs (resp. unimolecular CRNs). 


From a thermodynamic standpoint,
the work done in increasing the mass density and the change in free energy scale similarly in time like in unimolecular CRNs (compare Eqs.~\eqref{Eq:Gibbs_moiety_PL} and~\eqref{Eq:Gibbs_moeity_linear}).
Furthermore, the nonconservative work and the EPR scale similarly in time like in unimolecular CRNs (compare Eqs.~\eqref{Eq:EPR_work_nconc_PSL} and~\eqref{Eq:EPR_worknconc_linear}).
 However, pseudo-unimolecular CRNs grow out of equilibrium with the EPR linearly increasing in time (see Eq.~\eqref{Eq:PSL_EPR_growth}), 
whereas unimolecular CRNs grow close to equilibrium with the EPR monotonically decaying in time (see Eq.~\eqref{Eqn:EPR_linear_growth}). 

Unlike for unimolecular CRNs, the efficiency of growth~\eqref{Eq:PSL_FC_efficiency} goes to zero for pseudo-unimolecular CRNs, implying that growth is a thermodynamically irreversible process.

\section{Multimolecular CRNs} \label{Sec:nonlinear_full_SI}

\begin{figure*}[h]
    \centering
    \includegraphics[width = \textwidth]{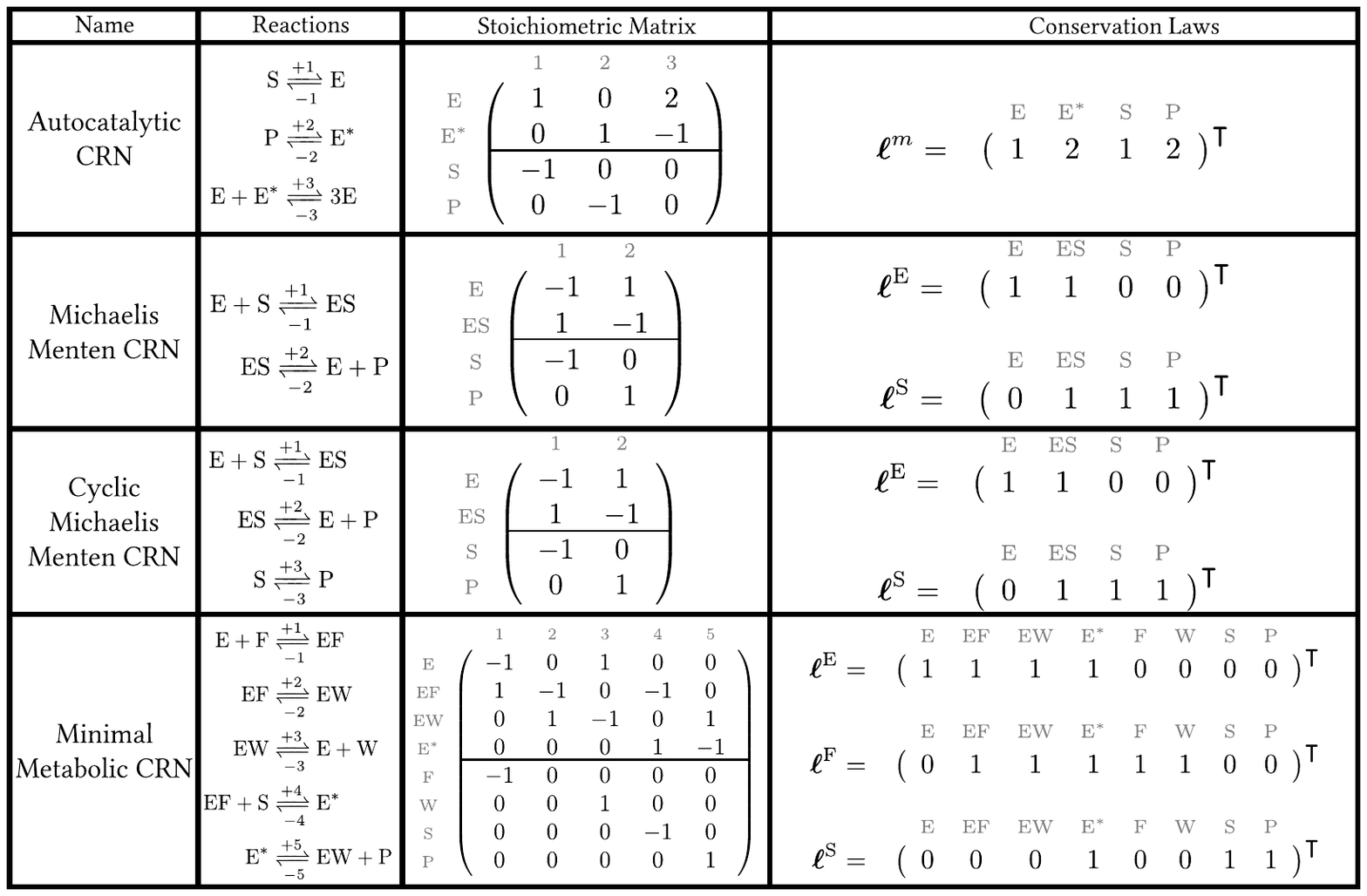}
    \caption{
    The CRNs studied in Sec.~\ref{Sec:nonlinear_full_SI}, 
    their stoichiometric matrices (with the horizontal line splitting $\mathbb S$ into $\mathbb S^X$ and $\mathbb S^Y$), 
    and their conservation laws.
    i) The autocatalytic CRN represents the interconversion of the substrate \ch{S} into the product \ch{P} promoted by the autocatalytic reaction between the enzymes \ch{E} and $\ch{E}^{*}$;
    ii) The Michaelis Menten CRN represents the interconversion of the substrate \ch{S} into the product \ch{P} promoted by the enzymes \ch{E} and \ch{ES};
    iii) The Cyclic Michaelis Menten CRN represents the same enzymatic mechanism as the Michaelis Menten CRN, but also includes a reaction directly interconverting \ch{S} into \ch{P};
    iv) The minimal metabolic CRN represents the enzymatic interconversion of the substrate \ch{S} into the product \ch{P} powered by the interconversion of the fuel \ch{F} into the waste \ch{W} and promoted by the enzymes \ch{E}, \ch{EF}, \ch{EW}, and $\ch{E}^{*}$.}
    \label{fig:eg_list}
\end{figure*}

CRNs are said to be multimolecular when at least one chemical reaction involves two or more species that do not undergo concentration control. Because of mass action kinetics~\eqref{Eq:mass_action_law}, the corresponding rate equation~\eqref{eq:req} is nonlinear and does not admit in genreal analytical solutions.  
Hence, we will focus on the four multimolecular CRNs defined in Fig.~\ref{fig:eg_list} and use both analytical methods and numerical simulations to investigate growth
under  the three different chemostatting procedures.
These CRNs have been chosen because of their different topological properties.
The autocatalytic CRN has one conservation law corresponding to the mass density, $L^{m} = [\ch{E}] + 2[\ch{E}^{*}] + [\ch{S}] + 2[\ch{P}]$. 
The Michaelis Menten CRN and the cyclic Michaelis Menten CRN have two conservation laws corresponding to the total concentration of the enzyme $L^{\ch{E}} = [\ch{E}] + [\ch{ES}]$ and total concentration of the substrate $L^{\ch{S}} = [\ch{ES}] + [\ch{S}] + [\ch{P}]$.
The minimal metabolic CRN has three conservation laws corresponding to 
the total concentration of enzyme, $L^{\ch{E}} = [\ch{E}] + [\ch{EF}] + [\ch{EW}] + [\ch{E}^{*}]$, the total concentration of fuel, $L^{\ch{F}} =  [\ch{EF}] + [\ch{EW}]+[\ch{E}^{*}] + [\ch{F}] + [\ch{W}]$, and the total concentration of substrate, $L^{\ch{S}} = [\ch{E}^{*}] + [\ch{S}] + [\ch{P}]$.

For all numerical simulations in the following, we rescale time, concentrations, energy/time respectively by i) $1/k_{+1}$, $k_{+1}/k_{+3}$, $RT k_{+1}^{2}/k_{+3}$ for the autocatalytic CRN, and by ii) $1/k_{-1}$, $k_{-1}/k_{+1}$, $RT k_{-1}^{2}/k_{+1}$ for the Michaelis Menten, the Cyclic Michaelis Menten and the Minimal Metabolic CRNs. Furthermore, we disregard initial transients and focus on long-time dynamics only.

\subsection{No Growth Under Concentration Control} \label{Sec:nonlinear_CC}

Growth in open CRNs under concentration control has been studied in the mathematical literature. 
It has been formally proven that the rate equation~\eqref{eq:req} (with fluxes satisfying mass action kinetics) does not admit solutions with unbounded concentrations for weakly reversible~\cite{feinberg_2019} single linkage class~\cite{anderson2011} CRNs, as well as strongly endotactic~\cite{manoj2014} CRNs.
This has also been conjectured to hold for any weakly reversible CRNs~\cite{feinberg_2019}, but a general proof is still missing.

We now confirm this conjecture for the autocatalytic CRN in Fig.~\ref{fig:eg_list}. The other three CRNs are pseudo-unimolecular CRNs under concentration control and, consequently, do not grow based on the results of Sec.~\ref{Sec:Pseudo_linear_closed}.

We start by showing the existence of a steady state for any concentration of the chemostatted species~$\ch{S}$ and~$\ch{P}$. 
According to Eq.~\eqref{Eqn:Open_internal_dynamics}, the concentrations of the internal species $[\ch{E}]$ and $[\ch{E}^{*}]$ follow 
\begin{subequations}\label{Eqn:TC_CC}
\begin{align}
    d_{t}[\ch{E}] &= \bar{k}_{+1} + 2k_{+3}[\ch{E}][\ch{E}^{*}] - {k}_{-1}[\ch{E}]-2k_{-3}[\ch{E}]^{3}\,, \label{Eqn:TC_CC_Dyn1}\\  
    d_{t}[\ch{E}^{*}] &= \bar{k}_{+2} + k_{-3}[\ch{E}]^{3} - k_{+3}[\ch{E}^{*}][\ch{E}] - k_{-2}[\ch{E}^{*}]\,,\label{Eqn:TC_CC_Dyn2}       \end{align}
\end{subequations}
where we introduced the effective kinetic constants $ \bar{k}_{+1} \equiv k_{+1}[\ch{S}]$ and $\bar{k}_{+2} \equiv k_{+2}[\ch{P}]$ since  $[\ch{S}]$ and $[\ch{P}]$ are constant in time. 
Correspondingly, the mass density $L^{m} = [\ch{E}] + 2[\ch{E}^{*}] + [\ch{S}] + 2[\ch{P}]$ follows
\begin{equation}\label{Eq:L_m_TC}
    d_{t}L^{m} = \bar{k}_{+1} - k_{-1}[\ch{E}] + 2\bar{k}_{+2} - 2k_{-2}[\ch{E}^{*}]\,.
\end{equation}
This implies that the steady-state concentrations $[\ch{E}]_{\text{ss}}$ and $[\ch{E}^{*}]_{\text{ss}}$ satisfy
\begin{equation}\label{Eq:EF_fp_TC}
    [\ch{E}^{*}]_{\text{ss}} = \frac{\bar{k}_{+1} + 2\bar{k}_{+2}}{2{k}_{-2}} - \frac{{k}_{-1}}{2k_{-2}}[\ch{E}]_{\text{ss}}\,.
\end{equation}
By substituting Eq.~\eqref{Eq:EF_fp_TC} into Eq.~\eqref{Eqn:TC_CC_Dyn2}, we obtain the cubic equation
\begin{align}
  h([\ch{E}]_{\text{ss}}) \equiv& [\ch{E}]_{\text{ss}}^{3} + \left(\frac{k_{-1}k_{+3}}{2k_{-2}k_{-3}}\right)[\ch{E}]_{\text{ss}}^{2}\label{Eq:E_Fp_TC_quad}\\ &+ \left(\frac{k_{-1}k_{-2}-\bar{k}_{+1}k_{+3}-2\bar{k}_{+2}k_{+3}}{2k_{-2}k_{-3}}\right)[\ch{E}]_{\text{ss}} -\frac{\bar{k}_{+1}}{2k_{-3}} = 0\notag\,,    
\end{align}
which admits a unique positive root, i.e., there is a unique steady state concentration $[\ch{E}]_{\text{ss}}$, because of the Descartes rule of sign~\cite{meserve_1953}.
Furthermore, the steady state concentration $[\ch{E}]_{\text{ss}}$ satisfies
\begin{equation}
    0<[\ch{E}]_{\text{ss}}<\frac{\bar{k}_{+1} + 2\bar{k}_{+2}}{k_{-1}}\,,
\end{equation} 
since $h(0)< 0$ and $h((\bar{k}_{+1} + 2\bar{k}_{+2})/k_{-1})>0$.
This, together with Eq.~\eqref{Eq:EF_fp_TC}, implies that $[\ch{E}^{*}]_{\text{ss}}>0$ and, therefore, the autocatalytic CRN has a well-defined steady state for every value of $[\ch{S}]$ and $[\ch{P}]$.

We then analyze the stability of the steady state.
By using Eqs.~\eqref{Eqn:TC_CC_Dyn1} and \eqref{Eqn:TC_CC_Dyn2}, the Jacobian matrix $\mathbb{J}$ of the steady state reads
\begin{equation}\label{Eqn:Jac_TC_CC}
    \mathbb{J} = \begin{pmatrix}
                2{k}_{+3}[\ch{E}^{*}]_{\text{ss}} - k_{-1} - 6k_{-3}[\ch{E}]_{\text{ss}}^{2} & 2{k}_{+3}[\ch{E}]_{\text{ss}}\\
                3k_{-3}[\ch{E}]^{2}_{\text{ss}}-{k}_{+3}[\ch{E}^{*}]_{\text{ss}} & -k_{-2}-k_{+3}[\ch{E}]_{\text{ss}}
                 \end{pmatrix}  \,.      
\end{equation}
Its determinant and trace can be written as
\begin{subequations}\small
\begin{align}
    &\mathrm{det}(\mathbb{J}) = \frac{4k_{-2}k_{-3}[\ch{E}]_{\text{ss}}^{3} + k_{-1}k_{+3}[\ch{E}]_{\text{ss}}^{2} + \bar{k}_{+1}k_{-2}}{[\ch{E}]_{\text{ss}}} > 0\,, \label{Eq:det_Jac_CC}\\   
    &\mathrm{tr}(\mathbb{J}) = -\bigg( 4k_{-3}[\ch{E}]_{\text{ss}}^{2} + k_{+3}[\ch{E}]_{\text{ss}} + k_{-2} + \frac{\bar{k}_{+1}}{[\ch{E}]_{\text{ss}}} \bigg) <0 \;.
    \label{Eq:tr_Jac_CC}   
\end{align}    
\end{subequations}
This implies that $\mathbb J$ has only negative eigenvalues and, therefore, the steady state is always locally stable for any values of $[\ch{S}]$ and $[\ch{P}]$.

\begin{figure}
    \centering
    \begin{minipage}{0.5\textwidth}
        \centering
        \includegraphics[width=\textwidth]{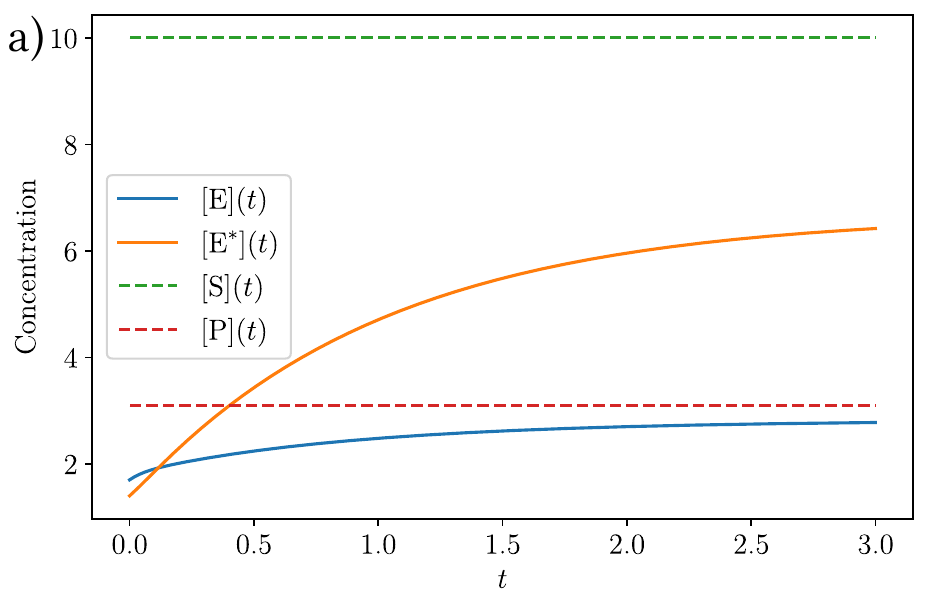}  
    \end{minipage}\hfill
    \begin{minipage}{0.5\textwidth}
        \centering
        \includegraphics[width=\textwidth]{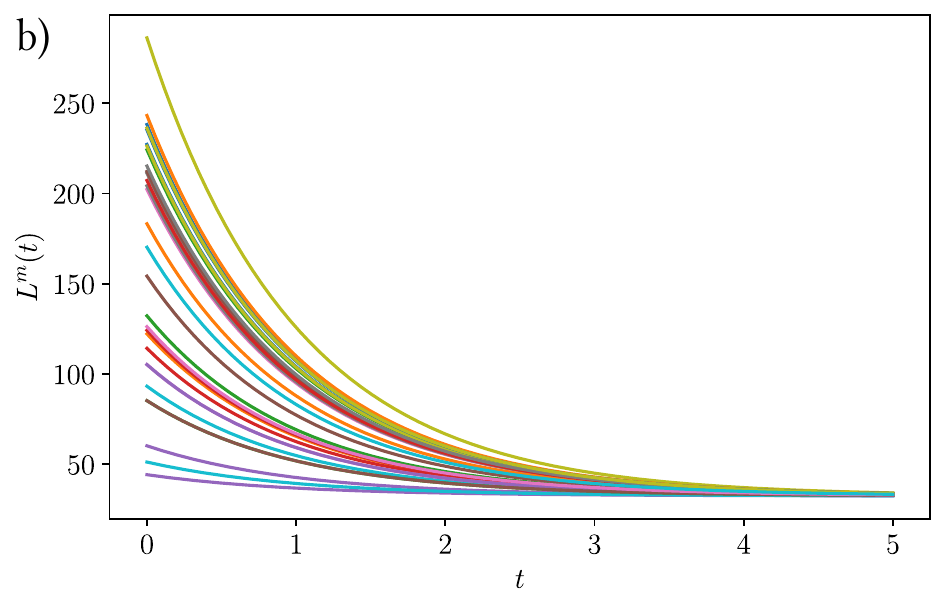}  
    \end{minipage}
\caption{
Dynamics of the autocatalytic CRN in Fig.~\ref{fig:eg_list} under concentration control. Evolution of 
a) Concentrations for the specific initial condition $[\ch{E}](0) = 1.7$ and $[\ch{E}^{*}](0)=1.4$.
b) Mass density for thirty randomly-selected initial conditions.
Here, $k_{\pm 1}=k_{\pm 2} = k_{\pm 3} =1$, $[\ch{S}](0)=10$, and $[\ch{P}](0)=3.1$.}
  
\label{fig:SI_TC_CC_my_bound}
\end{figure}

Finally, we numerically compute the dynamics of the autocalytic CRN.
Figure~\ref{fig:SI_TC_CC_my_bound}a shows the typical evolution the concentrations $[\ch{E}]$ and $[\ch{E}^*]$: after a transient, they reach steady state. 
Figure~\ref{fig:SI_TC_CC_my_bound}b shows the typical evolution of the mass density for different initial conditions.
Like the concentrations, the mass density relaxes towards a constant value.
We stress that the same behavior is observed for different values of the kinetic constants and of the concentrations of the chemostatted species.

\subsection{Growth Under Flux Control}\label{Sec:nonlinear_flux_control}
Flux control always leads to growth when $\boldsymbol{\ell}^{m}\cdot\Tilde{\boldsymbol{I}}> 0$, as already discussed in Sec.~\ref{sec:SI_growth_intro}.
In this subsection, we analyze the dynamics and thermodynamics of growth for the CRNs in Fig.~\ref{fig:eg_list} and compare them with the growth of unimolecular and pseudo-unimolecular CRNs examined in Secs.~\ref{Sec:linear_growth_SI}, \ref{subss:pseudo_dynamics} and~\ref{Sec:PSL_Flux_thermo}, respectively.

\subsubsection{Autocatalytic CRN}\label{Sec:AC_FC}
We consider the autocatalytic CRN in Fig.~\ref{fig:eg_list} when 
$\ch{S}$ is injected with flux $I_{\ch{S}}>0$ 
and $\ch{P}$ is extracted with flux $I_{\ch{P}}<0$
such that the mass density grows at the rate $d_{t}L^{m} =I_{\ch{S}} + 2I_{\text{P}} > 0$ (according to Eq.~\eqref{Eqn:Open_conserv_laws}).
\begin{figure}
    \centering
    \begin{minipage}{0.5\textwidth}
        \centering
        \includegraphics[width=\textwidth]{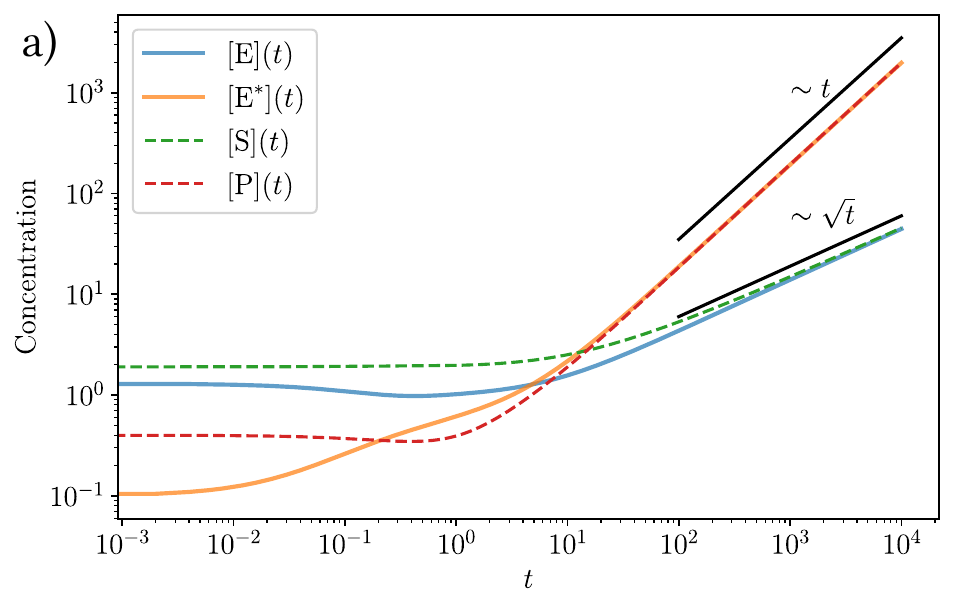}  
    \end{minipage}\hfill
    \begin{minipage}{0.5\textwidth}
        \centering
        \includegraphics[width=\textwidth]{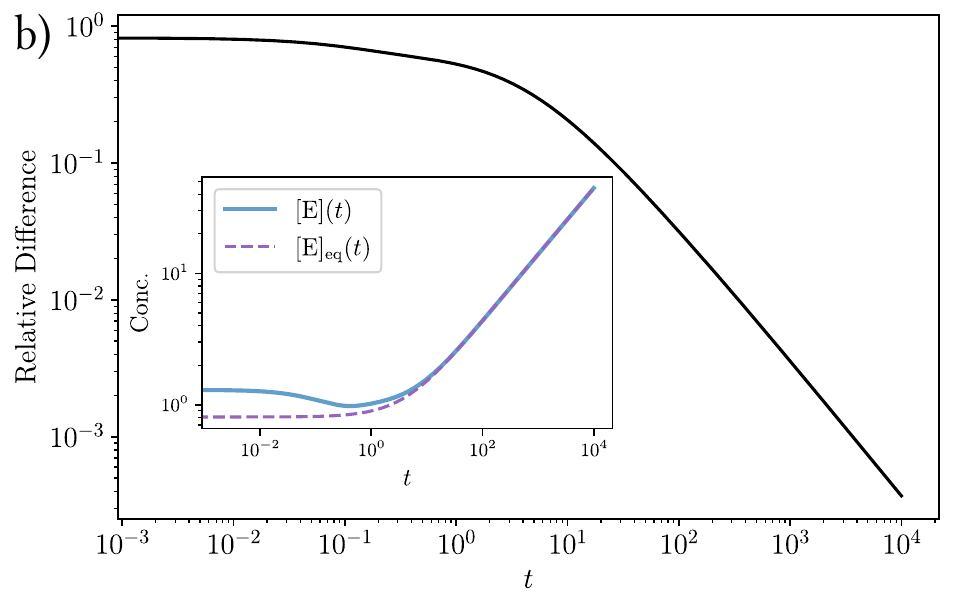}  
    \end{minipage}
\caption{Dynamics of the autocatalytic CRN under flux control. Evolution of
a) Concentrations for the initial condition 
$[\ch{E}](0) = 1.3$, $[\ch{E}^{*}]=0.1$, $[\ch{F}]=1.9$, and $[\ch{W}]=0.4$. 
b) Relative difference, $\lVert \boldsymbol{z}(t)-\boldsymbol{z}_{\text{eq}}(t)\rVert/\lVert\boldsymbol{z}_{\text{eq}}(t)\rVert$ (where the equilibrium is defined as the steady state to which the CRN would relax if it was closed and with mass density $L^m(t)$) and, in the inset, $[\ch{E}](t)$ and $[\ch{E}]_{\text{eq}}(t)$. 
Here, $k_{\pm 1}=k_{\pm 2} = k_{\pm 3} =1$,  $I_{\ch{S}} = 1.0$, and $I_{\ch{P}} = -0.1$.}
\label{fig:SI_TC_FC1}
\end{figure}
Figure~\ref{fig:SI_TC_FC1}a shows a typical evolution of the concentrations, while Fig.~\ref{fig:SI_TC_FC1}b compares it to the evolution of the equilibrium state to which the CRN would relax if it was closed and with mass density $L^{m}(t)$, labeled $\boldsymbol{z}_{\text{eq}}(L^{m}(t))$.
We first observe that all concentrations grow.
Second, the concentrations grow at different rates, 
in contrast to what happens in unimolecular and pseudo-unimolecular CRNs under flux control 
(see Eqs.~\eqref{Eq:flux_control_general_soln} and \eqref{Eq:flux_control_PSL_general_soln}).
In particular, ${[\ch{E}](t)\sim [\ch{S}](t) = \mathcal O(\sqrt t)}$ and
${[\ch{E}^{*}](t)\sim [\ch{P}](t) = \mathcal O(t)}$. Finally, the concentrations $\boldsymbol z(t)$ evolve close to the corresponding equilibrium concentrations $\boldsymbol z_{\text{eq}}(L^{m}(t))$
similarly to unimolecular CRNs (see Eq.~\eqref{Eq:flux_control_general_soln}).
Indeed, the relative difference $\lVert \boldsymbol{z}(t)-\boldsymbol{z}_{\text{eq}}(t)\rVert/\lVert\boldsymbol{z}_{\text{eq}}(t)\rVert$ decreases in time. 
Hence, the autocatalytic CRN under flux control undergoes equilibrium growth 
similar to unimolecular CRNs under flux control (see Secs.~\ref{sec:linear_flux_control_SI}).

This specific dynamics can be explained in terms of a time-scale separation 
between the slowly evolving mass density and the rapidly evolving concentrations of the chemical species.
On the one hand, the former increases at the constant rate $d_{t}L^{m} =I_{\ch{S}} + 2I_{\ch{P}} > 0$.
On the other hand, the latter evolves also because of the chemical reactions 
whose fluxes, according to mass-action kinetics~\eqref{Eq:mass_action_law}, increase when the concentrations increase. Hence, on a time scale in which the mass density $L^{m}$ remains almost constant, 
the concentrations $\boldsymbol z$ can dramatically change and relax towards the equilibrium $\boldsymbol{z}_{\text{eq}}(L^{m}(t))$. This can be captured mathematically by the ansatz
\begin{equation}\label{Eq:TC_TSS}
    \boldsymbol{z}(t) = \boldsymbol{z}_{\text{eq}}(L^{m}(t)) + \boldsymbol{c}(t)\,,
\end{equation}
with $|{c}_\alpha(t)| \ll [\alpha]_{\text{eq}}(L^{m}(t))$, 
resembling Eq.~\eqref{Eq:flux_control_general_soln}.

The ansatz~\eqref{Eq:TC_TSS} allows us to predict the rate of growth of the concentrations observed in Fig.~\ref{fig:SI_TC_FC1}.
Indeed, according to mass action kinetics~\eqref{Eq:mass_action_law},
the equilibrium concentrations $\boldsymbol{z}_{\text{eq}}(L^{m}(t))$ must satisfy
\begin{equation}
\begin{split}\label{Eq:Core_eq}
    [\ch{E}]_{\text{eq}}(t) &= \frac{k_{+1}}{k_{-1}}[\ch{S}]_{\text{eq}}(t)
    = \sqrt{ \frac{k_{+3}k_{+2}}{k_{-2}k_{-3}} [\ch{P}]_{\text{eq}}(t)}\\
    &= \sqrt{ \frac{k_{+3}}{k_{-3}} [\ch{E}^{*}]_{\text{eq}}(t)}\,,
\end{split}
\end{equation}
which, together with  $\boldsymbol{\ell}^{m}\cdot\boldsymbol{z}_{\text{eq}}(L^{m}(t)) = L^{m}(t)$, leads to
\small
\begin{equation}\label{Eqn:equilibrium_core_flux}
[\text{E}]^{2}_{\text{eq}}(t) + [\text{E}]_{\text{eq}}\underbrace{\left(\frac{1 + \frac{k_{-1}}{k_{+1}}}{\frac{2k_{-3}}{k_{+3}}+\frac{2k_{-2}k_{-3}}{k_{+2}k_{+3}}}\right)}_{=a_{1}}
-\underbrace{\left(\frac{L^{m}(t)}{\frac{2k_{-3}}{k_{+3}}+\frac{2k_{-2}k_{-3}}{k_{+2}k_{+3}}}\right)}_{=a_{2}(t)} = 0\,
\end{equation}
\normalsize
and therefore,
\begin{equation}\label{Eqn:equilibrium_conc_E_core}
   [\text{E}]_{\text{eq}}(t) = \sqrt{\left(\frac{a_1}{2}\right)^{2}+a_{2}(t)}- \left(\frac{a_1}{2}\right)\,.
\end{equation}
Since $a_{2}(t) \propto L^m(t) = \mathcal{O}(t)$, from Eqs.~\eqref{Eq:Core_eq} and ~\eqref{Eqn:equilibrium_conc_E_core}, we obtain that $[\ch{E}]_{\text{eq}}(t) \sim [\ch{S}]_{\text{eq}}(t) = \mathcal{O}(\sqrt{t})$ while $[\ch{P}]_{\text{eq}}(t) \sim [\ch{E}^{*}]_{\text{eq}}(t) = \mathcal{O}(t)$.

\begin{figure}
    \centering
    \begin{minipage}{0.5\textwidth}
        \centering
        \includegraphics[width=\textwidth]{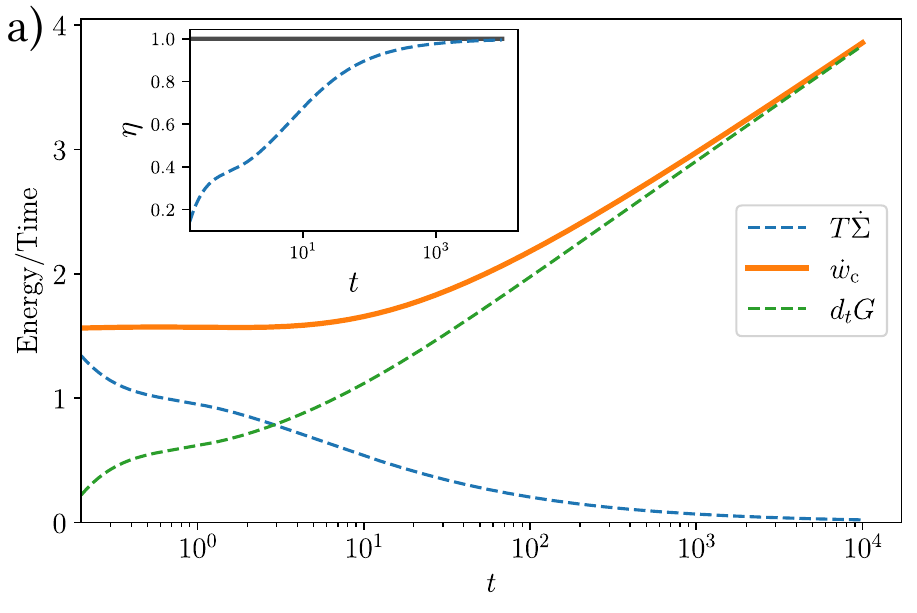} 
    \end{minipage}\hfill
    \begin{minipage}{0.5\textwidth}
        \centering
        \includegraphics[width=\textwidth]{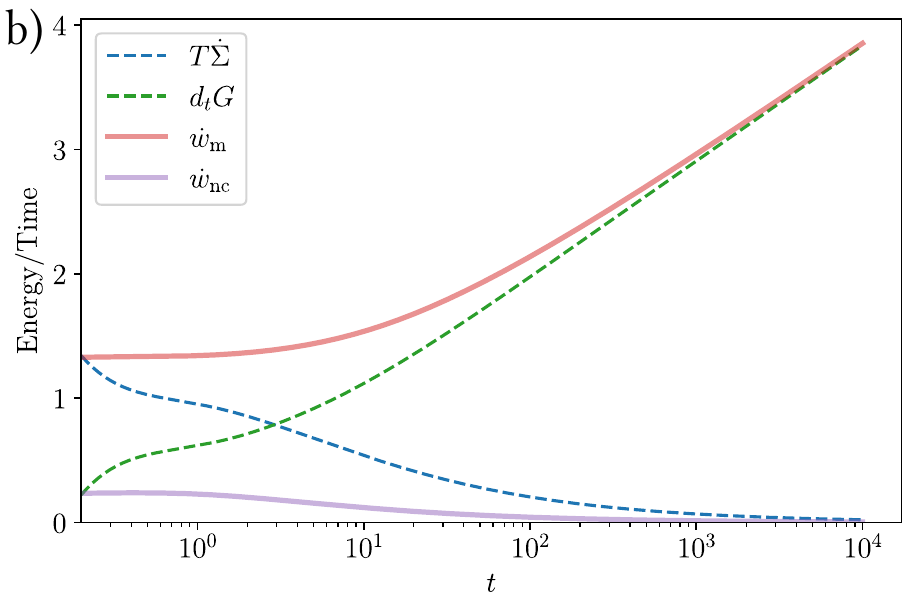} 
    \end{minipage}
\caption{Thermodynamics of the autocatalytic CRN under flux control during the dynamics in Fig.~\ref{fig:SI_TC_FC1}. Evolution of 
a) EPR, $d_{t}G$, chemical work rate, and efficiency of growth.
b) EPR, $d_{t}G$, nonconservative work rate, and the moiety work rate.
Here, $\mu^{0}_{\ch{E}} = 1$, $\mu^{0}_{\ch{E^{*}}} = 2$, $\mu^{0}_{\ch{S}} = 1$, $\mu^{0}_{\ch{P}} = 2$.
}
\label{fig:SI_TC_FC_thermo}
\end{figure}

We now examine the corresponding thermodynamics.
We start by considering the EPR~\eqref{Eqn:EPR_two_defn}, $d_{t}G$~\eqref{eq:gibbs} and chemical work rate~\eqref{Eqn:chemical_work_rate}
whose typical evolution is shown in Fig.~\ref{fig:SI_TC_FC_thermo}a.
The EPR continuously decreases until it vanishes.
On the other hand, $d_{t}G$ and the chemical work increase in time, 
and they coincide in the long time limit. 
This means that the chemical work is fully converted into Gibbs free energy at long times and the efficiency of growth, defined as in Eq.~\eqref{Eq:effiency_defn}, goes to one (see the inset in Fig.~\ref{fig:SI_TC_FC_thermo}a).
We then turn to the nonconservative work rate and the moiety work rate, defined in Eqs.~\eqref{Eqn:nonconserv_work_defn} and~\eqref{eqn:moiety_work_defn}, respectively.
Their typical evolution is shown in Fig.~\ref{fig:SI_TC_FC_thermo}b
(when $\ch{S}$ is chosen as the potential species):
in the long time limit, the nonconservative work vanishes and the moiety work rate balances the increase of Gibbs free energy.

This specific behavior of the thermodynamic quantities 
is similar to unimolecular CRNs under flux control (see Sec.~\ref{sec:linear_flux_control_SI}) 
and is consistent with the previous observation of equilibrium growth.
It can further be explained with the ansatz~\eqref{Eq:TC_TSS}. 
Indeed, by using Eq.~\eqref{Eq:TC_TSS} with $||\boldsymbol{c}(t)|| \ll ||\boldsymbol{z}_{\text{eq}}(L^{m}(t))||$ in Eq.~\eqref{Eqn:chem_pot_defn}, the chemical potentials $\boldsymbol{\mu}(t)$ read
\begin{equation}\label{Eq:chem_pot_Core}
       \boldsymbol{\mu}(t) \approx \boldsymbol{\mu}_{\text{eq}}(t) + RT\frac{\boldsymbol{c}(t)}{\boldsymbol{z}_{\text{eq}}(L^{m}(t))} = \mathcal{O}(\text{ln}(t))\,,
\end{equation}
where $\boldsymbol{c}(t)/\boldsymbol{z}_{\text{eq}}(L^{m}(t)) = (\dots,c_{\alpha}(t)/[\alpha]_{\text{eq}}(L^{m}(t)),\dots)^{\intercal}$ and $\boldsymbol{\mu}_{\text{eq}}(t) = \bar{\mu}(t)\boldsymbol{\ell}^{m}$ with $\bar{\mu}(t) = \mu^{0}_{\text{E}} + RT\text{ln}\left([\text{E}]_{\text{eq}}(t)\right) = \mathcal{O}(\text{ln}(t))$.
Then, by plugging Eq.~\eqref{Eq:chem_pot_Core} in Eq.~\eqref{Eqn:EPR_two_defn}, the EPR becomes
\begin{equation}\label{Eq:EPR_CORE}
\begin{split}
      T\dot{\Sigma} &= -RT \frac{\boldsymbol{c}(t)}{\boldsymbol{z}_{\text{eq}}(L^{m}(t))}\cdot \mathbb{S}\boldsymbol{j}\\ &\approx -RT \frac{\boldsymbol{c}(t)}{\boldsymbol{z}_{\text{eq}}(L^{m}(t))}\cdot \left(d_{t}\boldsymbol{z}_{\text{eq}}-\Tilde{\boldsymbol{I}}\right) \,,
\end{split} 
\end{equation}
where we used Eq.~\eqref{eq:req} and $d_{t}\boldsymbol{z} \approx d_{t}\boldsymbol{z}_{\text{eq}}$. 
At long times, EPR in Eq.~\eqref{Eq:EPR_CORE} vanishes since 
i) $d_{t}\boldsymbol{z}_{\text{eq}}$ approaches a constant value (see Eqs.~\eqref{Eq:Core_eq} and \eqref{Eqn:equilibrium_conc_E_core}),
and ii) $|{c}_\alpha(t)| \ll [\alpha]_{\text{eq}}(L^{m}(t))$. 
Furthermore, by plugging Eq.~\eqref{Eq:chem_pot_Core} in  Eqs.~\eqref{eq:gibbs} and~\eqref{Eqn:chemical_work_rate} and using $|{c}_\alpha(t)| \ll [\alpha]_{\text{eq}}(L^{m}(t))$, 
$d_{t}G$ and the chemical work rate read
\small
\begin{align}
    d_{t}G =& \bar{\mu}(t)(\boldsymbol{\ell}^{m} \cdot \Tilde{\boldsymbol{I}}) +  RT \frac{\boldsymbol{c}(t)}{\boldsymbol{z}_{\text{eq}}(L^{m}(t))}\cdot ({\Tilde{\boldsymbol{I}}}{ + }\mathbb{S}\boldsymbol{j}) = \mathcal{O}(\text{ln}(t))\\
    \dot{w}_{c} =& \bar{\mu}(t)(\boldsymbol{\ell}^{m} \cdot \Tilde{\boldsymbol{I}}) +  RT \frac{\boldsymbol{c}(t)}{\boldsymbol{z}_{\text{eq}}(L^{m}(t))}\cdot {\Tilde{\boldsymbol{I}}} = \mathcal{O}(\text{ln}(t))\,,\label{Eq:Chem_work_Core}
\end{align}
\normalsize
implying $d_{t}G\approx \dot{w}_{c}$ and $\eta\to 1$ in agreement with Fig.~\ref{fig:SI_TC_FC_thermo}a.

Similarly, by using Eq.~\eqref{Eq:chem_pot_Core} in Eqs.~\eqref{eqn:moiety_work_defn} and~\eqref{Eqn:nonconserv_work_defn}, the nonconservative work rate and the moiety work rate read
\begin{align}
\dot{w}_{\text{m}} &= \mu_{\ch{S}}d_{t}L^{m}= \bar{\mu}(t)(\boldsymbol{\ell}^{m} \cdot {\Tilde{\boldsymbol{I}}})+ RT\frac{c_{\ch{S}}(t)}{[\ch{S}]_{\text{eq}}(L^{m}(t))}(\boldsymbol{\ell}^{m} \cdot {\Tilde{\boldsymbol{I}}}) \nonumber\\
&= \mathcal{O}\left(\text{ln}(t)\right)\,, \label{Eq:Autocat_driving_work_flux_control}\\
\dot{w}_{\text{nc}} &= (\mu_{\ch{P}}-2\mu_{\ch{S}})I_{\ch{P}}\nonumber\\
         &= RT\left(\frac{c_{\ch{P}}(t)}{[\ch{P}]_{\text{eq}}(L^{m}(t))}-\frac{2c_{\ch{S}}(t)}{[\ch{S}]_{\text{eq}}(L^{m}(t))}\right)I_{\ch{P}}\label{Eq:Autocat_noncon_work_flux_control}\,,
\end{align}
implying that over long times
\begin{align}\label{Eq:Thermo_Second_law_FC_TC}
    &T\dot{\Sigma}\approx \dot{w}_{\text{nc}} \approx 0 \,,
    &d_{t}G\approx \dot{w}_{\text{m}} = \mathcal{O}(\text{ln}(t))\,.
\end{align}

\subsubsection{Michaelis Menten CRN}\label{Sec:MM_FLux_control}
We consider the Michaelis Menten CRN in Fig.~\ref{fig:eg_list} when the species $\ch{S}$ is injected with flux $I_{\ch{S}}>0$ and $\ch{P}$ is extracted with flux $I_{\ch{P}}<0$ such that the moiety concentration $L^{\text{S}}$ grows at rate $d_{t}L^{\ch{S}} = I_{\text{S}} + I_{\text{P}} > 0$. 

This chemostatting procedure has three main consequences. 
First, the conservation law $\ell^{\ch{E}}$ is unbroken, implying that the concentrations $[\ch{E}]$ and $[\ch{ES}]$ are bounded. 
Second, as $L^{\text{S}}$ grows, (at least one of) the concentrations $[\ch{S}]$ and $[\ch{P}]$ must grow as well.
Consequently, only a subset of species can grow in the Michaelis Menten CRN under flux control unlike 
unimolecular CRNs (Sec.~\ref{sec:linear_flux_control_SI}), 
pseudo-unimolecular CRNs (Sec.~\ref{subss:pseudo_dynamics}), 
and the autocatalytic CRN (Sec.~\ref{Sec:AC_FC}) under flux control.
Third, 
in the long time limit, when the concentrations $[\ch{S}]$ and $[\ch{P}]$ are much larger than the concentrations $[\ch{E}]$ and $[\ch{ES}]$,
a time scale separation emerges between the evolution of the chemostatted concentrations and the internal concentrations which allows us to coarse-grain the internal concentrations as discussed in Sec.~\ref{Sec:Supp_dynamics}.

\begin{figure}
    \centering
    \begin{minipage}{0.5\textwidth}
        \centering
        \includegraphics[width=\textwidth]{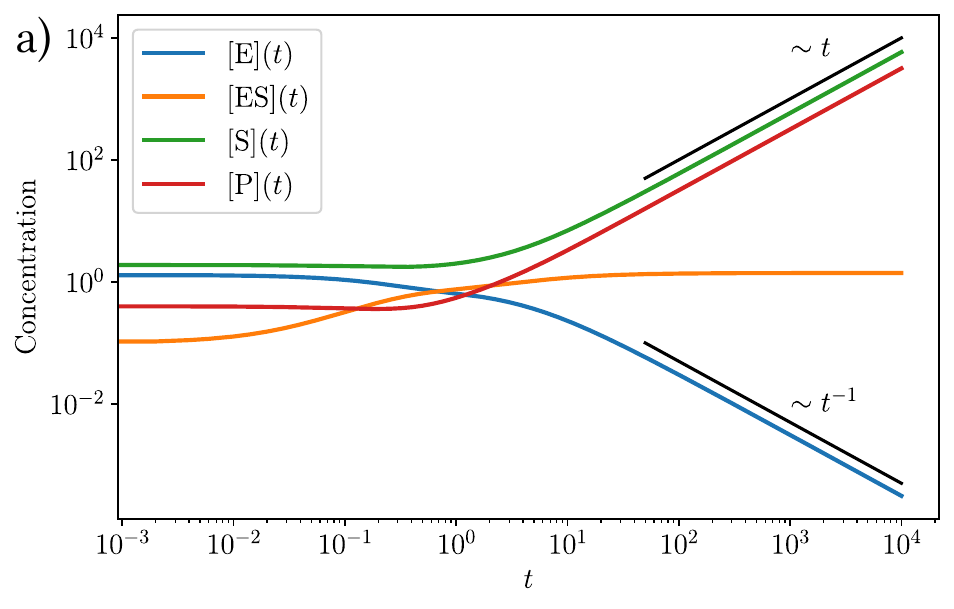}  
    \end{minipage}\hfill
    \begin{minipage}{0.5\textwidth}
        \centering
        \includegraphics[width=\textwidth]{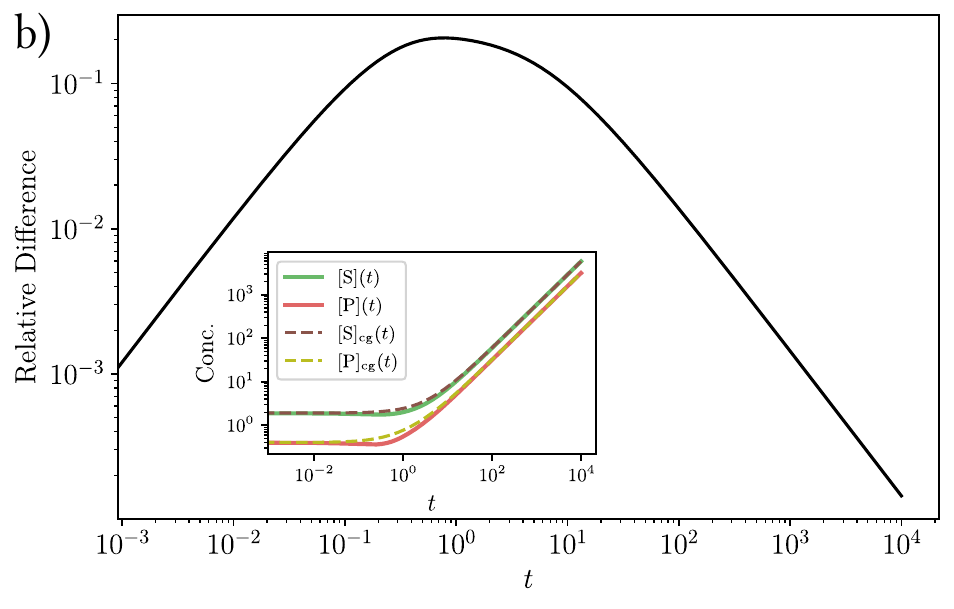}  
    \end{minipage}
\caption{Dynamics of the Michaelis Menten CRN under flux control. Evolution of
a) Concentrations. b)  Relative difference, given by $\lVert\boldsymbol{y} (t)-\boldsymbol{y}_{\text{cg}}(t)\rVert/\lVert\boldsymbol{y}_{\text{cg}}(t)\rVert$, between the concentrations $\boldsymbol{y}(t) = ([\ch{S}](t), [\ch{P}](t))^{\intercal}$ and the coarse-grained concentrations $\boldsymbol{y}_{\text{cg}}(t) = ([\ch{S}]_{\text{cg}}(t), [\ch{P}]_{\text{cg}}(t))^{\intercal}$ (given by Eq.~\eqref{Eq:Coarse_grained_dynamics}),  and comparison between the concentrations and the coarse-grained concentrations in the inset.
Here  $[\ch{E}](0) = 1.3$, $[\ch{ES}](0) = 0.1$, $[\ch{S}](0) = 1.9$, $[\ch{P}](0)=0.4$, $I_{\ch{S}} = 1.0$, $I_{\ch{P}} = -0.1$ and $k_{\pm 1} = k_{\pm 2} = 1$.
}
\label{fig:SI_MM_FC1}
\end{figure}

We show a typical trajectory of the concentrations in Fig.~\ref{fig:SI_MM_FC1}a and compare the dynamics of the actual concentrations $[\ch{S}]$ and $[\ch{P}]$ with the coarse-grained dynamics $[\ch{S}]_{\text{cg}}$ and $[\ch{P}]_{\text{cg}}$ obtained solving Eq.~\eqref{Eq:Coarse_grained_dynamics} in Fig. \ref{fig:SI_MM_FC1}b. 
We note that $[\ch{S}]$ and $[\ch{P}]$ grow linearly in time,
$[\ch{ES}]$ saturates to a constant value, while $[\ch{E}]$ decreases in time.  
Furthermore, the relative difference between the vectors $\boldsymbol{y}(t)=([\ch{S}](t),[\ch{P}](t))^\intercal$ and $\boldsymbol{y}_{\text{cg}}(t) = \left([\ch{S}]_{\text{cg}}(t),[\ch{P}]_{\text{cg}}(t)\right)^{\intercal}$ decreases in time.

We now use the coarse-grained dynamics discussed in Sec.~\ref{Sec:Supp_dynamics} to determine
i) the growth rate of $[\ch{S}]$ and $[\ch{P}]$,
and
ii) the evolution of $[\ch{E}]$ and $[\ch{ES}]$ in the long time limit.

First, by using the diagrammatic method~\cite{Hill1966} as done in Refs.~\cite{Wachtel2018,Avanzini2020b},
Eq.~\eqref{Eq:Coarse_grained_dynamics} for the Michaelis Menten CRN reads
\small
\begin{equation}
\begin{split}
    d_{t}[\text{S}]_{\text{cg}} - I_{\text{S}} &= -L^{\ch{E}}(0)\left[\frac{k_{+1}k_{+2}[\text{S}]_{\text{cg}}-k_{-1}k_{-2}[\text{P}]_{\text{cg}}}{k_{-1}+k_{+2}+k_{+1}[\text{S}]_{\text{cg}}+k_{-2}[\text{P}]_{\text{cg}}}\right]\\
    &=I_{\text{P}} - d_{t}[\text{P}]_{\text{cg}} \,.
\end{split} \label{Eqn:MM_flux_Conc_P}
\end{equation}
\normalsize
This, together with the ansatz 
${[\ch{S}]_{\text{cg}} =  v_{\ch{S}}t + c_{\ch{S}}}$ 
and ${[\ch{P}]_{\text{cg}} =  v_{\ch{P}}t + c_{\ch{P}}}$ 
(with constants $c_{\ch{S}}$ and $c_{\ch{P}}$) as well as the approximation ${d_{t}L^{\ch{S}} = I_{\ch{S}} + I_{\ch{P}}  \simeq v_{\ch{S}} + v_{\ch{P}}}$ (valid as long as time scale separation holds, i.e., $d_t [\ch{ES}]\simeq 0$),
leads to the quadratic equation
\small
\begin{equation}\label{Eqn:MM_flux_chemo_slope_S}
    v_{\ch{S}} =I_{\ch{S}}
    - L^{\ch{E}}(0) \frac{\left(k_{+1}k_{+2}+k_{-1}k_{-2}\right)v_{\ch{S}} - k_{-1}k_{-2}\left( I_{\ch{S}} +  I_{\ch{P}}\right)}{\left(k_{+1}-k_{-2}\right)v_{\ch{S}} + k_{-2}\left( I_{\ch{S}} +  I_{\ch{P}}\right)}\,,  
\end{equation}
\normalsize
once the terms of order $\mathcal O(1/t)$ are neglected. Equation~\eqref{Eqn:MM_flux_chemo_slope_S} admits the following solution

\small
\begin{equation}\label{Eqn:MM_flux_chemo_slope_S_sol}
\begin{split}
     v_{\ch{S}} &= \frac{k_{-2}I_{\ch P} + (2k_{-2} - k_{+1})I_{\ch S} + L^{\ch{E}}(0)\left(k_{+1}k_{+2}+k_{-1}k_{-2}\right)}{2(k_{-2}-k_{+1})}\\
     &+ \sqrt{\begin{aligned}
          &\frac{k_{-2}\left( I_{\ch{S}} +  I_{\ch{P}}\right)(I_{\ch S} + k_{-1} L^{\ch{E}}(0))}{k_{+1}-k_{-2}}+\\
          &~~\Bigg[\frac{k_{-2}I_{\ch P} + (2k_{-2} - k_{+1})I_{\ch S} + L^{\ch{E}}(0)\left(k_{+1}k_{+2}+k_{-1}k_{-2}\right)}{2(k_{-2}-k_{+1})}\Bigg]^{2} \,,\\
     \end{aligned}}
\end{split}
\end{equation}
\normalsize
which implies that the growth rates $v_{\ch{S}}$ and $v_{\ch{P}}$ depend on the value of $L^{\ch{E}}(0)$. This dependence results from the fact that the interconversion of $\ch{S}$ into $\ch{P}$ is mediated by the enzymes whose concentrations $[\ch{E}]$ and $[\ch{ES}]$ are bounded by $L^{\ch{E}}(0)$.

Second, in the coarse-grained dynamics, the concentrations of the internal species $[\ch{E}]$ and $[\ch{ES}]$ quickly relax to the steady state determined by $[\ch{S}]$ and $[\ch{P}]$ which, 
according to the diagrammatic method~\cite{Hill1966}, read
\begin{subequations}
    \begin{align}\label{Eqn:E_soln_MM_flux}
       [\ch{E}]_{\text{ss}} &=   \frac{(k_{-1}+k_{+2})L^{\ch{E}}(0)}{k_{-1}+k_{+2} + k_{+1}[\ch{S}]_{\text{cg}} + k_{-2}[\ch{P}]_{\text{cg}}}\,,\\
       [\ch{ES}]_{\text{ss}} &= [\ch{E}]_{\text{ss}}\left(\frac{k_{+1}[\ch{S}]_{\text{cg}}+k_{-2}[\ch{P}]_{\text{cg}}}{k_{-1}+k_{+2}}\right)\,.\label{Eqn:ES_soln_MM_flux} 
    \end{align}\label{Eqn:EandES_soln_MM_flux} 
\end{subequations}
Equations~\eqref{Eqn:E_soln_MM_flux} and \eqref{Eqn:ES_soln_MM_flux}, 
together with $[\ch{S}] \sim  v_{\ch{S}}t$ and $[\ch{P}] \sim  v_{\ch{P}}t$,
lead to the scaling $[\ch{E}] = \mathcal{O}(t^{-1})$ and $[\ch{ES}] = \mathcal{O}(1)$ observed in Figs.~\ref{fig:SI_MM_FC1}a.

\begin{figure}
    \centering
    \begin{minipage}{0.5\textwidth}
        \centering
        \includegraphics[width=\textwidth]{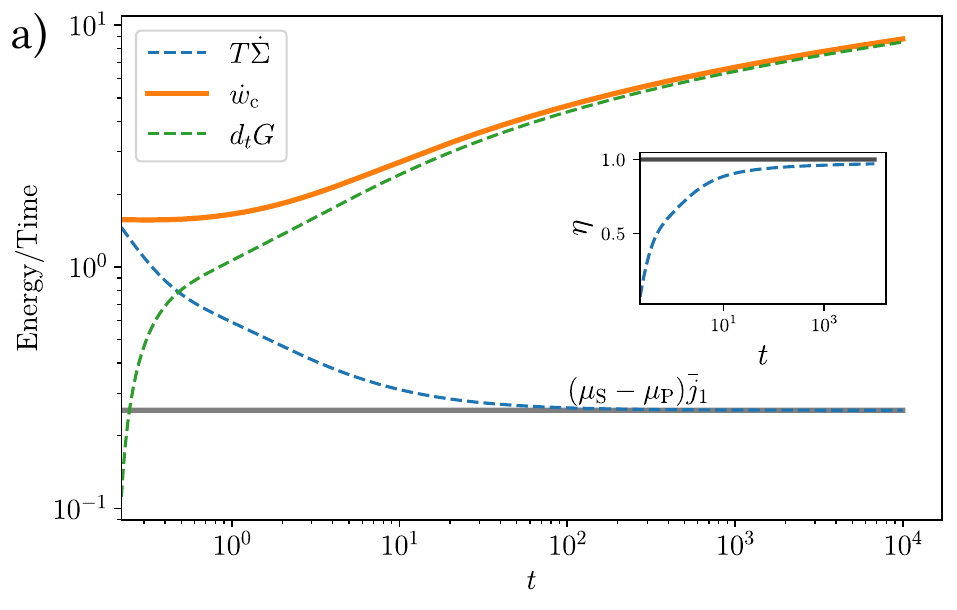}  
    \end{minipage}\hfill
    \begin{minipage}{0.5\textwidth}
        \centering
        \includegraphics[width=\textwidth]{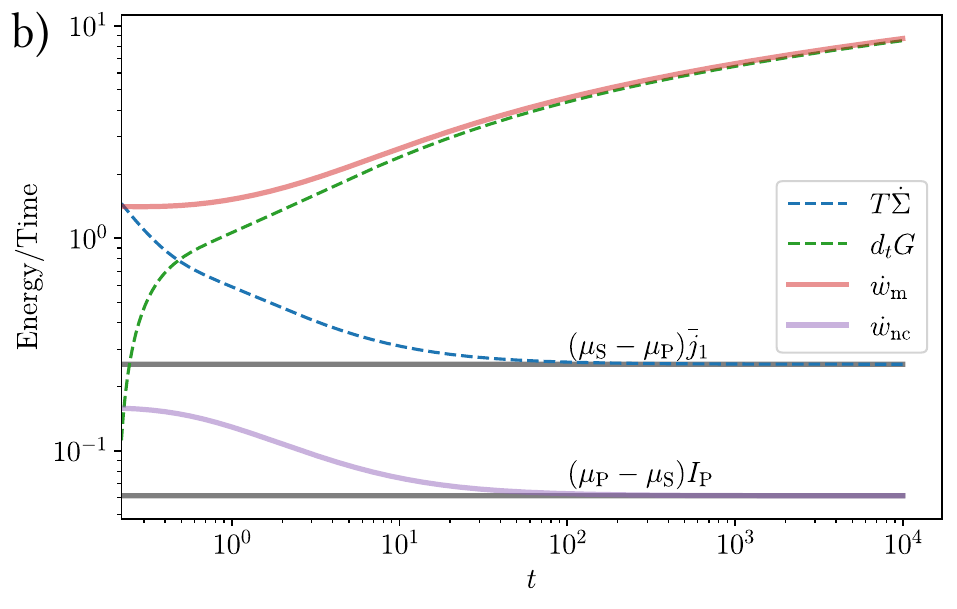}  
    \end{minipage}
\caption{Thermodynamics of the Michaelis Menten CRN under flux control during the dynamics in Fig.~\ref{fig:SI_MM_FC1}.  Evolution of a) EPR, $d_tG$, chemical work rate, and efficiency of growth (inset) b) EPR, $d_{t}G$, nonconservative work rate, and moiety work rate. Here, $\mu^{0}_{\ch{E}} = 1$, $\mu^{0}_{\ch{ES}} = 2$, $\mu^{0}_{\ch{S}} = 1$, $\mu^{0}_{\ch{P}} = 1$.}  
 \label{fig:SI_MM_FC2}
\end{figure}

We now turn to the corresponding thermodynamics. 
We start by considering the EPR~\eqref{Eqn:EPR_two_defn}, $d_{t}G$~\eqref{eq:gibbs} and chemical work rate~\eqref{Eqn:chemical_work_rate}, whose typical time evolution is plotted in Fig.~\ref{fig:SI_MM_FC2}a.
On the one hand, the EPR relaxes to a nonzero constant value and, therefore, 
the Michaelis Menten CRN under flux control grows out of equilibrium
like  pseudo-unimolecular CRNs under flux control (see Sec.~\ref{Sec:PSL_Flux_thermo}).
On the other hand, the chemical work rate and $d_{t}G$ increase in time approaching each other and, consequently, the efficiency of growth~\eqref{Eq:effiency_defn} approaches one
unlike in pseudo-unimolecular CRNs (see Sec.~\ref{Sec:PSL_Flux_thermo}).
We then examine the nonconservative work rate~\eqref{Eqn:nonconserv_work_defn} and the moiety work rate~\eqref{eqn:moiety_work_defn} whose typical evolution is shown in Fig.~\ref{fig:SI_MM_FC2}b
(when $\ch{S}$ is chosen as the potential species).
At long times, the nonconservative work saturates to a constant value while the moiety work rate increases with time and scales like  $d_tG$.

This specific behavior of the thermodynamic quantities can be explained using the coarse-grained dynamics described in Sec.~\ref{Sec:Supp_dynamics}.
By using the ansatz $[\ch{S}] = v_{\ch{S}}t + c_{\ch{S}}$ and $[\ch{P}] = v_{\ch{P}}t + c_{\ch{P}}$ with Eq.~\eqref{Eqn:chem_pot_defn} 
and neglecting terms of order $\mathcal{O}(1/t)$,
we first recognize that in the long time limit the chemical potentials of \ch{S} and \ch{P} read
\begin{equation}\label{Eq:MM_FC_chempot}
{\mu}_{\alpha} \approx RT \text{ln}(t){\ell}^{\ch{S}}_{\alpha} + RT\text{ln}(v_{\alpha}) + \mu^{0}_{\alpha} = \mathcal{O}(\text{ln}(t))\,.
\end{equation}
Then, the EPR can be expressed, according to Ref.~\cite{Avanzini2020b}, as
\begin{equation} \label{Eqn:EPR_MM_Flux}
     T\dot{\Sigma} = -\left(\mu_{\text{P}}-\mu_{\text{S}}\right) \bar{j}_{1}= \mathcal{O}(1)\,,
\end{equation}
where, using Eq~\eqref{Eq:Coarse_grained_dynamics},
the ansatz $[\ch{S}] = v_{\ch{S}}t + c_{\ch{S}}$ and $[\ch{P}] = v_{\ch{P}}t + c_{\ch{P}}$,
and Eq.~\eqref{Eq:MM_FC_chempot},

\begin{equation}\label{Eq:Currents_MM_FC}
    \bar{j}_{1} = I_{\ch{S}}-v_{\ch{S}} = \mathcal{O}(1) \,,
\end{equation}
and 
\begin{equation}\label{Eq:Chemical_pot_MM_flux}
    \mu_{\text{P}}-\mu_{\text{S}}
              = (\mu^{0}_{\ch{P}} - \mu^{0}_{\ch{S}}) + RT\text{ln}\left(\frac{v_{\text{P}}}{v_{\text{S}}}\right)   = \mathcal{O}(1)\,.
\end{equation}
Namely, we recover the constant (and positive) entropy production rate 
in Figs.~\ref{fig:SI_MM_FC2}a and~\ref{fig:SI_MM_FC2}b.
By using Eq.~\eqref{Eq:MM_FC_chempot}, $d_{t}G$~\eqref{eq:gibbs} and the chemical work rate~\eqref{Eqn:chemical_work_rate} become
\begin{align}
\begin{split}
     \dot{w}_{\text{c}} 
    &= RT\text{ln}(t)\left(d_{t}L^{\ch{S}}\right) + RT\sum_{\alpha \in Y}\left( \mu^{0}_{\alpha} + RT\text{ln}(v_{\alpha})\right){\Tilde{I}_{\alpha}} \\ &= \mathcal{O}(\text{ln}(t))\,,  \label{Eqn:chem_work_MM_Flux}
\end{split}\\  
\begin{split}
    d_{t}G  &= \text{RT}\text{ln}(t)\left(d_{t}L^{\ch{S}}\right)
    + RT\sum_{\alpha \in Y}\left( \mu^{0}_{\alpha} + RT\text{ln}(v_{\alpha})\right)v_{\alpha} \\&= \mathcal{O}(\text{ln}(t))\,,\label{Eq:dt_G_MM_flux}
\end{split}
\end{align}
namely, they grow logarithmically in time in agreement with Fig.~\ref{fig:SI_MM_FC2}a.
Crucially, the EPR in Eq.~\eqref{Eqn:EPR_MM_Flux} is subleading with respect to the chemical work in Eq.~\eqref{Eqn:chem_work_MM_Flux} and, consequently,
the efficiency of growth~\eqref{Eq:effiency_defn} approaches one
even if the CRN undergoes nonequilibrium growth.

We conclude by examining 
the moiety work rate and the nonconservative work rate.
Using Eq.~\eqref{Eq:Chemical_pot_MM_flux} 
in Eqs.\eqref{eqn:moiety_work_defn} and \eqref{Eqn:nonconserv_work_defn}, we obtain 
\begin{align}\label{Eqn:noncosnerv_work_MM_flux}
    \dot{w}_{\text{m}} &= \text{RT}\text{ln}(t)\left(d_{t}L^{\ch{S}}\right) + RT\text{ln}\left(v_{\text{S}}\exp\left(\frac{{\mu}^{0}_{\text{S}}}{RT}\right)\right)d_{t}L^{\ch{S}}\\
    &= \mathcal{O}(\text{ln}(t))\,,\nonumber\\
\dot{w}_{\text{nc}} &= (\mu_{\text{P}}-\mu_{\text{S}})I_{\text{P}} =  \mathcal{O}(1)\,,
\end{align}
implying 
\begin{equation}\label{Eqn:Second_Law_MM_FC}
\begin{split}
   d_{t}{G} &\sim \dot{w}_{\text{m}} = \mathcal{O}(\text{ln}(t))\,,\\
   \dot{w}_{\text{nc}} &\sim T\dot{\Sigma} =  \mathcal{O}(1)\,,\\ 
\end{split}
\end{equation}
in the long time limit. 
Note that the values predicted in Eqs.~\eqref{Eqn:noncosnerv_work_MM_flux} and \eqref{Eqn:EPR_MM_Flux} are shown in Fig.~\ref{fig:SI_MM_FC2}b and are consistent with the numerical simulations.

\subsubsection{Cyclic Michaelis Menten CRN}\label{Sec:CMM_FC}
We consider the Cyclic Michaelis Menten CRN in Fig.~\ref{fig:eg_list} when the species $\ch{S}$ is injected with flux $I_{\ch{S}}>0$ and $\ch{P}$ is extracted with flux $I_{\ch{P}}<0$ such that the moiety concentration $L^{\text{S}}$ grows at rate $d_{t}L^{\ch{S}} = I_{\text{S}} + I_{\text{P}} > 0$. 

\begin{figure}
    \centering
    \begin{minipage}{0.5\textwidth}
        \centering
        \includegraphics[width=\textwidth]{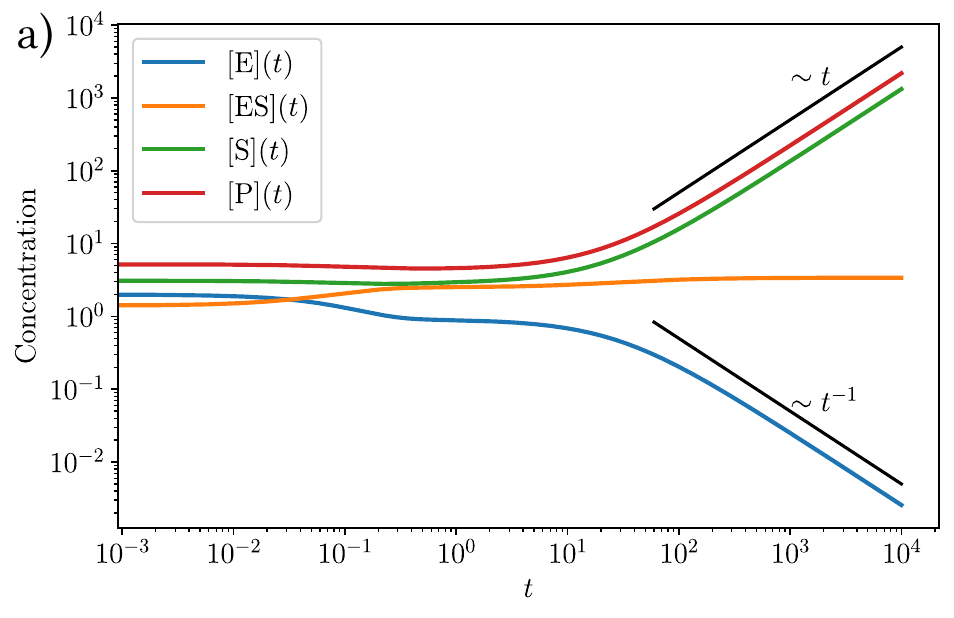}  
    \end{minipage}\hfill
    \begin{minipage}{0.5\textwidth}
        \centering
        \includegraphics[width=\textwidth]{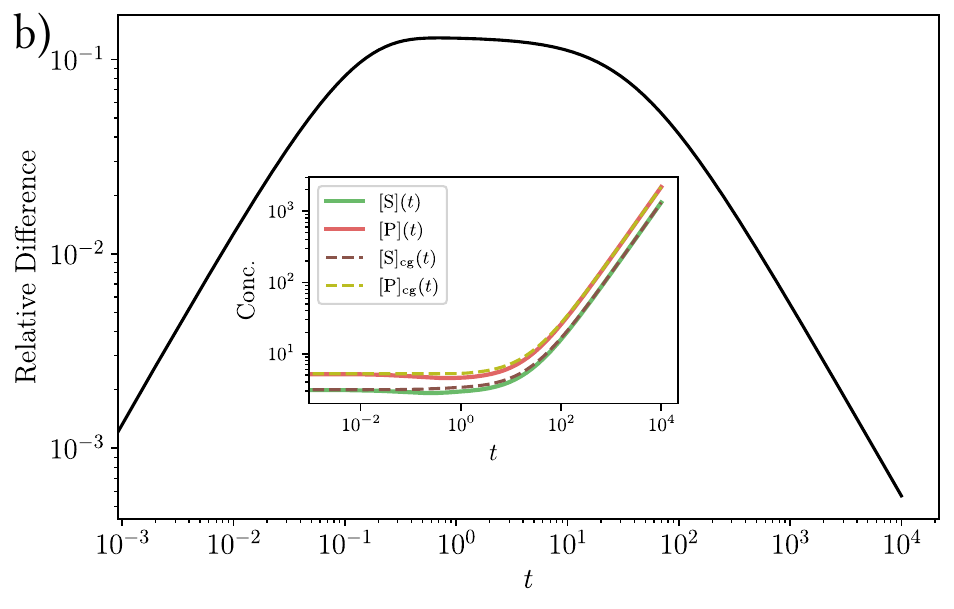}  
    \end{minipage}
\caption{Dynamics of the Cyclic Michaelis Menten CRN under flux control. Evolution of 
a) Concentrations b) Relative difference $\lVert\boldsymbol{y} (t)-\boldsymbol{y}_{\text{cg}}(t)\rVert/\lVert\boldsymbol{y}_{\text{cg}}(t)\rVert$, between the concentrations $\boldsymbol{y}(t) = ([\ch{S}](t), [\ch{P}](t))^{\intercal}$ and the coarse grained concentrations $\boldsymbol{y}_{\text{cg}}(t) = ([\ch{S}]_{\text{cg}}(t), [\ch{P}]_{\text{cg}}(t))^{\intercal}$ (given by Eq.~\eqref{Eq:Coarse_grained_dynamics}) and comparison between the concentrations and the coarse-grained concentrations in the inset.
Here  $[\ch{E}](0) = 2.0$, $[\ch{ES}](0) = 1.4$, $[\ch{S}](0) = 3.1$, $[\ch{P}](0)=5.2$, $I_{\ch{S}} = 0.45$, $I_{\ch{P}} = -0.10$ and $k_{\pm 1} = 1$, $ k_{+2} = k_{+3} = e^{0.25}$, $k_{-2}=k_{-3} = e^{-0.25}$.}
 \label{fig:SI_CMM_FC_dyn}
\end{figure}

From a dynamic point of view, 
the Cyclic Michaelis Menten CRN and 
the Michaelis Menten CRN (Sec.~\ref{Sec:MM_FLux_control})
have the same qualitative behavior.
The concentrations $[\ch{E}]$ and $[\ch{ES}]$ are bounded 
while $[\ch{S}]$ and $[\ch{P}]$ grow linearly in time 
(see their typical evolution in {Fig.~\ref{fig:SI_CMM_FC_dyn}a}).
At long times, 
because of a time scale separation between the chemostatted and internal species,
the evolution of $\boldsymbol{y}(t)=([\ch{S}](t),[\ch{P}](t))^\intercal$ 
converges towards
$\boldsymbol{y}_{\text{cg}}(t) = \left([\ch{S}]_{\text{cg}}(t),[\ch{P}]_{\text{cg}}(t)\right)^{\intercal}$ following Eq.~\eqref{Eq:Coarse_grained_dynamics} (see the typical evolution in Fig.~\ref{fig:SI_CMM_FC_dyn}b)

We use the coarse-grained dynamics to understand the long-time dynamics of the concentrations, as done in Sec.~\ref{Sec:MM_FLux_control}.
Using the diagrammatic method~\cite{Hill1966},
Eq.~\eqref{Eq:Coarse_grained_dynamics} for the Cyclic Michaelis Menten CRN reads
\begin{equation}
\begin{split}
d_{t}[\text{S}]_{\text{cg}} &+ k_{+3}[\text{S}]_{\text{cg}} - k_{-3}[\text{P}]_{\text{cg}} - I_{\text{S}} \\ 
   &=-L^{\ch{E}}(0)\left[\frac{k_{+1}k_{+2}[\text{S}]_{\text{cg}}-k_{-1}k_{-2}[\text{P}]_{\text{cg}}}{k_{-1}+k_{+2}+k_{+1}[\text{S}]_{\text{cg}}+k_{-2}[\text{P}]_{\text{cg}}}\right]  \\
   &=I_{\text{P}}+ k_{+3}[\text{S}]_{\text{cg}}-k_{-3}[\text{P}]_{\text{cg}}-d_{t}[\text{P}]_{\text{cg}}\,,
\end{split}\label{Eqn:CMM_flux_Conc_P}    
\end{equation}
which, together with the ansatz $[\ch{S}]_{\text{cg}} =  v_{\ch{S}}t + c_{\ch{S}}$ 
and $[\ch{P}]_{\text{cg}} =  v_{\ch{P}}t +c_{\ch{P}}$ 
(with $c_{\ch{S}}$ and $c_{\ch{P}}$ some constant parameters),
the approximation $d_{t}L^{\ch{S}} = I_{\ch{S}} + I_{\ch{P}}  \simeq v_{\ch{S}} + v_{\ch{P}}$ (valid as long as time scale separation holds, i.e., $d_t [\ch{ES}]\simeq 0$), 
leads to
\small
\begin{equation}\label{Eq:VS_CMM_Flux}
\begin{split}
        v_{\ch{S}} =& \left(k_{-3}\left(I_{\ch{S}}+I_{\ch{P}}\right)
        -\left(k_{+3}+k_{-3}\right)v_{\ch{S}}\right)t \\
        &+ I_{\ch{S}} + {\left(k_{-3}c_{\ch{P}}-k_{+3}c_{\ch{S}}\right)}\\
        &- L^{\ch{E}}(0)\frac{\left(k_{+1}k_{+2}+k_{-1}k_{-2}\right)v_{\ch{S}} - k_{-1}k_{-2}\left( I_{\ch{S}} +  I_{\ch{P}}\right)}{\left(k_{+1}-k_{-2}\right)v_{\ch{S}} + k_{-2}\left(I_{\ch{S}} +  I_{\ch{P}}\right)}\,.
\end{split}
\end{equation}
\normalsize
For consistency, the term proportional to time on the right-hand side of Eq.~\eqref{Eq:VS_CMM_Flux} must vanish.
Hence,
\begin{align}\label{Eqn:CMM_slope_S}
    v_{\text{S}} &= \frac{k_{-3}}{k_{+3}+k_{-3}}\left(I_{\text{S}}+I_{\text{P}}\right)\,,\\
    v_{\text{P}} &= \frac{k_{+3}}{k_{+3}+k_{-3}}\left(I_{\text{S}}+I_{\text{P}}\right)\,.\label{Eqn:CMM_slope_P}
\end{align}
We emphasize that unlike the Michaelis Menten CRN (see Eq.~\eqref{Eqn:MM_flux_chemo_slope_S_sol}),
the growth rates $v_{\ch{S}}$ and $v_{\ch{P}}$ are independent of $L^{\ch{E}}(0)$ because of the reaction directly interconverting \ch{S} into \ch{P}.
Finally, according to the diagrammatic method~\cite{Hill1966}, $[\ch{E}]_{\text{ss}}$ and $[\ch{ES}]_{\text{ss}}$ still follow Eq.~\eqref{Eqn:EandES_soln_MM_flux} and, therefore, 
$[\ch{E}] = \mathcal{O}(t^{-1})$ and $[\ch{ES}] = \mathcal{O}(1)$  (as observed in Figs.~\ref{fig:SI_CMM_FC_dyn}a).

\begin{figure}
    \centering
    \begin{minipage}{0.5\textwidth}
        \centering
        \includegraphics[width=\textwidth]{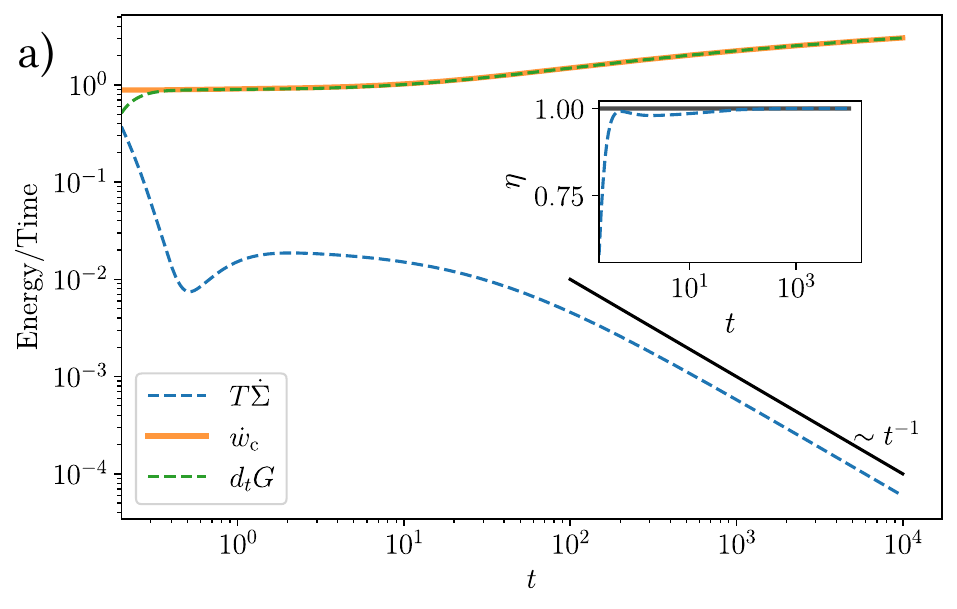}  
    \end{minipage}\hfill
    \begin{minipage}{0.5\textwidth}
        \centering
        \includegraphics[width=\textwidth]{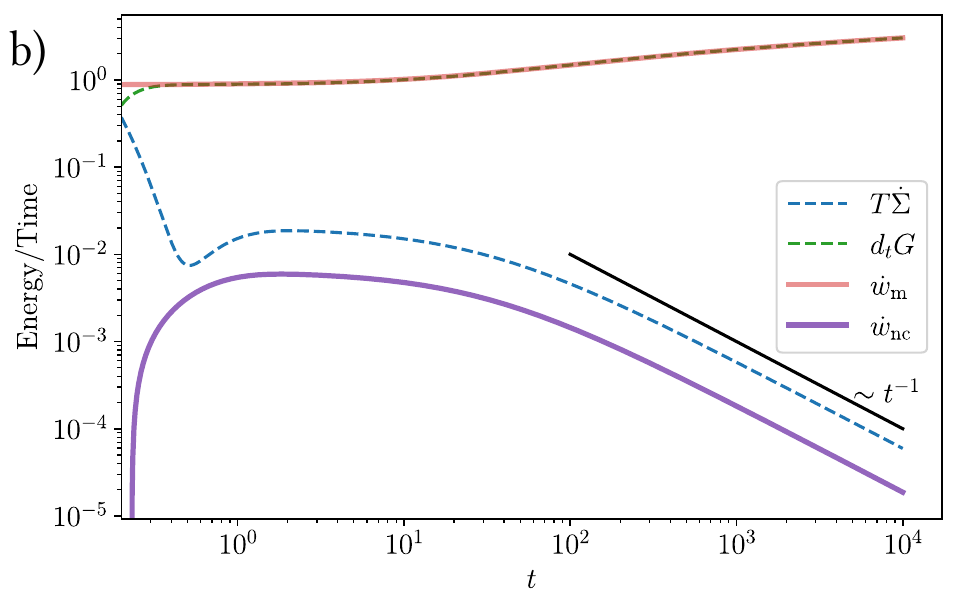}  
    \end{minipage}
\caption{Thermodynamics of the Cyclic Michaelis Menten CRN under flux control during the dynamics in Fig.~\ref{fig:SI_CMM_FC_dyn}.  Evolution of a) EPR, $d_tG$, chemical work rate, and efficiency of growth (inset) b) EPR, $d_{t}G$, nonconservative work rate, and moiety work rate. Here, $\mu^{0}_{\ch{E}} = 0.5$, $\mu^{0}_{\ch{ES}} = 2$, $\mu^{0}_{\ch{S}} = 1.5$, $\mu^{0}_{\ch{P}} = 1$.
} 
 \label{fig:SI_CMM_FC_thermo}
\end{figure}

The main difference between the Cyclic Michaelis Menten CRN and 
the Michaelis Menten CRN (see Sec.~\ref{Sec:MM_FLux_control}) under flux control
concerns the thermodynamics of growth.
We start by considering the EPR~\eqref{Eqn:EPR_two_defn}, $d_{t}G$~\eqref{eq:gibbs} and chemical work rate~\eqref{Eqn:chemical_work_rate}, whose typical time evolution is plotted in Fig.~\ref{fig:SI_CMM_FC_thermo}a.
Like the Michaelis Menten CRN (see Fig.~\ref{fig:SI_MM_FC2}a),
the $d_{t}G$ and the chemical work increase in time and coincide at long times,
while the efficiency of growth~\eqref{Eq:effiency_defn} goes to one.
Unlike the Michaelis Menten CRN,
EPR continuously decreases to zero.
This means that the Cyclic Michaelis Menten CRN under flux control undergoes equilibrium growth like unimolecular CRNs (Sec.~\ref{sec:linear_flux_control_SI}) and the autocatalytic CRN (Sec.~\ref{Sec:AC_FC}).
We then examine the nonconservative work rate~\eqref{Eqn:nonconserv_work_defn} and the moiety work rate~\eqref{eqn:moiety_work_defn} whose typical evolution is shown in Fig.~\ref{fig:SI_CMM_FC_thermo}b
(when $\ch{S}$ is chosen as the potential species).
Like the Michaelis Menten CRN (see Fig.~\ref{fig:SI_MM_FC2}b),
the moiety work rate increases with time and scales like  $d_tG$.
Unlike the Michaelis Menten CRN, 
the nonconservative work decays like the EPR.

This specific behavior of the thermodynamic quantities can be explained as a consequence of the timescale separation by using the coarse-grained dynamics discussed in Sec~\ref{Sec:Supp_dynamics}.
To do so, we start by recognizing that in the long time limit, the chemical potentials of the chemostatted species read
\begin{align}
{\mu}_{\alpha} &= RT \text{ln}(t){\ell}^{\ch{S}}_{\alpha} + RT\text{ln}\left({v}_{\alpha}\exp\left(\frac{{\mu}^{0}_{\alpha}}{RT}\right)\right)
     +RT\text{ln}\left(1 + \frac{c_\alpha}{v_{\alpha}t}\right)  
     \nonumber \\ &= \mathcal{O}(\text{ln}(t))\,,\label{Eq:chem_pot_CMM}
\end{align}
where  we used the ansatz $[\ch{S}] =  v_{\ch{S}}t + c_{\ch{S}}$ and $[\ch{P}] = v_{\ch{P}}t + c_{\ch{P}}$ in Eq.~\eqref{Eqn:chem_pot_defn}. 
Note that in Eq.~\eqref{Eq:chem_pot_CMM} we did not neglect the term of order $\mathcal{O}(1/t)$ like in Eq.~\eqref{Eq:MM_FC_chempot} because it explains the decrease of the EPR seen in Figs.~\ref{fig:SI_CMM_FC_thermo}a and~\ref{fig:SI_CMM_FC_thermo}b.
Indeed, according to Ref.~\cite{Avanzini2020b}, the EPR can be expressed as 
\begin{equation} \label{Eqn:EPR_CMM_Flux}
     T\dot{\Sigma} 
        =-\left(\mu_{\text{P}}-\mu_{\text{S}}\right) (\bar{j}_1+\bar{j}_{3}) 
        = \mathcal{O}(1/t),
\end{equation}
where, 
by using Eq~\eqref{Eq:Coarse_grained_dynamics} and
the ansatz $[\ch{S}] = v_{\ch{S}}t + c_{\ch{S}}$ and $[\ch{P}] = v_{\ch{P}}t + c_{\ch{P}}$,
\begin{equation}\label{Eq:Currents_CMM_FC}
    \bar{j}_1+\bar{j}_{3} = I_{\ch{S}}-v_{\ch{S}} = \mathcal{O}(1) \,,
\end{equation}
while, 
by using Eqs.~\eqref{Eqn:CMM_slope_S}, \eqref{Eqn:CMM_slope_P}, \eqref{Eq:chem_pot_CMM},
the approximation $\text{ln}(1+x)\approx 1+x$ for $x\ll 1$,
together with the local detailed balance~\eqref{eqn:LDB_SI} (implying $v_{\text{S}}\exp(\mu^{0}_{\text{S}}/RT) = v_{\text{P}}\exp(\mu^{0}_{\text{P}}/RT)$),
\begin{equation}\label{Eq:Chemical_pot_CMM_flux}
    \mu_{\ch{P}} - \mu_{\ch{S}} = {RT\frac{1}{t}\left(\frac{c_{\ch{P}}}{v_{\text{P}}}-\frac{c_{\ch{S}}}{v_{\text{S}}}\right)}=\mathcal{O}(1/t)\,.
\end{equation}

We now turn to the other thermodynamic quantities.
By using Eq.~\eqref{Eq:chem_pot_CMM} and neglecting terms of order $\mathcal O(1/t)$, the $d_{t}G$~\eqref{eq:gibbs} and the chemical work rate~\eqref{Eqn:chemical_work_rate} become
\begin{align}
\begin{split}
 \dot{w}_{\text{c}} 
    &= RT\text{ln}(t)\left(d_{t}L^{\ch{S}}\right)
    +RT\sum_{\alpha \in Y}\left( \mu^{0}_{\alpha} + RT\text{ln}(v_{\alpha})\right){\Tilde{I}_{\alpha}} \\&= \mathcal{O}(\text{ln}(t))  \label{Eqn:chem_work_CMM_Flux}\,,
\end{split}\\ 
\begin{split}
    d_{t}G  &= \text{RT}\text{ln}(t)\left(d_{t}L^{\ch{S}}\right)
    +RT\sum_{\alpha \in Y}\left( \mu^{0}_{\alpha} + RT\text{ln}(v_{\alpha})\right)v_{\alpha} \\&= \mathcal{O}(\text{ln}(t))\label{Eq:dt_G_CMM_flux} \,,
\end{split}    
\end{align}
namely, they grow logarithmically in time,
$d_{t}G\approx \dot{w}_{c}$,
and $\eta\to 1$ in agreement with  agreement with Fig.~\ref{fig:SI_CMM_FC_thermo}b. 
Furthermore, by using Eq.~\eqref{Eq:chem_pot_CMM},
the moiety work rate~\eqref{eqn:moiety_work_defn}  
and the nonconservative work rate~\eqref{Eqn:nonconserv_work_defn}
read
\begin{align}
 \begin{split}\label{Eqn:moiety_work_CMM_flux}
  \dot{w}_{\text{m}} &= \text{RT}\text{ln}(t)\left(d_{t}L^{\ch{S}}\right){}+RT\text{ln}\left(v_{\text{S}}\exp\left(\frac{{\mu}^{0}_{\text{S}}}{RT}\right)\right)d_{t}L^{\ch{S}} \\&= \mathcal{O}(\text{ln}(t))\,,
 \end{split}\\
  \dot{w}_{\text{nc}} &\approx RT\frac{1}{t}\left(\frac{c_{\ch{P}}}{v_{\text{P}}}-\frac{c_{\ch{S}}}{v_{\text{S}}}\right)I_{\text{P}} = \mathcal{O}(t^{-1})\,.\label{Eqn:noncosnerv_work_CMM_flux}
\end{align}
implying 
\begin{equation}\label{Eqn:Second_Law_CMM_FC}
\begin{split}
   d_{t}{G} &\sim \dot{w}_{\text{m}} = \mathcal{O}(\text{ln}(t))\,,\\
   \dot{w}_{\text{nc}} &\sim T\dot{\Sigma} =  \mathcal{O}(t^{-1})\,,\\ 
\end{split}
\end{equation}
in the long time limit.

Note that direct interconversion of \ch{S} into \ch{P} explains why the Cyclic Michaelis Menten CRN grows close to equilibrium unlike the Michaelis Menten CRN.
On the one hand,  the rate of interconversion of \ch{S} into \ch{P} (via the enzymatic mechanism) in the Michaelis Menten CRN is limited by the bounded concentration $[\ch{E}]$ and $[\ch{ES}]$. This prevents the equilibration between the chemostatted species, i.e., it
establishes a difference between the chemical potentials of \ch{S} and \ch{P} and creates the nonconservative force~\eqref{Eq:Chemical_pot_MM_flux}.
On the other hand, the rate of the direct interconversion of \ch{S} into \ch{P} in Cyclic Michaelis Menten CRN increases with the concentrations $[\ch{S}]$ and $[\ch{P}]$.
Their chemical potentials can thus rapidly equilibrate and the nonconservative force~\eqref{Eq:Chemical_pot_CMM_flux} decreases in time.

\subsubsection{Minimal Metabolic CRN}\label{Sec:Metabolic_FC}
The minimal metabolic CRN has already been numerically studied in Ref.~\cite{avanzini2022flux} when the species $\text{S}$ and $\text{F}$ are injected at constant rates $I_{\ch{S}}>0$ and $I_{\ch{F}}>0$ such that that the moiety concentrations $L^{\ch{S}}$ and $L^{\ch{F}}$ grow with rates $d_{t}L^{\ch{S}} = I_{\text{S}} > 0$ and $d_{t}L^{\ch{F}} = I_{\text{F}} > 0$.
Correspondingly, the concentrations $[\ch{S}]$, $[\ch{P}]$, $[\ch{F}]$ and $[\ch{W}]$ grow and the EPR decays in time, implying that the CRN grows close to equilibrium. 
This was explained by noting that $\dot{w}_{\text{nc}}= 0$ at all times since the chemostatting procedure does not create nonconservative forces (see Eq.~\eqref{Eqn:nonconserv_work_defn}).
 
Here, we study the minimal metabolic CRN when 
the species $\text{S}$ and $\text{F}$ are injected at constant rates $I_{\ch{S}}>0$ and $I_{\ch{F}}>0$ 
while the species $\text{P}$ and $\text{W}$ are extracted at constant rates $I_{\ch{P}}<0$ 
and $I_{\ch{W}}<0$ 
in such a way that the moiety concentrations $L^{\text{S}}$ and $L^{\text{F}}$ grow with rates $d_{t}L^{\ch{S}} = I_{\text{S}} + I_{\text{P}} > 0$ and $d_{t}L^{\ch{F}} = I_{\text{F}} + I_{\text{W}} > 0$.
This chemostatting procedure has four main consequences. First, it creates two nonconservative forces (see Eq.~\eqref{Eqn:nonconserv_work_defn}).
Second, the conservation law $\ell^{E}$ is unbroken and, therefore, 
the concentrations $[\ch{E}]$, $[\ch{ES}]$, $[\ch{EW}]$ and $[\ch{E}^{*}]$ are bounded.
Third, as $L^{\text{S}}$ (resp. $L^{\ch{F}}$) grows, (at least one of) the concentrations $[\ch{S}]$ and $[\ch{P}]$ (resp. $[\ch{F}]$ and $[\ch{W}]$) must grow as well.
Finally, in the long time limit, when the concentrations of the chemostatted species are much larger 
than the concentrations of the internal species,
a time scale separation emerges between the evolution of the former and the latter implying that the dynamics of the chemostatted species follow the coarse-grained dynamics discussed in Sec.~\ref{Sec:Supp_dynamics}

\begin{figure}
    \centering
    \begin{minipage}{0.5\textwidth}
        \centering
        \includegraphics[width=\textwidth]{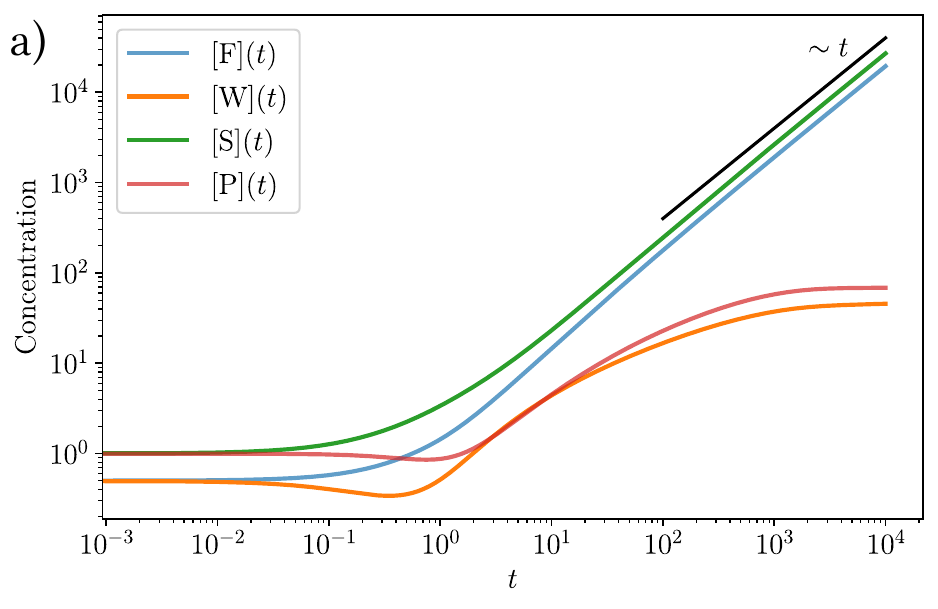}  
    \end{minipage}\hfill
    \begin{minipage}{0.5\textwidth}
        \centering
        \includegraphics[width=\textwidth]{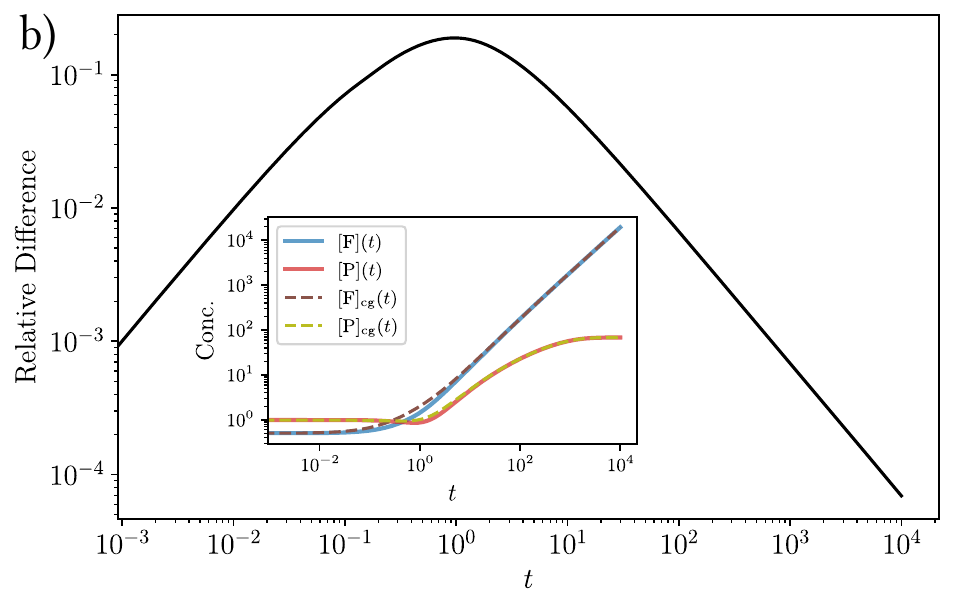}  
    \end{minipage}
\caption{
Dynamics of the minimal metabolic CRN under flux control. Evolution of
a) Concentrations of the chemostatted species $[\ch{F}](t)$, $[\ch{W}](t)$, $[\ch{S}](t)$ and $ [\ch{P}](t)$ b) Relative difference between the vectors $\boldsymbol{y}(t) = ([\ch{F}](t),[\ch{W}](t)[\ch{S}](t), [\ch{P}](t))^{\intercal}$ and $\boldsymbol{y}_{\text{cg}}(t) = ([\ch{F}]_{\text{cg}}(t),[\ch{W}]_{\text{cg}}(t),[\ch{S}]_{\text{cg}}(t), [\ch{P}]_{\text{cg}}(t))^{\intercal}$ (given by Eq.~\eqref{Eq:Coarse_grained_dynamics}), 
and comparison between the concentrations $[\ch{F}](t)$ and $[\ch{P}](t)$ with the coarse-grained concentrations $[\ch{F}]_{\text{cg}}(t)$ and $[\ch{P}]_{\text{cg}}(t)$ in the inset.
Here  $[\ch{E}](0) = 1.5$, $[\ch{EF}](0) = [\ch{EW}](0) = [\ch{E}^{*}](0) = 0.1$, $[\ch{F}](0) = [\ch{W}](0) = 0.5$, $[\ch{S}](0) = [\ch{P}](0)= 1$, $I_{\ch{F}} = 2.8$, $I_{\ch{W}} = -0.1$ , $I_{\ch{S}} = 2.0$, $I_{\ch{P}} = -0.05$ and $k_{\pm 1} = k_{-2} = k_{\pm3} = k_{\pm 4} = k_{\pm 5} = 2$ while $k_{+2} = 2e$.}
\label{fig:Met_FC1}
\end{figure}

We now show a typical evolution of the concentrations of the chemostatted species in Fig.~\ref{fig:Met_FC1}a and compare it with the coarse-grained dynamics obtained from Eq.~\eqref{Eq:Coarse_grained_dynamics} in Fig.~\ref{fig:Met_FC1}b.
We note that concentrations $[\ch{F}]$ and $[\ch{S}]$ grow linearly in time while the concentrations $[\ch{P}]$ and $[\ch{W}]$ saturate to constant values. 
Furthermore, the relative distance between the vectors $\boldsymbol{y}(t)=([\ch{F}](t),[\ch{W}](t),[\ch{S}](t),[\ch{P}](t)[\ch{P}](t))^\intercal$ and $\boldsymbol{y}_{\text{cg}}(t) = \left([\ch{F}]_{\text{cg}}(t),[\ch{W}]_{\text{cg}}(t),[\ch{S}]_{\text{cg}}(t),[\ch{P}]_{\text{cg}}(t)\right)^{\intercal}$ increases initially and then decreases at long times. 

\begin{figure}
    \centering
    \begin{minipage}{0.5\textwidth}
        \centering
        \includegraphics[width=\textwidth]{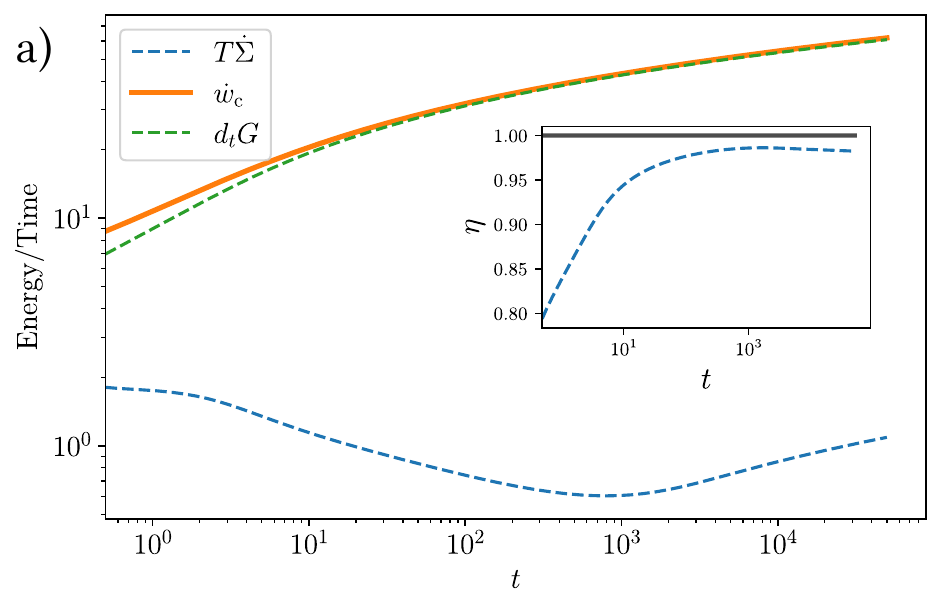}  
    \end{minipage}\hfill
    \begin{minipage}{0.5\textwidth}
        \centering
        \includegraphics[width=\textwidth]{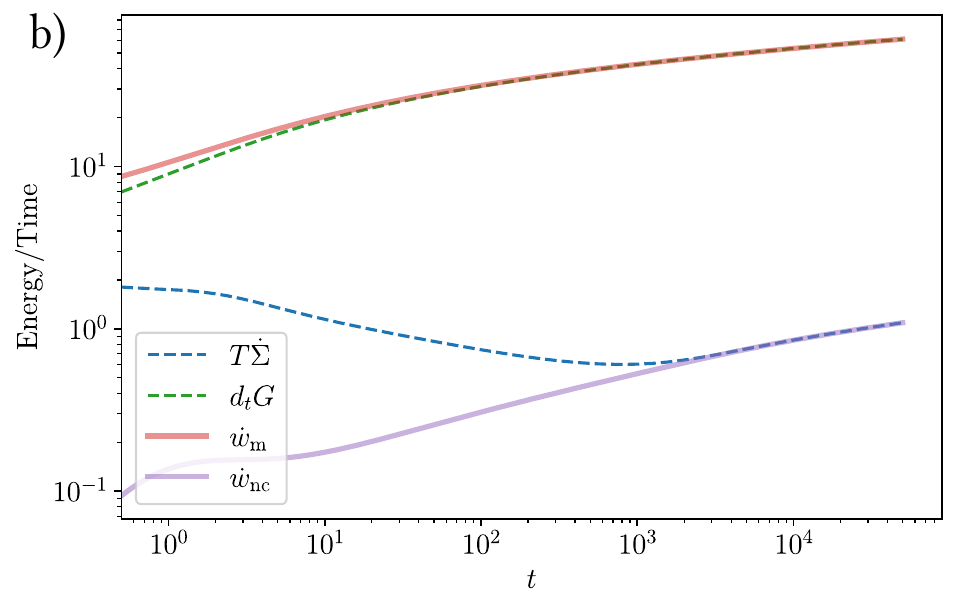}  
    \end{minipage}
\caption{
Thermodynamics of the minimal metabolic CRN under flux control during the dynamics in Fig.~\ref{fig:Met_FC1}. Evolution of a) EPR, $d_tG$, chemical work rate, and efficiency of growth (inset) b) EPR, $d_{t}G$, nonconservative work rate, and moiety work rate. Here, $\mu^{0}_{\ch{E}} = 1$, $\mu^{0}_{\ch{EF}} = 3$, $\mu^{0}_{\ch{EW}} = 2$, $\mu^{0}_{\ch{E}^{*}} = 4$, $\mu^{0}_{\ch{F}} = 2$, $\mu^{0}_{\ch{W}} = 1$, $\mu^{0}_{\ch{S}} = 1$, $\mu^{0}_{\ch{P}} = 2$.
} 
 \label{fig:Met_FC2}
\end{figure}

We now turn to thermodynamics. 
The typical time evolution of the EPR~\eqref{Eqn:EPR_two_defn}, $d_{t}G$~\eqref{eq:gibbs} and chemical work rate~\eqref{Eqn:chemical_work_rate} is plotted in Fig.~\ref{fig:Met_FC2}a.
The chemical work rate and $d_{t}G$ increase in time.
However, they do not coincide at long times, unlike in the Autocatalytic CRN in Sec.~\ref{Sec:AC_FC}, the Michaelis Menten CRN in Sec.~\ref{Sec:MM_FLux_control} and the Cyclic Michaelis Menten CRN in Sec.~\ref{Sec:CMM_FC}, since the EPR increases in time too.
Consequently, the minimal metabolic CRN under flux control grows out of equilibrium, like the Michaelis Menten CRN (see Sec.~\ref{Sec:MM_FLux_control}),
and the efficiency of growth~\eqref{Eq:effiency_defn}, which is initially close to one, decreases at long times, unlike the Michaelis Menten CRN (see Fig.~\ref{fig:SI_MM_FC2}a).

We then examine the nonconservative work rate~\eqref{Eqn:nonconserv_work_defn} and the moiety work rate~\eqref{eqn:moiety_work_defn} whose typical evolution is shown in Fig.~\ref{fig:Met_FC2}b
(when $\ch{S}$ and $\ch{F}$ are chosen as the potential species).
At long times, there is a splitting between the magnitudes of the EPR and the nonconservative work rate on the one hand and the moiety work rate and $d_{t}G$ on the other hand. 
Furthermore, 
\begin{equation}\label{Eqn:Second_Law_Met_FC}
\begin{split}
   d_{t}{G} &\sim \dot{w}_{\text{m}}\,,\\
   \dot{w}_{\text{nc}} &\sim T\dot{\Sigma} > 0\,,\\ 
\end{split}
\end{equation}
similar to the Autocatalytic CRN \eqref{Eq:Thermo_Second_law_FC_TC}, the Michaelis Menten CRN \eqref{Eqn:Second_Law_MM_FC}, and the Cyclic Michaelis Menten CRN , \eqref{Eqn:Second_Law_CMM_FC}.

\subsubsection{Summary}

From a dynamical standpoint,
growth in multimolecular CRNs under flux control can significantly differ from unimolecular and pseudo-unimolecular CRNs.
For instance, 
i) the concentrations of some species scale as $\mathcal{O}(\sqrt{t})$ in the Autocatalytic CRN (see Sec.~\ref{Sec:AC_FC}), while the concentrations of all species scale as $\mathcal{O}({t})$ in unimolecular and pseudo-unimolecular CRNs;
ii) the concentrations of only a subset of species grow in the Michaelis Menten CRN (see Sec.~\ref{Sec:MM_FLux_control}).


From a thermodynamic standpoint,
some multimolecular CRNs grow close to equilibrium 
(e.g., the Autocatalytic CRN and the cyclic Michaelis Menten CRN), 
while other multimolecular CRNs grow out ot equilibrium 
(e.g., the Michaelis Menten CRN and the minimal metabolic CRN).
We noticed that the presence of an unbroken conservation law is a necessary 
but not sufficient condition for nonequilibrium growth.
On the one hand,  it can prevent the equilibration of the chemostatted species and create nonconservative forces like in the Michaelis Menten CRN and the minimal metabolic CRN.
On the other hand, if there are other reaction pathways leading to equilibration, 
like the direct interconversion of \ch{S} into \ch{P} in the cyclic Michaelis Menten CRN, the CRN grow at equilibrium (see discussion at the end of Subs.~\ref{Sec:CMM_FC}).

\subsection{Growth under Mixed Control}\label{Sec:SI_mixed_full}
Unimolecular and pseudo-unimolecular CRNs cannot grow under mixed control, as we proved in Secs.~\ref{Sec:SI_Linear_Mixed chemo} and Sec.~\ref{Sec:PSL_open_dynamics},
respectively. On the other hand, multimolecular CRNs can grow under mixed control 
as proven in Ref.~\cite{avanzini2022flux} for the minimal metabolic CRN.

Here, we first prove in general that the concentrations of extracted chemostatted species are bounded for CRNs growing under mixed control with a monotonically increasing mass density, i.e., $d_{t}L^{m} > 0$. Indeed, from Eq.~\eqref{eq:mass_evol_mean} and the definition of mixed control, the mass density $L^{m}$ follows
\begin{equation}\label{Eqn:Mass_density_evol_mixed}
       d_{t}L^{m} = -\ell^{m}_{Y}\cdot\Tilde{\mathbb{D}}\boldsymbol{y} + \ell^{m}_{Y}\cdot\Tilde{\boldsymbol{I}}\,,
\end{equation} 
which, together with $d_{t}L^{m} > 0$, implies
\begin{equation}\label{Eq:Bound_chemo_mixed}
    [\alpha] \leq \frac{\ell^{m}_{Y}\cdot\Tilde{\boldsymbol{I}}}{\ell^{m}_{\alpha}\Tilde{\mathbb{D}}_{\alpha,\alpha}}\,,
\end{equation}
 for every chemostatted species $\alpha$ such that $\Tilde{\mathbb{D}}_{\alpha,\alpha} > 0$,
i.e., every extracted chemostatted species. Note that this has the obvious implication that Continuous-flow Stirred Tank Reactors (CSTR) cannot grow as already proven in Sec.~\ref{sec:SI_growth_intro}.

We now examine the specific dynamics and thermodynamics of the multimolecular CRNs listed in Fig.~\ref{fig:eg_list} under mixed control.
\subsubsection{Autocatalytic CRN}\label{Sec:Autocat_MC}
\paragraph*{Case I.} We start by considering the autocatalytic CRN when $\text{S}$ (resp. $\text{P}$) is injected with  constant rate $I_{\text{S}} > 0$ (resp. $I_{\text{P}} > 0$) 
and extracted with rate ${k}^{e}_{\ch{S}}[\text{S}]$ (resp. ${k}^{e}_{\ch{P}}[\text{P}]$).
We first analytically prove that the autocatalytic CRN cannot grow with a monotonically increasing mass density,
and we then use numerical simulations to show that the mass density does not grow.

\begin{figure*}
    \centering
    \includegraphics{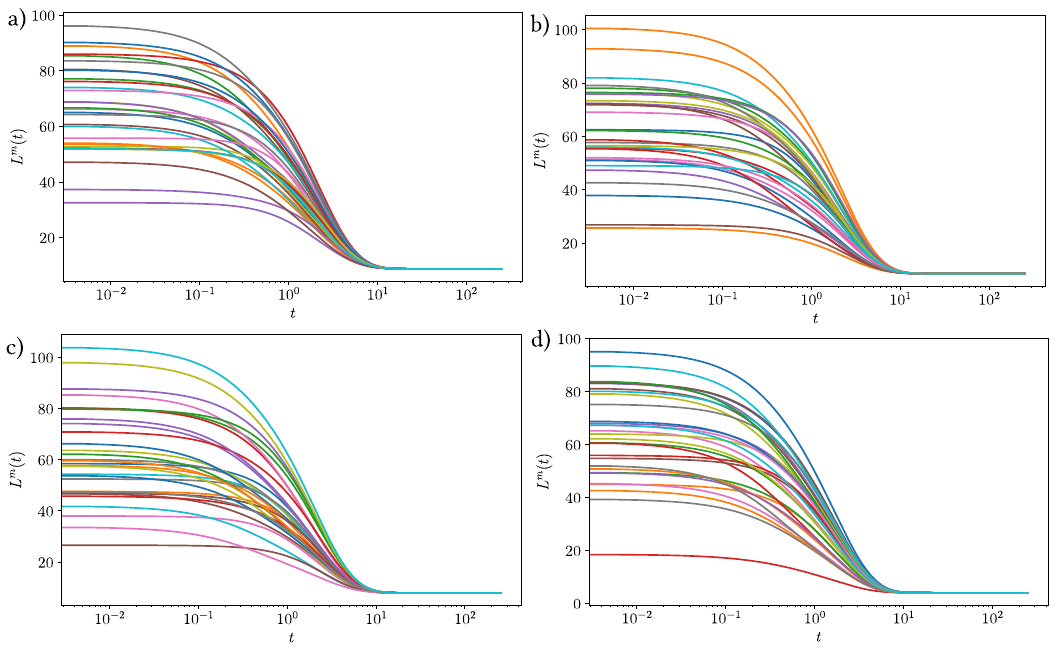}
    \caption{Evolution of the mass density of the Autocatalytic CRN under mixed control where the species $\text{S}$ and $\text{P}$ have influx rates $I_{\text{S}} = 1.0$ and $I_{\text{P}} = 1.5$ and outflux rates $k^{e}_{\ch{S}}[\text{S}]$ and $k^{e}_{\ch{P}}[\text{P}]$ for thirty randomly chosen initial conditions and a) ${I}/{\Tilde{k}} = 750$, b)  ${I}/{\Tilde{k}} = 75$, c) ${I}/{\Tilde{k}} = 15$, d) ${I}/{\Tilde{k}} = 0.75$, where the parameter ${I}/{\Tilde{k}} \equiv \text{max}\{I_{\text{S}},I_{\text{P}}\}/ \text{min}\{k^{e}_{\ch{S}},k^{e}_{\ch{P}}\}$.
Here, $k_{\pm 1}=k_{\pm 2} = k_{\pm 3} =1$ and $k^{e}_{\ch{P}} = 2.0$.} 
\label{fig:SI_TC_MC_bound1} 
\end{figure*}

Let us assume that the autocatalytic CRN grows with $d_{t}L^{m} > 0$ which, 
according to Eq.~\eqref{Eq:Bound_chemo_mixed}, implies 
\begin{equation}\label{Eqn:Bound_chemo_mixed_core}
        [\text{S}],[\text{P}] \leq \frac{I_{\text{S}}+I_{\text{P}}}{\text{min}({k}^{e}_{\ch S},{k}^{e}_{\ch P})}\,.
\end{equation}
Hence, at least one of the concentrations $[\text{E}]$ and $[\text{E}^{*}]$ grow. 
However, according to Eq.~\eqref{eq:req} and the definition of mixed control, the evolution equation of $[\text{S}]$,
\begin{equation}\label{Eq:S_evol_MC}
    d_{t}[\text{S}] + (k_{+1}+{k}^{e}_{\ch{S}})[\text{S}] = k_{-1}[\text{E}] + I_{\text{S}}\,,
\end{equation}
has the formal solution
\begin{align}
    [\text{S}](t) =&  [\text{S}](0)e^{-(k_{+1}+k^{e}_{\ch{S}})t}
    + \frac{I_{\text{S}}}{k_{+1}+k^{e}_{\ch{S}}}\left(1-e^{-(k_{+1}+k^{e}_{\ch{S}})t}\right)\nonumber \\
    &+ k_{-1}\int_{0}^{t} ds  [\text{E}](s)e^{-(k_{+1}+k^{e}_{\ch{S}})(t-s)}\,,\label{Eq:S_evol_MC_sol}
\end{align}
implying that $[\text{E}](t)$ must also be bounded. 
By applying exactly the same reasoning to the evolution equation of $[\ch{P}]$, we find that $[\ch{E}^{*}]$ must be bounded too.
Thus, the autocatalytic CRN does not grow with $d_{t}L^{m} > 0$.

We now turn to numerical simulations. 
Figure~\ref{fig:SI_TC_MC_bound1} shows that the mass density does not grow
for different initial conditions and different values of the parameter ${I}/{\Tilde{k}} \equiv {\text{max}\big(I_{\text{S}},I_{\text{P}}\big)}/\text{min}\big({k}^{e}_{\ch{S}},k^{e}_{\ch{P}}\big)$, representing the ratio between influx and outflux.
Note that the autocatalytic CRN does not grow even for large ${I}/{\Tilde{k}}$, namely, large influx rates.

\paragraph*{Case II.} We now consider the autocatalytic CRN when $\text{S}$ is being injected with rate $I_{\text{S}}>0$ while $\ch{P}$ is extracted with rate ${k^{e}_{\ch{P}}}[\text{P}]$. 
As for the previous case, 
we first analytically prove that the autocatalytic CRN cannot grow with a monotonically increasing mass density
and then we use numerical simulations to show that the mass density does not grow.

Let us assume that $d_{t}L^{m} > 0$ which, 
according to Eq.~\eqref{Eq:Bound_chemo_mixed}, implies 
\begin{equation}\label{Eqn:Bounded_chemo_mixed_core2}
       [\text{P}] \leq \frac{I_{\text{S}}}{k^{e}_{\ch P}}\,.
\end{equation}
\begin{figure*}
    \centering
    \includegraphics{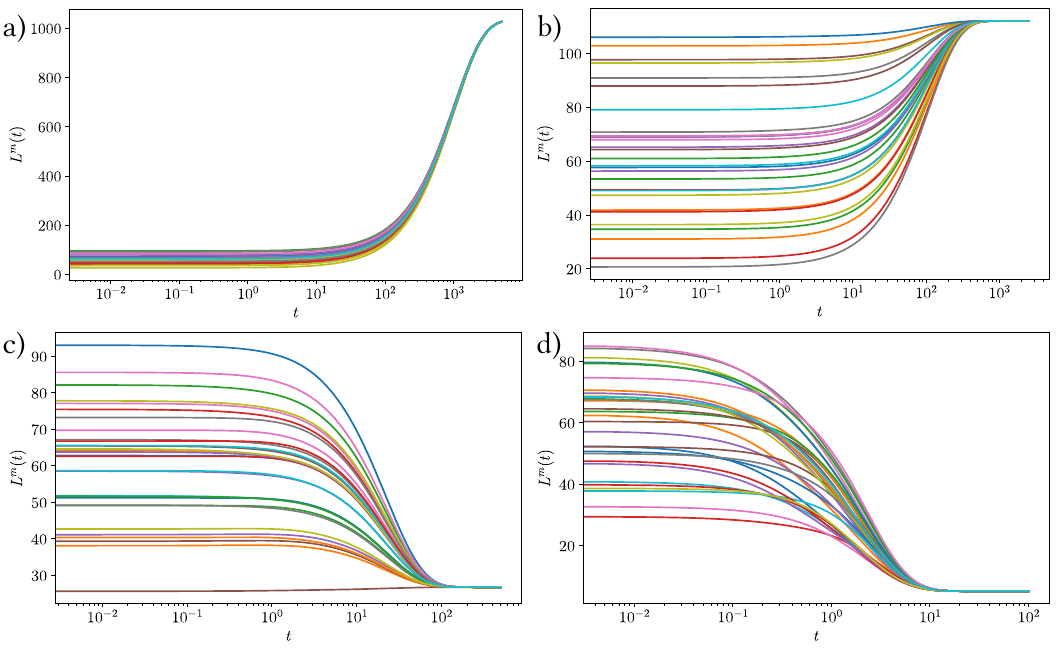}
    \caption{
  Evolution of the mass density of autocatalytic CRN under mixed control when the species $\text{S}$ has influx rate $I_{\text{S}} > 0$ while $\ch{P}$ is extracted with a rate $k^{e}_{\ch{P}}[\text{P}]$ for thirty randomly chosen initial conditions is plotted and a) ${I}/{\Tilde{k}} = 500$, b) ${I}/{\Tilde{k}} = 50$, c) ${I}/{\Tilde{k}} = 10$, d) ${I}/{\Tilde{k}} = 0.5$, where ${I}/{\Tilde{k}} = I_{\text{S}}/k^{e}_{\ch P}$. 
  Here, $k_{\pm 1}=k_{\pm 2} = k_{\pm 3} =1$ and $I_{\text{S}} = 1.0$.
}
    \label{fig:SI_TC_MC_bound2}
\end{figure*}
Then, the rate equation~\eqref{eq:req} for \ch{P} under mixed control
\begin{equation}\label{Eq:P_evol_MC}
    d_{t}[\text{P}] + (k_{+2}+k^{e}_{\ch P})[\text{P}] = k_{-2}[\text{E}^{*}]\,,
\end{equation}
has the formal solution,
\begin{equation}\label{Eq:P_evol_MC_sol}
\begin{split}
    [\text{P}](t) =&  [\text{P}](0)e^{-(k_{+2}+k^{e}_{\ch P})t}\\
    &+ k_{-2}\int_{0}^{t} ds  [\text{E}^{*}](s)e^{-(k_{+2}+k^{e}_{\ch P})(t-s)}\,,
\end{split}
\end{equation}
which implies that $[\text{E}^{*}]$ is bounded (according to the exact same reasoning as below Eq.~\eqref{Eq:S_evol_MC_sol}). Hence, only $[\ch{S}]$ and $[\ch{E}]$ can grow. 
Let us consider the sum $\mathbb{A} = [\ch{S}] + [\ch{E}]$ which is unbounded if the CRN grows. From Eq.~\eqref{eq:req} and the definition of mixed control, $\mathbb{A}$ follows
\begin{equation}\label{Eq:A_evol_MC}
    d_{t}\mathbb{A} = I_{\text{S}} + 2k_{+3}[\ch{E}][\ch{E}^{*}]-2k_{-3}[\ch{E}]^{3}\,.
\end{equation}
By using now that $[\text{E}^{*}]$ is bounded and assuming that $d_{t}\mathbb{A}>0$, 
$[\ch{E}](t)$ satisfies $2k_{-3}[\ch{E}]^{3}< I_{\text{S}} + 2k_{+3}[\ch{E}][\ch{E}^{*}]< I_{\text{S}} + 2C[\ch{E}]$ (with $C$ being a positive number such that $C>k_{+3}[\ch{E}^{*}]$), or, equivalently, $[\ch{E}]$ is bounded. 
Finally, rate equation~\eqref{eq:req} for $\ch{S}$ satisfies
\begin{equation}\label{Eq:S_evol_MC2}
     d_{t}[\text{S}] + k_{+1}[\text{S}] = k_{-1}[\text{E}] + I_{\text{S}} < C'\,,
\end{equation}
where $C'$ is an appropriately chosen positive number, with the formal solution
\begin{equation}\label{Eq:S_evol_MC2_sol}
    [\text{S}](t) \leq [\text{S}](0)e^{-k_{+1}t} + \frac{C'}{k_{+1}}\left(1-e^{-k_{+1}t}\right)<\infty\,,
\end{equation}
implying that $[\ch{S}]$ is also bounded. 
Thus, the autocatalytic CRN does not grow with $d_tL^m>0$ and $d_t\mathbb{A}>0$.

 
We now turn to numerical simulations. 
Figure~\ref{fig:SI_TC_MC_bound2} shows that the mass density does not grow
for different initial conditions and different values of the parameter ${I}/{\Tilde{k}} = I_{\text{S}}/k^{e}_{\ch P}$.
Note that also in this case the autocatalytic CRN does not grow even for large ${I}/{\Tilde{k}}$, namely, large influx rates.

\subsubsection{Michaelis Menten}
\label{Sec:MM_Mixed}

We consider the Michaelis Menten CRN when the substrate $\text{S}$ is injected at the rate $I_{\ch{S}}>0$ and the product $\text{P}$ is extracted at the rate $k^{e}_{\ch P}[\text{P}]$. 

Note that, as in Sec.~\ref{Sec:MM_FLux_control}, this chemostatting procedure does not break the conservation law $\ell^{\text{E}}$ and the concentrations $[\text{E}]$ and $[\text{ES}]$ are thus bounded. 
Furthermore, according to Eq.~\eqref{Eq:Bound_chemo_mixed},  if the CRN grows with $d_{t}L^{\ch S} > 0$, the concentration $[\ch{P}]$ is bounded above by
\begin{equation}\label{Eq:MM_MC_P_growth}
    [\text{P}] \leq \frac{I_{\text{S}}}{k^{e}_{\ch P}}\,.
\end{equation}
Hence, 
only the concentration $[\ch{S}]$ can grow. 

\begin{figure}
    \centering
    \begin{minipage}{0.5\textwidth}
        \centering
        \includegraphics[width=\textwidth]{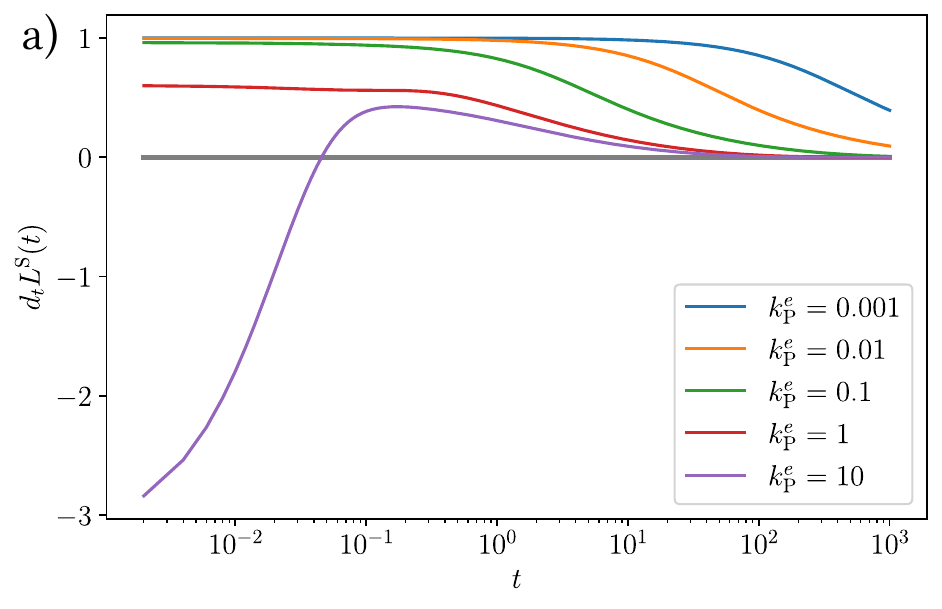}  
    \end{minipage}\hfill
    \begin{minipage}{0.5\textwidth}
        \centering
        \includegraphics[width=\textwidth]{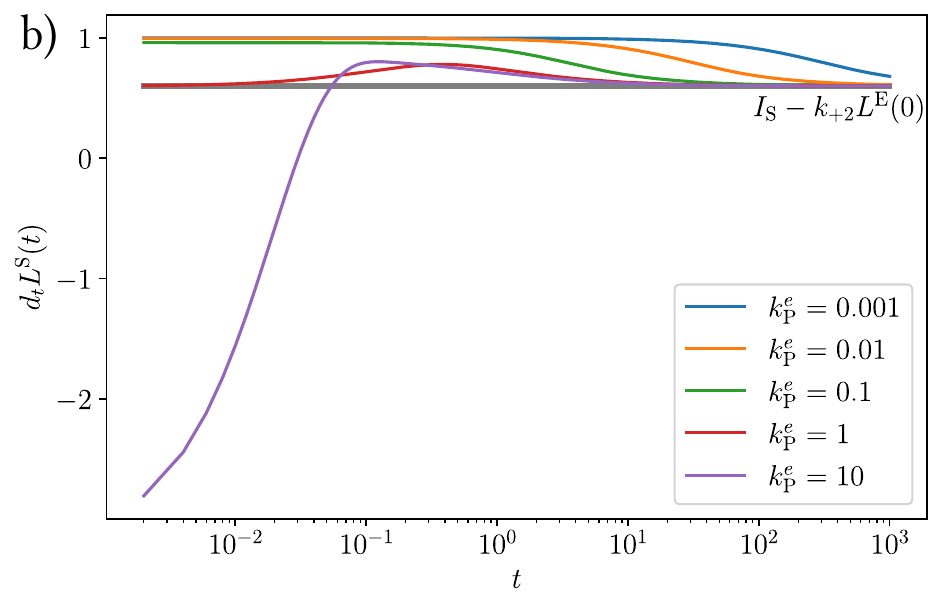}  
    \end{minipage}
\caption{Evolution of $d_tL^{\ch S}$ for the Michaelis Menten CRN under mixed control when $\text{S}$ is injected at the constant rate  $I_{\text{S}} = 1.0$ and \ch{P} is extracted at the rate $I_{\text{P}} = - k^{e}_{\ch{P}}[\text{P}]$. 
a) Initial condition $[\text{E}](0)=0.3$, $[\text{ES}](0)=0.8$, $[\text{S}](0)=1.9$, and $[\text{P}](0)=0.4$ 
b) Initial condition $[\text{E}](0)=0.3$, $[\text{ES}](0)=0.1$, $[\text{S}](0)=1.9$, and $[\text{P}](0)=0.4$. 
Here $k_{\pm 1} = k_{\pm 2} = 1$.}
\label{fig:SI_MM_MC1}
\end{figure}

\begin{figure}
    \centering
    \begin{minipage}{0.5\textwidth}
        \centering
        \includegraphics[width=\textwidth]{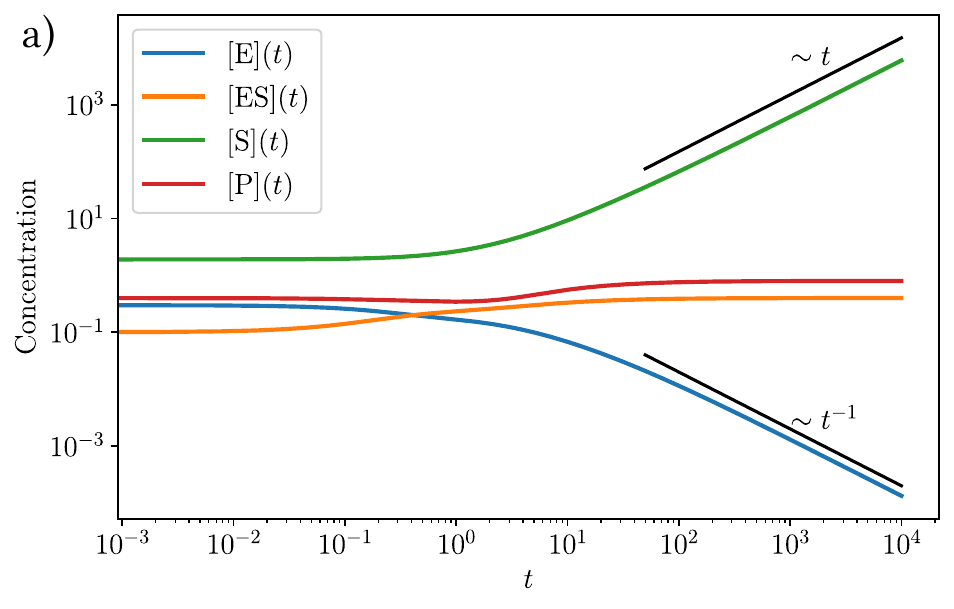}  
    \end{minipage}\hfill
    \begin{minipage}{0.5\textwidth}
        \centering
        \includegraphics[width=\textwidth]{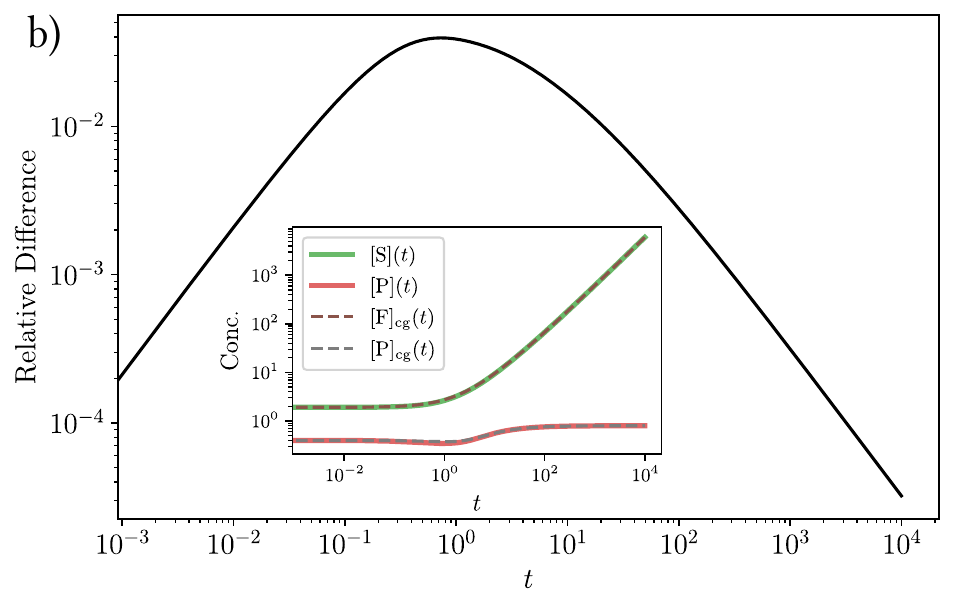}  
    \end{minipage}
\caption{Dynamics of the Michaelis Menten CRN under mixed control. Evolution of 
a) Concentrations b) Relative difference, $\lVert\boldsymbol{y} (t)-\boldsymbol{y}_{\text{cg}}(t)\rVert/\lVert\boldsymbol{y}_{\text{cg}}(t)\rVert$, between the concentrations $\boldsymbol{y}(t) = ([\ch{S}](t), [\ch{P}](t))^{\intercal}$ and the coarse grained concentrations $\boldsymbol{y}_{\text{cg}}(t) = ([\ch{S}]_{\text{cg}}(t), [\ch{P}]_{\text{cg}}(t))^{\intercal}$ (given by Eq.~\eqref{Eq:Coarse_grained_dynamics}) and comparison between the concentrations and the coarse-grained concentrations in the inset.
Here, $[\text{E}](0)=0.3$, $[\text{ES}]{0}=0.1$, $[\text{S}](0)=1.9$, $[\text{P}](0)=0.4$, $I_{\ch{S}} = 1.0$, $k^{e}_{\ch{P}} = 0.5$, $k_{\pm 1} = k_{\pm 2} = 1$. } 
\label{fig:SI_MM_MC2}
\end{figure}

Strikingly, as shown in Fig.~\ref{fig:SI_MM_MC1}, depending on the initial conditions but independently of the value of $k^{e}_{\ch P}$, $d_{t}L^{\ch{S}}$ can either vanish or reach a constant positive value.
In the latter case, the typical evolution of the concentrations is shown in  Fig. \ref{fig:SI_MM_MC2}a: 
$[\ch{S}]$ grows linearly in time,
$[\ch{P}]$  and $[\ch{ES}]$ saturate to constant values, while $[\ch{E}]$ decreases in time. 
Note that, as shown in Fig. \ref{fig:SI_MM_MC2}b, 
the evolution of the concentrations $[\ch{S}]$ and $[\ch{P}]$ is well approximated by the coarse-grained dynamics obtained solving Eq.~\eqref{Eq:Coarse_grained_dynamics}.

When $[\ch{S}]$ grows, a time scale separation emerges between the evolution of the concentrations $[\ch{S}]$ and $[\ch{P}]$, on the one hand, and the concentrations of the internal species, on the other hand. Thus, following Sec.~\ref{Sec:Supp_dynamics}, we now use the coarse-grained dynamics~\eqref{Eq:Coarse_grained_dynamics} to determine
i) when the Michaelis Menten CRN grows,
ii) the growth rate of $[\ch{S}]$, and 
iii) the evolution of $[\ch{P}]$, $[\ch{E}]$ and $[\ch{ES}]$ in the long time limit.
By using the diagrammatic method~\cite{Hill1966} as done in Refs.~\cite{Wachtel2018,Avanzini2020b}, Eq.~\eqref{Eq:Coarse_grained_dynamics} for the Michaelis Menten CRN under mixed control reads
\small
\begin{align}\label{Eqn:Coarse_grained_MM_MC_S}
    d_{t}[\text{S}] &= I_{\text{S}}-L^{E}(0)\left[\frac{k_{+1}k_{+2}[\text{S}]-k_{-1}k_{-2}[\text{P}]}{k_{-1}+k_{+2}+k_{+1}[\text{S}]+k_{-2}[\text{P}]}\right]\,,\\
    d_{t}[\text{P}] &= L^{\ch{E}}(0)\left[\frac{k_{+1}k_{+2}[\text{S}]-k_{-1}k_{-2}[\text{P}]}{k_{-1}+k_{+2}+k_{+1}[\text{S}]+k_{-2}[\text{P}]}\right]-k^{e}_{\ch P}[\text{P}]\,,\label{Eqn:Coarse_grained_MM_MC_P}
\end{align}
\normalsize
which, together with the ansatz $[\text{S}] = v_{\text{S}}t + c_{\ch S}$ (for some constant parameters $v_{\text{S}}$ and $c_{\ch S}$) leads (in the long time limit) to
\begin{align}
     v_{\text{S}}   &= I_{\text{S}}-k_{+2}L^{E}(0)\label{Eq:Growth_rate_S_MM_MC}\,,\\
    d_{t}[\text{P}] &= k_{+2}L^{E}(0) - k^{e}_{\ch P}[\text{P}]\,,\label{Eq:Dyn_P_MM_MC}
\end{align}
by neglecting terms of order $\mathcal{O}(1/t)$.
Here, $k_{+2}L^{E}(0)$ represents the rate of interconversion of \ch{S} into \ch{P} in the long time limit.
On the one hand, Eq.~\eqref{Eq:Growth_rate_S_MM_MC} implies that the concentration 
$[\ch S]$ (and, correspondingly, the moiety $L^{\ch S}$) grows, i.e., $v_{\text{S}} >0$, 
if $I_{\text{S}} > k_{+2}L^{\ch{E}}(0)$ independently of $k^{e}_{\ch P}$.
This physically means that growth happens when \ch{S} is injected faster 
than its (enzyme-dependent) interconversion into \ch{P}, which is then extracted.
Note that the numerical simulations shown in Fig.~\ref{fig:SI_MM_MC1}a (resp. Fig.~\ref{fig:SI_MM_MC1}b) satisfy $I_{\text{S}} < k_{+2}L^{\ch{E}}(0)$ (resp. $I_{\text{S}} > k_{+2}L^{\ch{E}}(0)$).
On the other hand, Eq.~\eqref{Eq:Dyn_P_MM_MC} implies that $[\ch{P}]$ reaches the limiting value $k_{+2}L^{E}(0)/k^{e}_{\ch P}$, consistent with Eq.~\eqref{Eq:MM_MC_P_growth}.
Finally, according to the diagrammatic method~\cite{Hill1966}, $[\ch{E}]_{\text{ss}}$ and $[\ch{ES}]_{\text{ss}}$ still follow Eqs~\eqref{Eqn:E_soln_MM_flux} and, therefore, using Eqs.~\eqref{Eq:Growth_rate_S_MM_MC} and \eqref{Eq:Dyn_P_MM_MC},  
$[\ch{E}] = \mathcal{O}(t^{-1})$ and $[\ch{ES}] = \mathcal{O}(1)$  (as observed in Fig.~\ref{fig:SI_MM_MC2}a).

\begin{figure}
    \centering
    \begin{minipage}{0.5\textwidth}
        \centering
        \includegraphics[width=\textwidth]{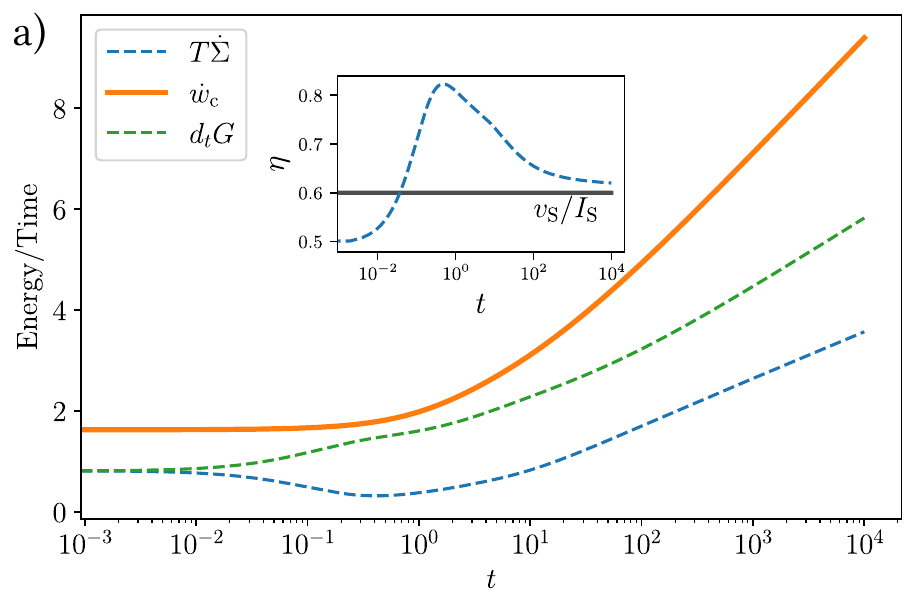}  
    \end{minipage}\hfill
    \begin{minipage}{0.5\textwidth}
        \centering
        \includegraphics[width=\textwidth]{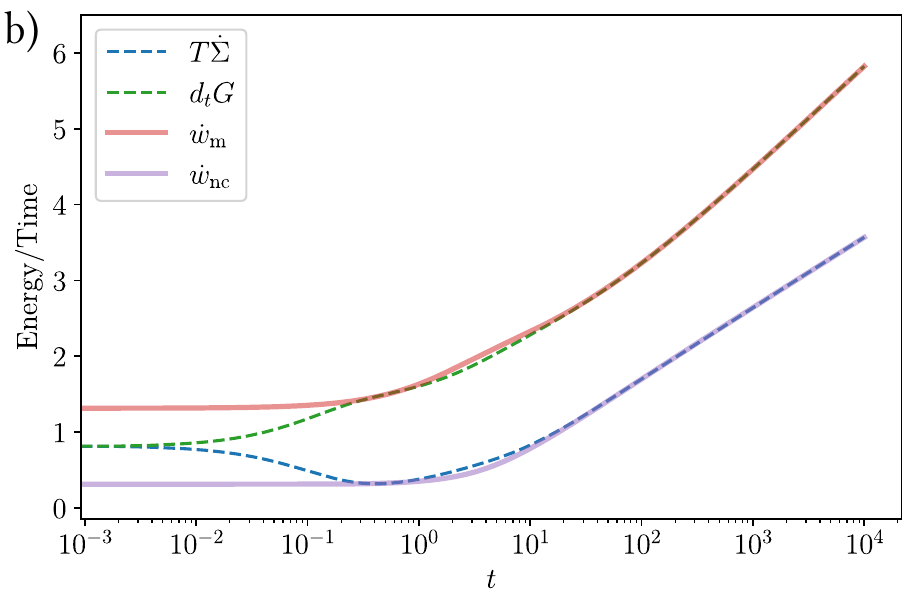}  
    \end{minipage}
\caption{Thermodynamics of the Michaelis Menten CRN under mixed control under the dynamics in Fig.~\ref{fig:SI_MM_MC2}. Evolution of a)  EPR, the chemical work rate, and $d_{t}G$ and efficiency of growth (inset). b) EPR, moiety work rate, nonconservative work rate, and $d_{t}G$. Here, $\mu^{0}_{\ch{E}} = 1$, $\mu^{0}_{\ch{ES}} = 2$, $\mu^{0}_{\ch{S}} = 1$, $\mu^{0}_{\ch{P}} = 1$. } 
\label{fig:SI_MM_MC3}
\end{figure}

We now examine the thermodynamics of the Michaelis Menten CRN under mixed control when  $I_{\text{S}} > k_{+2}L^{\ch{E}}(0)$. 
We start by considering the EPR~\eqref{Eqn:EPR_two_defn}, $d_{t}G$~\eqref{eq:gibbs} and chemical work rate~\eqref{Eqn:chemical_work_rate}, whose typical time evolution is plotted in Fig.~\ref{fig:SI_MM_MC3}a.
The EPR increases in time implying the Michaelis Menten CRN under mixed control grows out of equilibrium. 
The chemical work rate and $d_{t}G$ increase in time at the same rate as the EPR. Consequently, the efficiency of growth~\eqref{Eq:effiency_defn} saturates to a value smaller than one (see the inset in Fig.~\ref{fig:SI_MM_MC3}a).
We then examine the nonconservative work rate~\eqref{Eqn:nonconserv_work_defn} and the moiety work rate~\eqref{eqn:moiety_work_defn} whose typical evolution is shown in Fig.~\ref{fig:SI_MM_MC3}b
(when $\ch{S}$ is chosen as the potential species). Both the nonconservative work rate and the moiety work rate increase with time.
Crucially, there is a splitting between the magnitudes of the EPR and the nonconservative work rate on the one hand, and the moiety work rate and $d_{t}G$ on the other hand. 

We explain this behavior of the thermodynamic quantities by using the coarse-grained dynamics Eqs.~\eqref{Eqn:Coarse_grained_MM_MC_S} and~\eqref{Eqn:Coarse_grained_MM_MC_P} and the ansatz $[\ch{S}] = v_{\ch{S}}t + c_{\ch{S}}$.
We start by recognizing that the chemical potential~\eqref{Eqn:chem_pot_defn} 
of \ch{S} and \ch{P} in the long time limit read 
\begin{subequations}
\begin{equation}\label{Eq:MM_MC_Schem}
\mu_{\text{S}} = \mu^{0}_{\text{S}} + RT\text{ln}\left(v_{\text{S}}\right) + RT\text{ln}(t) =  \mathcal{O}(\text{ln}(t))\,,
\end{equation}
\begin{equation}\label{Eq:MM_MC_Pchem}
\mu_{\text{P}} =  \mu^{0}_{\ch{P}} + RT\text{ln}\left(\frac{k_{+2}L^{E}(0)}{k^{e}_{\ch P}}\right) = \mathcal{O}(1)\,,
\end{equation}    
\end{subequations}
respectively, by neglecting terms of order $\mathcal{O}(1/t)$.
Then, the EPR can be expressed, according to Ref.~\cite{Avanzini2020b}, as
\begin{equation} \label{Eqn:EPR_MM_MC}
     T\dot{\Sigma} = \left(\mu_{\text{S}}-\mu_{\text{P}}\right) \bar{j}_{1}= \mathcal{O}(\text{ln}(t))\,,
\end{equation}
where, by using Eq~\eqref{Eq:Coarse_grained_dynamics},
the ansatz $[\ch{S}] = v_{\ch{S}}t + c_{\ch{S}}$,
and Eqs.~\eqref{Eq:MM_MC_Schem} and \eqref{Eq:MM_MC_Pchem},
\begin{equation}\label{Eq:Currents_MM_MC}
    \bar{j}_{1} = k_{+2}L^{E}(0)\,,
\end{equation}
and 
\begin{equation}\label{Eq:Chemical_pot_MM_MC}
\begin{split}
    \mu_{\text{S}}-\mu_{\text{P}}
              &= (\mu^{0}_{\ch{S}} - \mu^{0}_{\ch{P}}) + RT\text{ln}\left(\frac{k^{e}_{\ch P}v_{\ch{S}}}{k_{+2}L^{E}(0)}\right)+ RT\text{ln}(t)\\&= \mathcal{O}(\text{ln}(t))\,.
\end{split}              
\end{equation}
Namely, we recover the positive increase in the entropy production rate observed
in Figs.~\ref{fig:SI_MM_MC3}a and ~\ref{fig:SI_MM_MC3}b.
By using Eqs.~\eqref{Eq:MM_MC_Schem} and \eqref{Eq:MM_MC_Pchem}, $d_{t}G$~\eqref{eq:gibbs} and the chemical work rate~\eqref{Eqn:chemical_work_rate} become
\begin{align}
\begin{split}
    \dot{w}_{c} 
    &= RT\text{ln}(t)I_{\ch{S}} + (\mu^{0}_{\ch{S}}I_{\ch{S}}-\mu^{0}_{\ch{P}}k_{+2}L^{E}(0)) + RT\text{ln}\left(v_{\text{S}}\right)I_{\ch{S}}\\
    &{}~~~-RT\text{ln}\left(\frac{k_{+2}L^{E}(0)}{k^{e}_{\ch P}}\right)k_{+2}L^{E}(0)  = \mathcal{O}(\text{ln}(t))\label{Eqn:chem_work_MM_MC}\,,\end{split}\\   
    d_{t}G  &= RT\text{ln}(t)v_{\ch{S}} + \mu^{0}_{\ch{S}}v_{\text{S}} + RT\text{ln}\left(v_{\text{S}}\right)v_{\ch{S}} = \mathcal{O}(\text{ln}(t))\label{Eq:dt_G_MM_MC} \,,
\end{align}
namely, they grow logarithmically in time like the EPR in agreement with Fig.~\ref{fig:SI_MM_MC3}a. Furthermore, plugging Eqs. \eqref{Eq:dt_G_MM_MC} and \eqref{Eqn:chem_work_MM_MC} in Eq.~\eqref{Eq:effiency_defn}, the efficiency saturates in the long time limit to the value 
\begin{equation}\label{Eq:efficiency_MM_MC}
        \eta = 1-\frac{k_{+2}L^{E}(0)}{I_{\text{S}}} = \frac{v_{\ch{S}}}{I_{\ch{S}}} < 1\,,
\end{equation}
as shown in Fig.~\ref{fig:SI_MM_MC3}a.
 
We conclude by examining  the moiety work rate and the nonconservative work rate.
By using Eq.~\eqref{Eq:MM_MC_Schem}
in Eqs.\eqref{eqn:moiety_work_defn} and \eqref{Eqn:nonconserv_work_defn}, we obtain 
\begin{align}
\label{Eqn:moeity_work_MM_MC}
   \dot{w}_{\text{m}} &= RT\text{ln}(t)v_{\ch{S}} + RT\text{ln}\left(v_{\text{S}}\right)v_{\ch{S}} +  \mu^{0}_{\ch{S}}v_{\text{S}}  = \mathcal{O}(\text{ln}(t))\,,\\
\begin{split}   
  \dot{w}_{\text{nc}} &= RT\text{ln}(t)\bar{j}_{1} + (\mu^{0}_{\ch{S}} - \mu^{0}_{\ch{P}})\bar{j}_{1}\\ &{}~~~~+  RT\text{ln}\left(\frac{k^{e}_{\ch P}v_{\ch{S}}}{k_{+2}L^{E}(0)}\right)\bar{j}_{1}   = \mathcal{O}\left(\text{ln}(t)\right) \,,\label{Eqn:noncosnerv_work_MM_MC}\end{split}
\end{align}
implying 
\begin{equation}\label{Eqn:Second_Law_MM_MC}
\begin{split}
   d_{t}{G} &\sim \dot{w}_{\text{m}} = \mathcal{O}(\text{ln}(t))\,,\\
   \dot{w}_{\text{nc}} &\sim T\dot{\Sigma}  = \mathcal{O}(\text{ln}(t))\,.\\ 
\end{split}
\end{equation}
in the long time limit.

\subsubsection{Cyclic Michaelis Menten}
\label{Sec:CMM_MC}
We consider the Cyclic Michaelis Menten CRN under mixed control when $\text{S}$ is injected at the rate $I_{\text{S}} > 0$ and $\text{P}$ is extracted at the rate $k^{e}_{\ch P}[\ch{P}]$. 
As in the case of the Michaelis Menten CRN under mixed control, Sec.~\ref{Sec:MM_Mixed}, the conservation law $\ell^{\text{E}}$ is unbroken implying that the concentrations $[\text{E}]$ and $[\text{ES}]$ are bounded in time.
Furthermore, if the CRN grows with $d_{t}L^{\ch S} > 0$, Eq.~\eqref{Eq:Bound_chemo_mixed} becomes 
\begin{equation}\label{Eq:CMM_MC_P_growth}
    [\text{P}] \leq \frac{I_{\text{S}}}{k^{e}_{\ch P}}\,,
\end{equation}
namely, $[\ch{P}]$ is also bounded. Furthermore, by using Eqs.~\eqref{eq:req},~\eqref{Eq:CMM_MC_P_growth} and $[\ch{ES}] \leq L^{\ch{E}}(0)$,
$[\ch{S}]$ satisfies
\begin{equation}\label{Eq:CMM_MC_S_dyn1}
    d_{t}[\ch{S}] + k_{+3}[\ch{S}] \leq k_{-1}L^{\ch{E}}(0) + I_{\text{S}}\left(1 + \frac{k_{-3}}{k^{e}_{\ch P}}\right)\,.
\end{equation}
implying that $[\ch S]$ is bounded too. We conclude that the  Cyclic Michaelis Menten CRN does not grow with $d_{t}L^{\ch S} > 0$.

\begin{figure*}
    \centering
    \includegraphics[scale = 1.0]{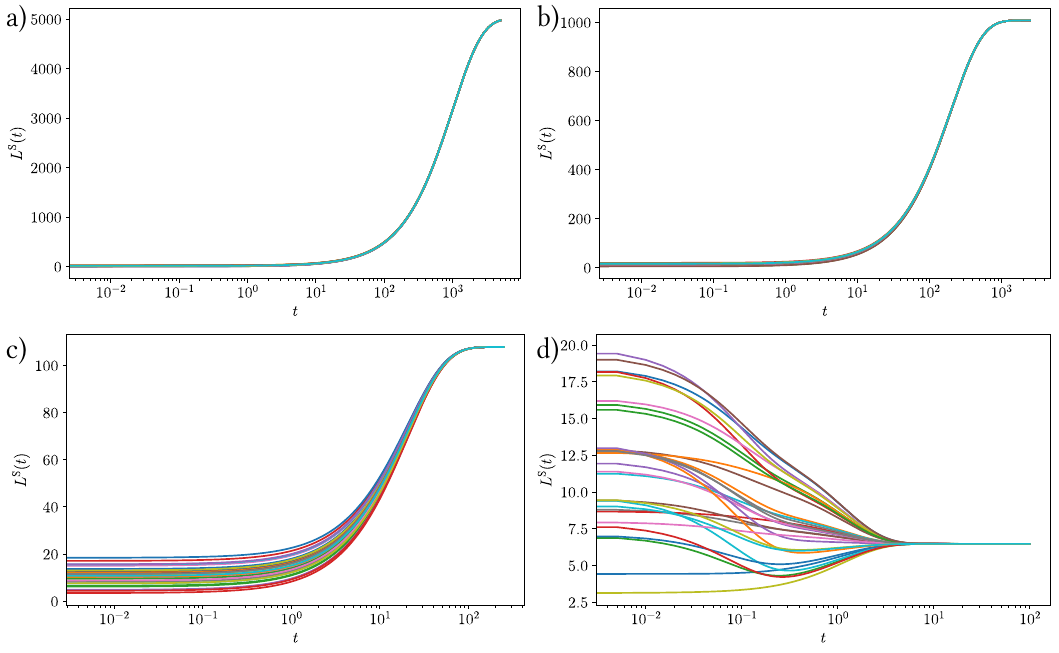}
    \caption{ 
Evolution of $L^{\ch S}$ of the Cyclic Michaelis Menten CRN under mixed control when the species $\text{S}$ is injected at rate $I_{\text{S}} = 5.0$ while $\ch{P}$ is extracted at rate $I_{\text{P}} = -k^{e}_{\ch{P}}[\text{P}]$ for thirty randomly chosen initial conditions (consistent with $L^{E}(0) = 3.0$) and a) ${I}/{\Tilde{k}} = 2500$, b) ${I}/{\Tilde{k}} = 500$, c) ${I}/{\Tilde{k}} = 50$, d) ${I}/{\Tilde{k}} = 0.5$  where ${I}/{\Tilde{k}} = I_{\text{S}}/{k^{e}_{\ch{P}}}$.
Here,  $k_{\pm 1}=k_{\pm 2} = k_{\pm 3} =1$.
}
\label{Fig:SI_CMM_MC_bound1}
\end{figure*}

We now verify this result numerically. Figure~\ref{Fig:SI_CMM_MC_bound1} shows the typical evolution of the moiety concentration $L^{\ch{S}}$ for thirty different initial conditions and for four values of the parameter ${I}/{\Tilde{k}} = I_{\text{S}}/{k^{e}_{\ch{P}}}$ (which represents the ratio of influx to outflux rates). We see that the CRN always relaxes to a steady state, even for large ${I}/{\Tilde{k}}$, namely large influx rates.

The direct interconversion of \ch{S} into \ch{P} is the reason why the Cyclic Michaelis Menten CRN does not grow under mixed control unlike the Michaelis Menten CRN (see Sec.~\ref{Sec:MM_Mixed}).
On the one hand, in the Michaelis Menten CRN
the rate of interconversion of \ch{S} into \ch{P} (via the enzymatic mechanism) is limited by $k_{+2}L^{\ch E}(0)$ (see discussion below Eq.~\eqref{Eq:Growth_rate_S_MM_MC}). 
Hence, when $I_{\ch S} > k_{+2}L^{\ch E}(0)$, $\ch{S}$ accumulates and the CRN grows.
On the other hand, the rate of the direct interconversion of \ch{S} into \ch{P} in the Cyclic Michaelis Menten CRN increases with the concentration $[\ch{S}]$, implying that \ch S is rapidly interconverted into \ch P which is rapidly extracted as the concentrations $[\ch S]$ increases.
Hence, $\ch{S}$ cannot accumulate and the CRN does not grow.

\subsubsection{Minimal Metabolic CRN}\label{Sec:Met_MC}
We consider the minimal metabolic CRN under mixed control when the species $\text{S}$ and $\text{F}$ are injected at rates $I_{\ch{S}}>0$ and $I_{\ch{F}}>0$, respectively, while the species $\text{P}$ and $\text{W}$ are extracted at rates ${k^{e}_{\ch P}}[\ch{P}]$ and ${k^{e}_{\ch W}}[\ch{W}]$, respectively. 

\begin{figure}
    \centering
    \begin{minipage}{0.5\textwidth}
        \centering
        \includegraphics[width=\textwidth]{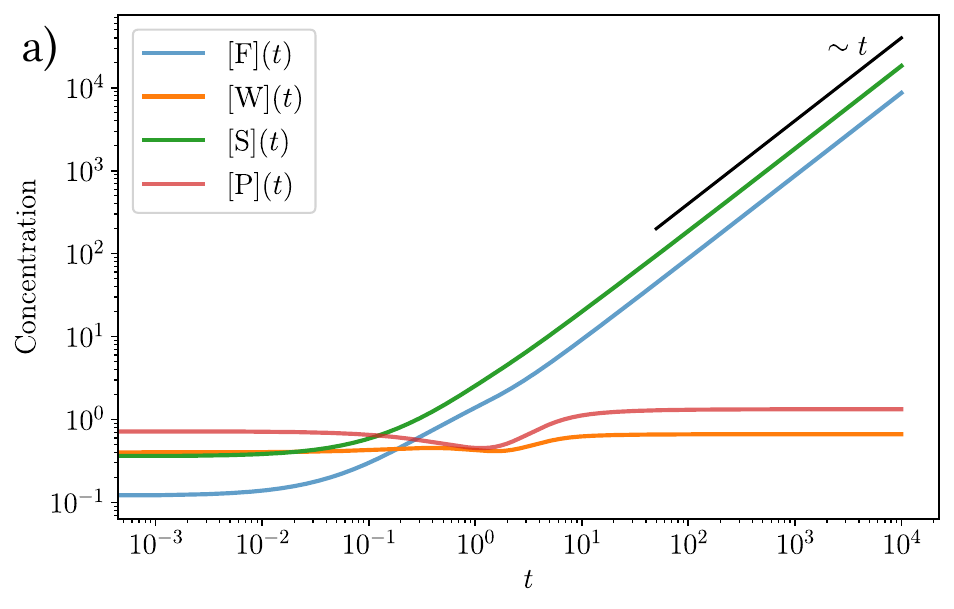}  
    \end{minipage}\hfill
    \begin{minipage}{0.5\textwidth}
        \centering
        \includegraphics[width=\textwidth]{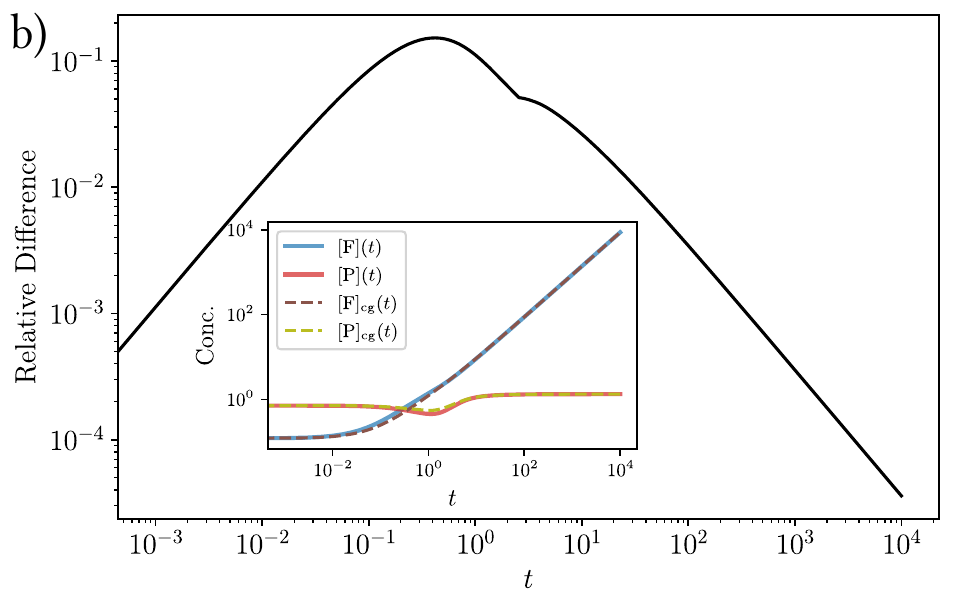}  
    \end{minipage}
\caption{Dynamics of the minimal metabolic CRN under mixed control when $\ch{F}$ and $\ch{S}$ are injected at rates $I_{\ch{F}} = 1.2$, $I_{\ch{S}} = 2.5$ and $\ch{W}$ and $\ch{P}$ are extracted at rates $k^{e}_{\ch{W}}[\ch{W}]$ and ${k^{e}_{\ch{P}}}[\ch{P}]$ with ${k^{e}_{\ch{W}} =  k^{e}_{\ch{P}}} = 0.5$ . Evolution of
a) Concentrations of the chemostatted species $[\ch{F}](t)$, $[\ch{W}](t)$, $[\ch{S}](t)$ and $ [\ch{P}](t)$ b) Relative difference between the vectors $\boldsymbol{y}(t) = ([\ch{F}](t),[\ch{W}](t)[\ch{S}](t), [\ch{P}](t))^{\intercal}$ and $\boldsymbol{y}_{\text{cg}}(t) = ([\ch{F}]_{\text{cg}}(t),[\ch{W}]_{\text{cg}}(t),[\ch{S}]_{\text{cg}}(t), [\ch{P}]_{\text{cg}}(t))^{\intercal}$ (given by Eq.~\eqref{Eq:Coarse_grained_dynamics}), and comparison between the concentrations $[\ch{F}](t)$ and $[\ch{P}](t)$ with the coarse-grained concentrations $[\ch{F}]_{\text{cg}}(t)$ and $[\ch{P}]_{\text{cg}}(t)$ in the inset.
Here, $k_{\pm 1} = k_{\pm2} = k_{\pm3} = k_{\pm 4} = k_{\pm 5} = 1$, 
$[\ch{E}](0) = 0.12$, $[\ch{ES}](0) = 0.68$,$[\ch{EW}](0) = 0.6$, $[\ch{E}^{*}](0) = 0.04$, $[\ch{F}](0) =0.12$, $[\ch{W}](0) = 0.4$, $[\ch{S}](0) = 0.36$,$[\ch{P}](0) = 0.72$.}

\label{fig:SI_Met_MC1}
\end{figure}

The minimal metabolic CRN under this chemostatting procedure has also been studied numerically in Ref.~\cite{avanzini2022flux}. It has been observed that depending on the values of the parameters $I_{\ch{S}}$, $I_{\ch{F}}$, $k^{e}_{\ch{P}}$ and $k^{e}_{\ch{W}}$, the CRN either relaxes to a non-equilibrium steady state or grows. 
From a dynamical standpoint, only $[\ch{S}]$ and $[\ch{F}]$ grow while $[\ch{P}]$ and $[\ch{W}]$ saturate to constant values. 
From a thermodynamic standpoint, when the CRN grows, a positive increasing EPR balanced by $\dot{w}_{\text{nc}}$ was observed.  

Here, we recreate and explain the above observations when the CRN grows. Figure \ref{fig:SI_Met_MC1}a shows the  typical evolution of the chemostatted species:
the concentrations $[\ch{F}]$ and $[\ch{S}]$ grow while both $[\ch{P}]$ and $[\ch{W}]$ saturate. 
Furthermore, as in the case of the Michaelis Menten CRN (see Sec.~\ref{Sec:MM_Mixed}), the chemostatted concentrations are well approximated by the coarse-grained dynamics obtained from Eq.~\eqref{Eq:Coarse_grained_dynamics}, as shown by Fig.~\ref{fig:SI_Met_MC1}b.

This can be explained by recognizing that the chemostatting procedure
does not break the conservation law $\ell^{E}$ (and the concentrations $[\ch{E}]$, $[\ch{ES}]$, $[\ch{EW}]$ and $[\ch{E}^{*}]$ are bounded), while
it breaks the conservation laws $\ell^{\ch{F}}$ and $\ell^{\ch{S}}$  with the moiety concentrations $L^{\text{S}}$ and $L^{\text{F}}$ following as $d_{t}L^{\ch{S}} = I_{\text{S}} - k^{e}_{\ch{P}}[\ch{P}]$ and $d_{t}L^{\ch{F}} = I_{\text{F}} - k^{e}_{\ch{W}}[\ch{W}]$.
However, 
\begin{equation}
    [\ch{P}],[\ch{W}] \leq \frac{I_{\ch{S}}+ I_{\ch{F}}}{\text{min}(k^{e}_{\ch{P}},k^{e}_{\ch{W}})}\,,
\end{equation}
according to Eq.~\eqref{Eq:Bound_chemo_mixed}. Hence, only the concentrations $[\ch{F}]$ and $[\ch{S}]$ can grow, as seen in Fig.~\ref{fig:SI_Met_MC1}a.
The condition for growth has the same origin as in the Michaelis Menten CRN under mixed control (see Sec.~\ref{Sec:MM_Mixed}). 
The rate of enzyme-mediated interconversion of $\ch{F}$ (resp. $\ch{S}$) into $\ch{W}$ (resp. $\ch{P}$) is bounded by some function of $L^{\ch{E}}(0)$. 
Thus, when $I_{\ch{F}}$ (resp. $I_{\ch{S}}$) is larger than this rate, $\ch{F}$ (resp. $\ch{S}$) accumulates and the CRN grows. 

\begin{figure}
    \centering
    \begin{minipage}{0.5\textwidth}
        \centering
        \includegraphics[width=\textwidth]{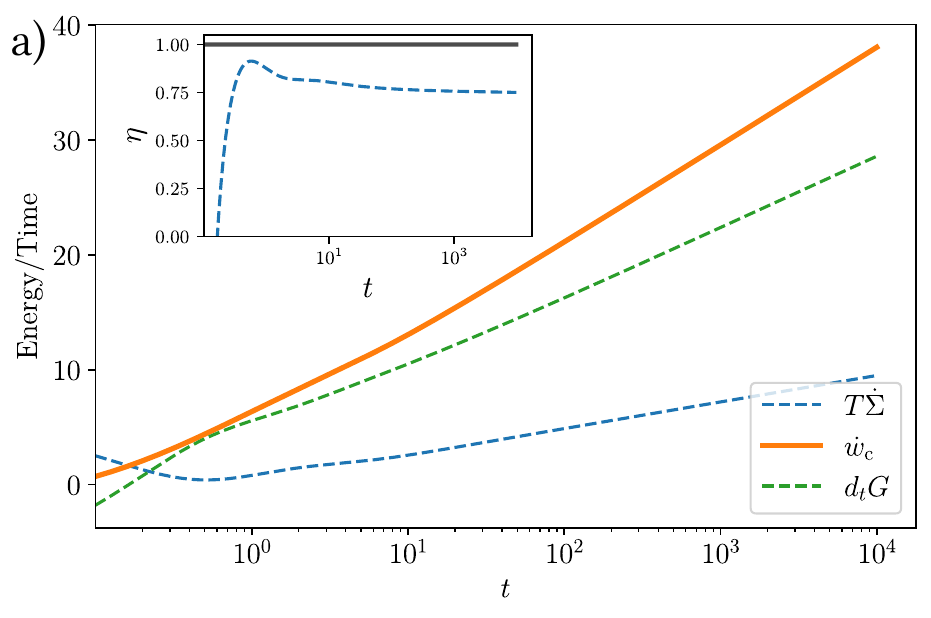}  
    \end{minipage}\hfill
    \begin{minipage}{0.5\textwidth}
        \centering
        \includegraphics[width=\textwidth]{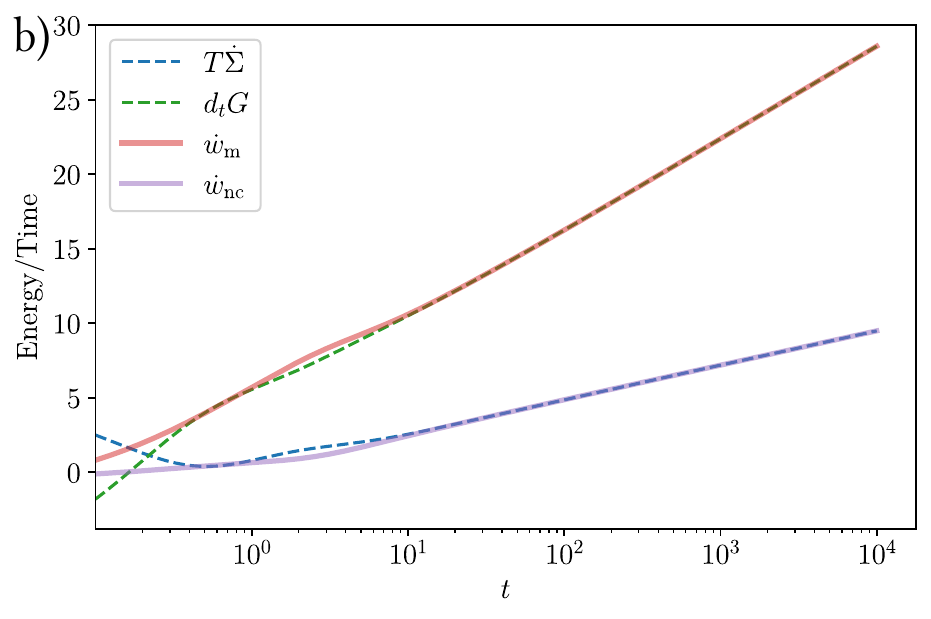}  
    \end{minipage}
\caption{
Thermodynamics of the metabolic CRN under mixed control under the dynamics in Fig.~\ref{fig:SI_Met_MC1}. Evolution of a)  EPR, the chemical work rate, and $d_{t}G$ and efficiency of growth (inset). b) EPR, moiety work rate, nonconservative work rate, and $d_{t}G$. Here, $\mu^{0}_{\ch{E}} = 1$, $\mu^{0}_{\ch{EF}} = 2$, $\mu^{0}_{\ch{EW}} = 2$, $\mu^{0}_{\ch{E^{*}}} = 3$, $\mu^{0}_{\ch{F}} = 1$, $\mu^{0}_{\ch{W}} = 1$  $\mu^{0}_{\ch{S}} = 1$ and $\mu^{0}_{\ch{P}} = 1$. 
} 
\label{fig:SI_Met_MC2}
\end{figure}

We now examine the corresponding thermodynamics.
The typical evolution of  the EPR~\eqref{Eqn:EPR_two_defn}, $d_{t}G$~\eqref{eq:gibbs} and chemical work rate~\eqref{Eqn:chemical_work_rate} are plotted in Fig.~\ref{fig:SI_Met_MC2}a. 
All the quantities increase with time and consequently, the efficiency of growth Eq.~\eqref{Eq:effiency_defn} saturates to a constant smaller than one (see inset in Fig.~\ref{fig:SI_Met_MC2}a), similar to the Michaelis Menten CRN in Fig.~\ref{fig:SI_MM_MC3}a.  

Figure~\ref{fig:SI_Met_MC2}b shows the typical evolution of the nonconservative work rate~\eqref{Eqn:nonconserv_work_defn}, the moiety work rate~\eqref{eqn:moiety_work_defn} along with the EPR and $d_{t}G$
(when $\ch{S}$ and $\ch{F}$ are chosen as the potential species): both the nonconservative work rate and moiety work rate increase with time.
Furthermore, we observe a splitting between the magnitudes of the EPR and the nonconservative work rate, on the one hand, and the moiety work rate and $d_{t}G$, on the other hand, implying that
\begin{equation}\label{Eqn:Second_Law_Met_MC}
\begin{split}
   d_{t}{G} &\sim \dot{w}_{\text{m}} \,,\\
   \dot{w}_{\text{nc}} &\sim T\dot{\Sigma} > 0\,,\\ 
\end{split}
\end{equation}
in the long time limit.
We note the similarity between the thermodynamics of the Metabolic CRN in Figs.~\ref{fig:SI_Met_MC2}a, \ref{fig:SI_Met_MC2}b and of the Michaelis Menten CRN in Figs.~\ref{fig:SI_MM_MC3}a, ~\ref{fig:SI_MM_MC3}b.

\subsubsection{Summary}
From the results of Sec.~\ref{Sec:Autocat_MC} - \ref{Sec:Met_MC}, we note the following requirements for growth under mixed control.

First, from Eq.~\eqref{Eq:Bound_chemo_mixed}, the CRN needs at least one chemostatted species that is solely injected (not extracted) to grow. 
Second, by comparing the autocatalytic CRN, which does not grow under mixed control (see Sec.~\ref{Sec:Autocat_MC}), with both the Michaelis Menten CRN and the minimal metabolic CRN, which grow under mixed control (see Sec.~\ref{Sec:MM_Mixed} and \ref{Sec:Met_MC}), we expect that CRNs need to have at least one unbroken conservation law to grow. 
Finally, given that the Cyclic Michaelis Menten CRN does not grow under mixed control (see Sec.~\ref{Sec:CMM_MC}), we expect that CRNs cannot have any reactions interconverting only the chemostatted species to grow. 

\section{Conclusions}\label{Sec:Discussion}

\begin{figure*}
    \centering
    \includegraphics[width = \textwidth]{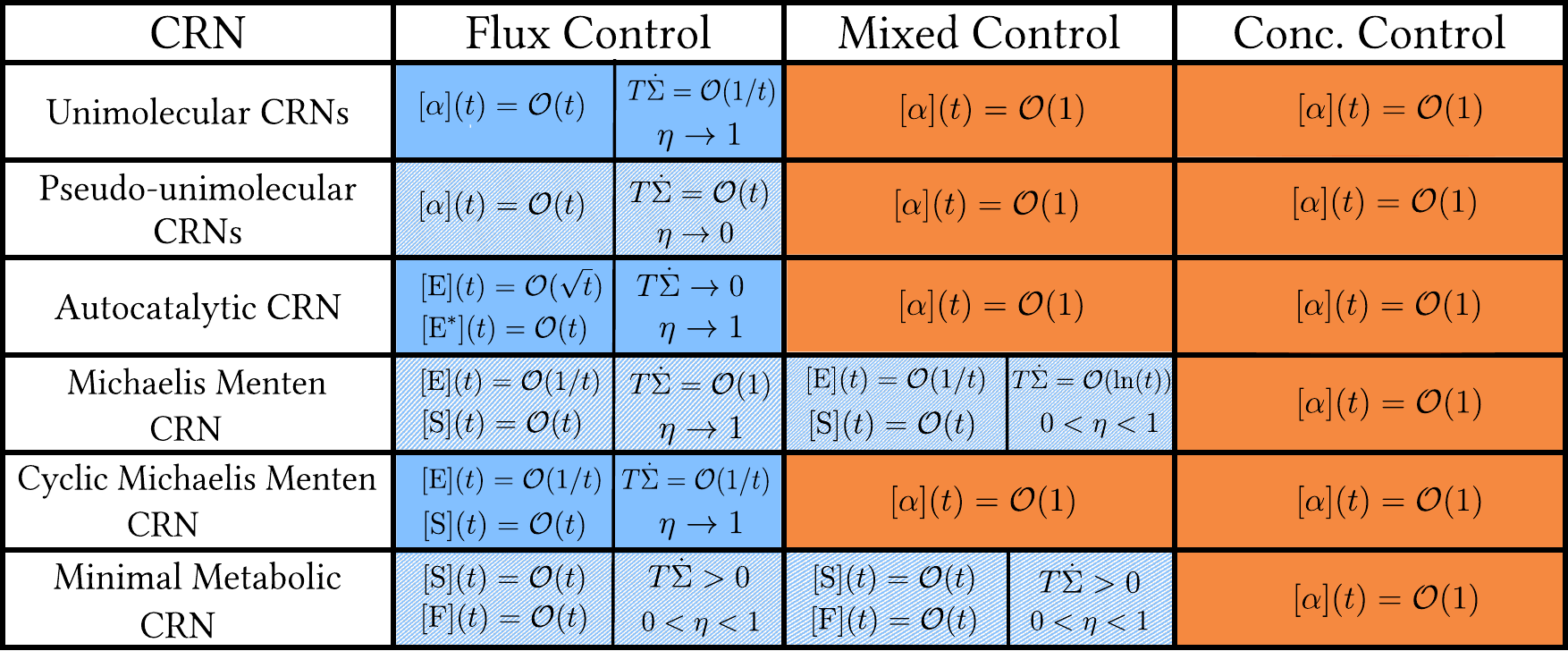}  
    \caption{Long time behaviour of the concentrations, EPR and efficiency under the three different chemostatting procedures for all CRNs. Conditions in orange show no growth. Conditions in striped blue (resp. solid) show nonequilibrium (resp. equilibrium) growth. For multimolecular CRNs, scalings of only a subset of concentrations are depicted. For cases where the exact scaling of the EPR is not known, nonequilibrium (resp. equilibrium) growth is indicated by $T\dot{\Sigma} > 0$ (resp. $T\dot{\Sigma} \to 0$).
} 
\label{fig:results_table}
\end{figure*}

We analyzed growth in open CRNs under three different chemostatting procedures.
A summary of our results is provided in Fig.~\ref{fig:results_table}. 

We first showed the importance of the mass conservation law in analyzing growth. By investigating the dynamics of the mass density, we proved that no CRN can grow in a Continuous-flow Stirred Tank Reactor setup while any CRN can grow under flux control 

We then analyzed unimolecular and pseudo-unimolecular CRNs.
We proved that these CRNs cannot grow under concentration and mixed control and only grow under flux control. 
On one hand, unimolecular CRNs grow close to equilibrium. 
We proved that the dissipation scales as $\mathcal{O}(1/t)$ and that the efficiency of free-energy conversion goes to one.
On the other hand, pseudo-unimolecular CRNs grow far from equilibrium. 
We proved that the dissipation scales as  $\mathcal{O}(t)$ and that the efficiency goes to zero. 

We then studied four multimolecular CRNs with different topological features.
We confirmed numerically and via a fixed point analysis the conjecture that no growth is possible in the Autocatalytic CRN under concentration control. 
We then analyzed growth under flux control.
We showed that multimolecular CRNs display different growth features compared to unimolecular and pseudo-unimolecular CRNs. For instance, some concentrations scale as $\mathcal{O}(\sqrt{t})$ in the Autocatalytic CRN and only a subset of concentrations grow in the Michaelis Menten CRN.
Furthermore, the Autocatalytic CRN and the Cylic Michaelis Menten CRN  grow at equilibrium, while the Michaelis Menten CRN and the Minimal Metabolic CRN grow out of equilibrium.
We finally considered mixed control and found that only the Michaelis Menten CRN and the Minimal Metabolic CRN can grow.  Strikingly, in both  CRNs, some initial conditions relax to steady states while others grow.  When they grow, both CRNs grow out of equilibrium with finite efficiencies.

The analysis of the of the four multimolecular CRNs shows the importance of unbroken conservation laws for nonequilibrium growth.
Indeed,
the presence of unbroken conservation laws is a necessary but not sufficient condition leading to nonequilibrium growth.
On the one hand, they can prevent the equilibration between the chemostatted species and, consequently, create nonconservative forces (like in the Michaelis Menten CRN and the Minimal Metabolic CRN).
On the other hand, 
if there are no unbroken conservation laws 
(like in the Autocatalytic CRN) or
if there are reaction pathways not involving the unbroken conservation laws 
(like in the Cyclic Michaelis Menten CRN), the chemostatted species can equilibrate leading to equilibrium growth.

Our results on unimolecular and pseudo-unimolecular CRNs rigorously support the considerations made in the companion paper \cite{Comp2}. We also analyzed in much greater depth the Autocatalytic CRN and the Michaelis Menten CRN compared to Ref.\cite{Comp2}. 

Our work could be extended in various directions. 
We only considered time-independent chemostatting procedures but  time-dependent ones could produce interesting results.
Sughiyama et al~\cite{Kobayashi2022} studied the thermodynamics of growth in detailed balanced CRNs (which always relax to equilibrium), but their analysis included the effect of volume which we neglected. 
Adding this effect in our analysis can in principle be easily done by introducing an equation of state coupling volume and species concentrations, as done for instance in Refs.~\cite{bigan2015,Kondo2011}. 
Autocatalysis did not play an important role in our analysis of growth. It remains to be seen if autocatalysis may play a more important role on growth in finite time~\cite{unterberger2021}.
Our approach to growth considered a finite number of species and reactions. Exploring chemical growth in CRNs where the number of species can grow as the dynamics evolves (see e.g. \cite{RaoLacoste2015}) is certainly very important.


\begin{acknowledgments}
The authors would like to thank Danilo Forastiere and Gianmaria Falasco for helpful discussions. 
This research was supported by the Luxembourg National
Research Fund (FNR), via the research funding schemes PRIDE (Grant No. 19/14063202/ACTIVE), CORE project ChemComplex (Grant No. C21/MS/16356329), and by project INTER/FNRS/20/15074473 funded by F.R.S.-FNRS (Belgium) and FNR (Luxembourg).
\end{acknowledgments}

\appendix

\section{Eigenvalues of matrix $\mathbb V$}\label{sec:spectral}
The rate equation of unimolecular and pseudo-unimolecular CRNs under mixed and concentration control can be recast in the form of Eq.~\eqref{Eq:linear_general_evol} (see Secs.~\ref{sec:SI_linear_chemo_general} and \ref{Sec:PSL_open_dynamics})
where $\mathbb{V} = \mathbb{W} - \mathbb{D} $ with $\mathbb{W}$ being the rate matrix  and $\mathbb{D}$ being a nonegative diagonal matrix. In this subsection, we prove that all the eigenvalues of $\mathbb{V}$ have negative real parts under mixed and concentration control. The proof is split into two parts.

First, we prove that $\mathbb{V}$ has no null eigenvalue.
Since $\mathbb W$ is a rate matrix, namely, 
$\mathbb W_{\alpha,\beta}\geq0$ and $\mathbb W_{\alpha,\alpha} = - \sum_{\beta\neq\alpha}\mathbb W_{\beta,\alpha} <0$ (see Eqs.~\eqref{Eq:W_def} 
and~\eqref{Eq:W_def_diagonal}) and $\mathbb D$ is a nonnegative diagonal matrix, 
then $\mathbb{V}$ is irreducible as ${\mathbb{W}}$ is irreducible, and 
$\sum_{\beta\neq \alpha} |\mathbb{V}_{\beta,\alpha}| \leq |\mathbb{V}_{\alpha,\alpha}|$ for all columns of $\mathbb{V}$ with a strict inequality for at least one $\alpha$ (where $\mathbb{D}$ is nonzero).
Because of these properties, $\mathbb{V}$ is an irreducible weakly diagonally dominant matrix that is known to be nonsingular~\cite{hornjohnson, Taussky}. Hence, matrix $\mathbb{V}$ has no zero eigenvalues unlike $\mathbb{W}$.

Second, we prove that $\mathbb{V}$ can only admit eigenvalues with  non-positive real parts.
To do so, we use Gershgorin's theorem~\cite{hornjohnson} which states that the eigenvalues of $\mathbb{V}$ can always be found in the union of discs in the complex plane defined as follows. For every $\alpha$, there is a disc centered on the negative real axis at $\mathbb{V}_{\alpha,\alpha}<0$ with radius $\sum_{\beta \neq \alpha} |\mathbb{V}_{\beta,\alpha}|$. Since $\sum_{\beta\neq \alpha} |\mathbb{V}_{\beta,\alpha}| \leq |\mathbb{V}_{\alpha,\alpha}|$, the discs are either fully contained in the left half of the complex plane or can be (at most) tangential to the origin. Therefore all eigenvalues of $\mathbb{V}$ have nonpositive real parts.

The final implication of the two arguments is that all the eigenvalues of $\mathbb{V}$ have negative real parts. Note that this proof did not need $\mathbb{W}$ to be detailed balanced. 

\section{Solutions of Eq.~\eqref{Eq:linear_general_evol}}
For unimolecular and pseudo-unimolecular CRNs, under any chemostatting procedure, the rate equation Eq.~\eqref{eq:req} can be recast in the general form
of Eq.~\eqref{Eq:linear_general_evol} (see Secs.~\ref{sec:SI_linear_chemo_general} and \ref{Sec:PSL_open_dynamics}) where $\boldsymbol{a}$ is the vector of dynamical variables, $\mathbb{V}$ is an appropriately chosen (real) matrix and $\bar{\boldsymbol{I}}_{a}$ is a constant vector. Formally, the solution of Eq.~\eqref{Eq:linear_general_evol} with the initial condition $\boldsymbol{a}_{0}$ is given by
\begin{equation}\label{Eq:linear_general_sol_matrix_exp}
    \boldsymbol{a}(t) = e^{\mathbb{V}t} \boldsymbol{a}_{0} + \int_{0}^{t} ds\, e^{\mathbb{V}(t-s)}\bar{\boldsymbol{I}}_{a}\,,
\end{equation}
 where the integral on the right-hand side of Eq.~\eqref{Eq:linear_general_sol_matrix_exp} satisfies
\begin{equation}\label{Eq:matrix_int_val}
    \mathbb{V} \int_{0}^{t} ds\, e^{\mathbb{V}(t-s)} = e^{\mathbb{V}t} - \mathbb{1}\,.
\end{equation}


The specific dynamical behavior described by the formal solution~\eqref{Eq:linear_general_sol_matrix_exp} depends on the spectral properties of $\mathbb V$.
First, 
under mixed or concentration control, $\mathbb{V}$ is invertible since all eigenvalues have negative real part (see App.~\ref{sec:spectral}).
This implies that $\boldsymbol{a}(t)$ relaxes to a steady state as we prove in App.~\ref{Sec:Appendix_I}.
Second, 
under flux control, $\mathbb{V}$ coincides with the rate matrix $\mathbb W$ and is therefore non-invertible with a unique zero eigenvalue (see Secs.~\ref{sec:SI_linear_chemo_general} and \ref{Sec:PSL_open_dynamics}). 
If $\mathbb{V}$ is diagonalizable (which is always the case if $\mathbb W$ is detailed balanced), $\boldsymbol{a}(t)$ will grow as shown in Secs.~\ref{sec:linear_flux_control_SI} and \ref{subss:pseudo_dynamics}.
If $\mathbb{V}$ is nondiagonalizable {(which is possible when $\mathbb W$ is nondetailed balanced~\cite{andrieux2011spectral,Polettini2014fisher})}, $\boldsymbol{a}(t)$ still grows as we prove in App.~\ref{Sec:Appendix_NI}.

\subsection{Invertible Case}\label{Sec:Appendix_I}
We consider here the case of invertible $\mathbb V$ 
(corresponding to unimolecular and pseudo-unimolecular CRNs under mixed or concentration control). 
Since all the eigenvalues of $\mathbb{V}$ have negative real parts (see App.~\ref{sec:spectral}),
\begin{equation}\label{Eq:lim_mat_V}
    \lim_{t \to \infty}e^{\mathbb{V}t} \boldsymbol{a}_{0} = \boldsymbol{0}\,,
\end{equation}
for any initial condition $\boldsymbol{a}_{0}$ \cite{Perko2014}. Since $\mathbb{V}$ is invertible, Eq.~\eqref{Eq:matrix_int_val} implies that the general solution~\eqref{Eq:linear_general_sol_matrix_exp} reduces to
\begin{equation}
   \boldsymbol{a}(t) = e^{\mathbb{V}t} \boldsymbol{a}_{0} + \mathbb{V}^{-1}\left(e^{\mathbb{V}t} - \mathbb{1}\right) \bar{\boldsymbol{I}}_{a} \xrightarrow[t \to \infty]{} - \mathbb{V}^{-1} \bar{\boldsymbol{I}}_{a}  \,,
\end{equation}
where the second limit uses Eq.~\eqref{Eq:lim_mat_V}.
Hence, in the long time limit, $\boldsymbol{a}(t)$  relaxes to the steady state $-\mathbb{V}^{-1}\bar{\boldsymbol{I}}_{a}$.

\subsection{Noninvertible Case}\label{Sec:Appendix_NI}
We consider here the case of noninvertible nondigaonalizable $\mathbb V$ 
(corresponding to pseudo-unimolecular CRNs under flux control). 
From the Perron Frobenius theorem, its largest eigenvalue is unique and equal to $0$. The corresponding right and left null eigenvectors will be denoted as $\boldsymbol{\pi}$ and $\ell^{m} = (1,1,\dots)^{\intercal}$, respectively (as done in Sec.~\ref{Sec:linear_closed}). All other eigenvalues have negative real parts.

We start by converting $\mathbb V$ into its Jordan normal form by means of the invertible matrix $\mathbb{P}$ of \textit{generalized} eigenvectors~\cite{Perko2014}: 
\begin{equation}
    \mathbb \Lambda = \mathbb{P}^{-1}\mathbb V \mathbb P\,, 
\end{equation}
where the matrix $\mathbb \Lambda$ has the block diagonal form
\begin{equation}\label{Eq:block_form_mat}
\Lambda = \left( \begin{array}{ccccc}
0 &  &  &  &  \\
 & \mathbb B_{2} &  &  &  \\
 &  & \ddots & &  \\
 &  &  & \mathbb B_{r-1} &  \\
 &  &  &  & \mathbb B_{r} \\
\end{array} \right)\,.
\end{equation}
The block matrices $\mathbb B_{k}$ take the form
\begin{equation}\label{Eq:Block_form_1}
  \mathbb B_{k} = \left( \begin{array}{ccccc}
\lambda_{k} & 1  &  &  &  \\
 & \lambda_{k} & 1  &  &  \\
 &  & \ddots & &  \\
 &  &  & \lambda_{k} & 1 \\
 &  &  &  & \lambda_{k} \\
\end{array} \right)\,,
\end{equation}
when the corresponding eigenvalue $\lambda_{k}$ is real, and the form
\begin{equation}\label{Eq:Block_form_2}
  \mathbb B_{k} = \left( \begin{array}{ccccc}
\Omega_{k}
& \mathbb{1}_{2}
&  &  &  \\
 &  \Omega_{k} & \mathbb{1}_{2}  &  &  \\
 &  & \ddots & &  \\
 &  &  & \Omega_{k} & \mathbb{1}_{2} \\
 &  &  &  & \Omega_{k} \\
\end{array} \right)\,,
\end{equation}
when the corresponding eigenvalue is complex $\lambda_{k} = \kappa_{k} + \iota\omega_{k}$. Here, $\mathbb {1}_{2}$ is the 2x2 identity matrix, 
\begin{equation}
    \Omega_{k} = \begin{pmatrix}
    \kappa_{k} & -\omega_{k}\\
    \omega_{k} & \kappa_{k} 
\end{pmatrix}\,,
\end{equation} 
and $\lambda_{k} < 0$ and $\kappa_{k}<0$. Note that the matrix $\mathbb{B}_{k}$ in Eq.~\eqref{Eq:Block_form_1} (resp. Eq.~\eqref{Eq:Block_form_2}) is invertible with all eigenvalues being $\lambda_{k}$ (resp. $\kappa_{k} + \iota \omega_{k}$).

We now make a linear transformation into the space of generalized eigenvectors $\boldsymbol{u} = \mathbb{P}^{-1}\boldsymbol{a}$. Correspondingly, Eq.~\eqref{Eq:linear_general_evol} becomes
\begin{equation}\label{Eq:gen_eigen_ref2}
   d_{t}\boldsymbol{u} = \Lambda \boldsymbol{u} + \mathbb{P}^{-1}\bar{\boldsymbol{I}}_{a}\,.
\end{equation}
The first component $u_{1}(t)$ grows linearly in time as
\begin{equation}\label{Eq:first_component}
    d_{t}u_{1} = (\mathbb{P}^{-1}\bar{\boldsymbol{I}}_{a})_{1} = \sum_{m}(\bar{{I}}_{a})_{m}\,,
\end{equation}
where the second equality used the fact that the left null eigenvector $\ell^{m} = (1,1,\dots)^{\intercal}$ is the first row of $\mathbb{P}^{-1}$.
The remaining components of $\boldsymbol{u}$ can be grouped into (sub)vectors $\boldsymbol{u}_{k}$ such that each (sub)vector evolves as
\begin{equation}\label{Eq:component_gen_eigen2}
    d_{t}\boldsymbol{u}_{k} = \mathbb{B}_{k}\boldsymbol{u}_{k} + (\mathbb{P}^{-1}\bar{\boldsymbol{I}}_{a})_{k}\,.
\end{equation}
Note the similarity between Eqs.~\eqref{Eq:component_gen_eigen2} and \eqref{Eq:linear_general_evol}. 
Furthermore, since each $\mathbb{B}_{k}$ is invertible with all eigenvalues having negative real parts, the results of Appendix~\ref{Sec:Appendix_I} apply. Thus, the components $\boldsymbol{u}_{k}$ in Eq.~\eqref{Eq:component_gen_eigen2} relax to the values
\begin{equation}\label{Eq:other_components}
    \boldsymbol{u}_{k}(t) \xrightarrow[t \to \infty]{}  - \mathbb{B}_{k}^{-1}(\mathbb{P}^{-1}\bar{\boldsymbol{I}}_{a})_{k}\,.
\end{equation}
By combining Eqs.~\eqref{Eq:first_component} and~\eqref{Eq:other_components} with $\boldsymbol{a} = \mathbb{P}\boldsymbol{u}(t)$, at sufficiently long times,
\begin{equation}\label{Eq:final_sol_JB}
    \boldsymbol{a}(t) = \boldsymbol{\pi}u_{1}(t) + \underbrace{\sum_{k>1} \mathbb{P}_{k}\left(-\mathbb B_{k}^{-1}\left((\mathbb{P}^{-1}\bar{\boldsymbol{I}}_{a})_{k}\right)\right)}_{= \boldsymbol{c}(\overline{\boldsymbol I})}\,,
\end{equation}
where we used the fact that the first column of $\mathbb{P}_{1} = \boldsymbol{\pi}$ and $\mathbb{P}_{k}$ refers to the (sub)matrix formed from the columns of $\mathbb{P}$ corresponding to the $k$-th Jordan Block. By comparing Eqs.~\eqref{Eq:final_sol_JB} and \eqref{Eq:flux_control_PSL_general_soln}, we observe that the only change due to the  nondiagonalizability of $\mathbb V$ is in the form of the offset $\boldsymbol{c}(\overline{\boldsymbol I})$.
Hence, in the long time limit, $\boldsymbol a(t)$ grows.


\section{Broken Conservation Laws in Fully Open Pseudo-unimolecular CRNs}
\label{Sec:PSL_open_topology}
When one of the $Z_l$ species is chemostatted the conservation law $\boldsymbol{\ell}^{ml}$ is broken.
Thus, the set of potential $Y_p$ (resp. force $Y_f$) species includes both dynamically linear and hidden species, i.e.,  $Y_p = {Y_{pl}} \cup Y_{ph}$ (resp. $Y_f = {Y_{fl}} \cup Y_{fh}$).
As for unimolecular CRNs (see Subs.~\ref{sec:linear_thermo_growth_SI}) the set $Y_{pl}$ includes only one species.
Hence,  matrix $\mathbb L$ in Eq.~\eqref{Eqn:CLaw:PSL} { is equivalent to the matrix $\mathbb{L}^{b}$ (introduced in Sec.~\ref{sec:SI_thermo_intro}) as it }collects now only broken conservation laws and can be written as
\begin{equation}\label{Eq:Broken_claw_PSL}
\renewcommand*{\arraystretch}{1.5}
    \mathbb{L}^{b} = \kbordermatrix{
    &\color{g}X & \color{g}{Y}_{{pl}} & \color{g}{Y}_{fl} &\color{g}{Y}_{ph} &\color{g}{Y}_{fh} \\   
     &\boldsymbol{\ell}^{ml}_{X} & \ell^{ml}_{pl} &\boldsymbol{\ell}^{ml}_{fl}  & \boldsymbol{0} & \boldsymbol{0}\\
    &\Tilde{\mathbb{L}}_{X} & \Tilde{\boldsymbol{\ell}}_{{p{l}}} & 
    \Tilde{\mathbb{L}}_{{f{l}}} &
    \Tilde{\mathbb{L}}_{{p{h}}} & 
    \Tilde{\mathbb{L}}_{{f{h}}}}\,.  
\end{equation}
From Eq.~\eqref{Eq:Broken_claw_PSL}, the matrix $\mathbb{L}^{b}_{Y_p}$ and its inverse~\cite{Meyer2001} read
\begin{subequations}
 \begin{equation}
\renewcommand*{\arraystretch}{1.5}
     \mathbb{L}^{b}_{Y_{p}} = 
     \kbordermatrix{    
   & \color{g}{Y}_{{pl}} &\color{g}{Y}_{ph}\\
   &  \ell^{ml}_{pl} & \boldsymbol{0}\\
   &\Tilde{\boldsymbol{\ell}}_{{p{l}}} &
    \Tilde{\mathbb{L}}_{{p{h}}}}  \,,\label{Eq:Yp_PSL}\\
\end{equation}
\begin{equation}
\renewcommand*{\arraystretch}{1.5}
      (\mathbb{L}^{b}_{Y_{p}})^{-1} = \kbordermatrix{
   & \color{g}{Y}_{pl} &\color{g}{Y}_{ph}\\
   &  (\ell^{ml}_{pl})^{-1} & \boldsymbol{0}\\
   & -(\Tilde{\mathbb{L}}_{{p{h}}})^{-1}\Tilde{\boldsymbol{\ell}}_{{p{l}}}(\ell^{ml}_{pl})^{-1}  &
    (\Tilde{\mathbb{L}}_{{p{h}}})^{-1} 
    }\,.\label{Eq:Yp_inv_PSL}    
\end{equation}
\end{subequations}
By plugging Eqs.~\eqref{Eq:Yp_inv_PSL} and  \eqref{Eq:Broken_claw_PSL} in Eq.~\eqref{Eq:moiety_defn}, the concentration vector of the moieties reads
\begin{equation}\label{Eq:Moiety_PSL_opensetup}
\renewcommand*{\arraystretch}{1.5}
    \boldsymbol{m}
    = \kbordermatrix{
    &\color{g}X & \color{g}{Y}_{{pl}} & \color{g}{Y}_{fl} &\color{g}{Y}_{ph} &\color{g}{Y}_{fh} \\
    & (\ell^{ml}_{pl})^{-1} \boldsymbol{\ell}^{ml}_{X} & 1 & (\ell^{ml}_{pl})^{-1} \boldsymbol{\ell}^{ml}_{fl}  & \boldsymbol{0} & \boldsymbol{0}\\
    &\overline{\mathbb{L}}_{X} & \boldsymbol{0} & 
    \overline{\mathbb{L}}_{fl} &
    \mathbb{I} & 
    \overline{\mathbb{L}}_{fh}} \cdot \boldsymbol{z}\,,
\end{equation}
with
\begin{align}
&\overline{\mathbb{L}}_{X} =  (\Tilde{\mathbb{L}}_{{p{h}}})^{-1}\left(\Tilde{\mathbb{L}}_{X} - (\ell^{ml}_{pl})^{-1}\Tilde{\boldsymbol{\ell}}_{{p{l}}}\boldsymbol{\ell}^{ml}_{X}\right)\,,\\
&\overline{\mathbb{L}}_{fl} =  (\Tilde{\mathbb{L}}_{{p{h}}})^{-1}\left(\Tilde{\mathbb{L}}_{fl} - (\ell^{ml}_{pl})^{-1}\Tilde{\boldsymbol{\ell}}_{{p{l}}}\boldsymbol{\ell}^{ml}_{fl}\right)\,,\\
&\overline{\mathbb{L}}_{fh} = (\Tilde{\mathbb{L}}_{{p{h}}})^{-1}\Tilde{\mathbb{L}}_{{fh}}\;.
\end{align}

\bibliography{biblio}

\end{document}